\documentclass[11pt]{article}

\usepackage[T1]{fontenc}
\usepackage[utf8]{inputenc}

\usepackage{fancyhdr}
\newcommand{\blind}{1}

\usepackage[inline]{enumitem}
\usepackage{multicol,multirow,booktabs}

\usepackage{times}

\usepackage[hang,small,labelfont=bf,up,textfont=it,up]{caption}
\usepackage{subcaption}
\usepackage{amsmath,amsfonts,amsthm,amssymb,bm,dsfont}	
\setlist{nolistsep} 

\usepackage{graphicx}
\usepackage[a4paper,top=2.75cm,bottom=2.75cm,left=1.75cm,right=1.75cm]{geometry}

\usepackage{numprint}
\npthousandsep{,}\npthousandthpartsep{}\npdecimalsign{.}

\newcommand{\titledoc}{NIRVAR: Network Informed Restricted Vector Autoregression}
\newcommand{\titleshort}{NIRVAR: Network Informed Restricted Vector Autoregression}

\providecommand{\keywords}[1]{{\small{\textbf{\textit{Keywords ---}} #1}}}

\usepackage{fontawesome}
\usepackage{natbib}
\usepackage{setspace}
\usepackage{diagbox}
\usepackage{authblk}

\usepackage{mathtools,bigints}
\mathtoolsset{showonlyrefs}

\usepackage[ruled,linesnumbered,noend]{algorithm2e}
\SetKwInput{KwInput}{Input}
\let\oldnl\nl
\newcommand{\nonl}{\renewcommand{\nl}{\let\nl\oldnl}}
\makeatother

\usepackage{authblk}
\usepackage{natbib}

\usepackage{graphicx}
\graphicspath{{Fig/}}

\usepackage{amsmath,amssymb,amsthm,amsfonts}
\allowdisplaybreaks
\usepackage{url} 

\newtheorem{theorem}{Theorem}[section]
\newtheorem{proposition}[theorem]{Proposition}%
\newtheorem{remark}{Remark}%
\newtheorem{definition}[theorem]{Definition}

\usepackage{bm}
\usepackage{bbm}
\usepackage{caption, subcaption}
\usepackage{mathtools}
\usepackage{placeins}
\usepackage[T1]{fontenc}
\usepackage{pgfplots}
\pgfplotsset{compat=newest}
\usepackage{standalone}
\usepackage{graphicx}
\usetikzlibrary{pgfplots.dateplot}
\mathtoolsset{showonlyrefs}
\usepackage{booktabs}
\usepackage{appendix}
\usepackage[colorlinks=true,allcolors=blue]{hyperref}
\DeclareMathOperator{\supp}{supp}
\DeclareMathOperator{\plim}{plim}
\DeclareMathOperator{\vecc}{vec}
\DeclareMathOperator{\rank}{rk}
\newcommand{\indfun}{\mathbbm{1}}
\DeclareMathOperator*{\argmin}{arg\,min}

\title{NIRVAR: Network Informed Restricted Vector Autoregression}
\author[1,2]{Brendan Martin}
\author[2]{Francesco Sanna Passino}
\author[1,3,4,5]{Mihai Cucuringu}
\author[2,6]{Alessandra Luati}

\affil[1]{Department of Mathematics, University of California, Los Angeles, United States}
\affil[2]{Department of Mathematics, Imperial College London, United Kingdom}
\affil[3]{Department of Statistics, University of Oxford, United Kingdom}
\affil[4]{Oxford-Man Institute of Quantitative Finance, University of Oxford, United Kingdom}
\affil[5]{The Alan Turing Institute, London, United Kingdom}
\affil[6]{Department of Statistical Sciences, University of Bologna, Italy}

\date{}

\title{\LARGE\textbf{\titledoc}}

\allowdisplaybreaks

\newcommand\blfootnote[1]{%
  \begingroup
  \renewcommand\thefootnote{}\footnote{#1}%
  \addtocounter{footnote}{-1}%
  \endgroup
}

\usepackage{multibib}
\newcites{SM}{Supplementary references}

\begin{document}

\maketitle


\blfootnote{
\noindent Corresponding author: Brendan Martin -- \faEnvelopeO\ \texttt{brendan@math.ucla.edu}
}

\begin{abstract}
High-dimensional panels of time series often arise in finance and macroeconomics, where co-movements within groups of panel components occur. 
Extracting these groupings from the data provides a coarse-grained description of the complex system in question and can inform subsequent prediction tasks. We develop a novel methodology to model such a panel as a restricted vector autoregressive process, where the coefficient matrix is the weighted adjacency matrix of a stochastic block model. This network time series model, which we call the Network Informed Restricted Vector Autoregression (NIRVAR) model, yields a coefficient matrix that has a sparse block-diagonal structure. We propose an estimation procedure that embeds each panel component in a low-dimensional latent space and clusters the embedded points to recover the blocks of the coefficient matrix. Crucially, the method allows for network-based time series modelling when the underlying network is unobserved. We derive the bias, consistency and asymptotic normality of the NIRVAR estimator. Simulation studies suggest that the NIRVAR estimated embedded points are Gaussian distributed around the ground truth latent positions. On three applications to finance, macroeconomics, and transportation systems, NIRVAR outperforms competing models in terms of prediction and provides interpretable results regarding group recovery.


\end{abstract}


\section{Introduction}
Panels of stochastic processes, $\{(X_{1,t},\dots,X_{N,t})^{\prime} \}_{t \in \mathbb{Z}}$, which exhibit co-movements between components are central to many scientific disciplines, including econometrics and macroeconomics. Often, $X_{i,t}$ depends not only on its own past values, but also on the past values of a subset of other panel components,  $\{X_{j,s} : j \subseteq \{1,\dots,N\}, s <t\}$. For large $N$, modelling via the time series vector autoregression (VAR) framework becomes prohibitive as the number of model parameters grows as $O(N^{2})$ and can quickly exceed the number of observations. Techniques from high-dimensional statistics that introduce some form of sparsity or dimensionality reduction are therefore required in this setting. 

This paper proposes a method for inferring a graph structure from multivariate time series data and using it to impose restrictions on the coefficient matrix of a 
VAR model. Since the network determines the restricted VAR, we call the model the Network Informed Restricted VAR (NIRVAR) model. NIRVAR models a panel of multivariate time series as a VAR(1) in which the VAR coefficient matrix $\Phi$ is the weighted adjacency matrix of some random graph $\mathcal{G} = (\mathcal{V},\mathcal{E})$ 
where each of the $N$ vertexes in the node set $\mathcal{V}$ has an associated $d$-dimensional latent position $\bm{\theta}_{i} \in \mathcal{H},\ i \in \mathcal{V}$, with $\mathcal{H}\subseteq\mathbb{R}^d$ being some latent space, and where $\mathcal{E} \subseteq \mathcal{V} \times \mathcal{V}$ is a set of random {edges}.
In particular, $\mathcal{G}$ will be modelled as a stochastic block model (SBM). 
Despite their simplicity, SBMs constitute a powerful class of consistent estimators of any exchangeable random graph model \citep{airoldi2013stochastic}.

Typically, in the random graphs literature, the latent positions are estimated through spectral embedding of the observed adjacency matrix. However, in the settings we consider, the realised graph and its adjacency matrix are unobserved. We therefore propose to construct latent positions directly from the observed time series rather than from an observed adjacency matrix. Once latent positions have been constructed, they are clustered into $K$ communities and these communities are used to enforce zero constraints on the VAR coefficients. In particular, if the constructed latent positions of two panel components $i$ and $j$ belong to  different clusters, then $\Phi_{ij}$ is set to 0. In practice, the network implies a subset VAR model  \citep{lutkepohl2005new} whose unrestricted parameters will be estimated via least-squares. 

The NIRVAR estimation procedure draws inspiration from the workflow rationalised in \citet{whiteley2022statistical}: linear dimensionality reduction via principal component analysis, embedding, clustering, and graph construction. 
In particular, we compute the singular value decomposition (SVD) of concatenated sample covariance matrices, consider the corresponding left singular vectors embedding, and cluster the embedded points via a Gaussian mixture model with $K$ components. The embedding dimension is chosen as the number of eigenvalues of the sample covariance matrix that are larger than the upper bound of the support of the Mar\v{c}enko-Pastur distribution \citep{marchenko1967distribution}.  
A graph with $K$ cliques is then constructed based on a binary allocation of each embedded point to its most probable Gaussian mixture component. 
The estimated communities are reconstructions of the blocks of the SBM. The NIRVAR estimator assumes that the probability $p^{(\text{in})}$ of an edge forming within a block is 1, and the probability $p^{(\text{out})}$ of an edge forming between blocks is 0. If $p^{(\text{out})} > p^{(\text{in})}$, the data generating SBM will have a large proportion of inter-block edges and the NIRVAR estimation framework will not capture any of these edges. We therefore restrict our attention to assortative SBMs \citep[see][for a review of SBMs]{lee2019review}. 
Since the NIRVAR estimator does not recover edges between vertices in different blocks of the data generating SBM, the estimator  will be biased whenever such inter-block edges are present. We derive the closed-form expression for the bias,  prove consistency and asymptotic normality of the estimator, and discuss its asymptotic efficiency. Estimation bias can be mitigated by introducing an $\ell_{1}$ penalty for inter-block parameters. The trade-off with this soft-thresholding approach is an increase in variance of the estimator as well as the need for hyperparameter tuning.
%
%

The core motivation for introducing NIRVAR is twofold: to improve forecasts in high-dimensional settings, and to uncover assortative group structure. First, NIRVAR improves forecasting by positioning itself between two bias-variance extremes: a univariate autoregressive prediction model (high bias) and an unrestricted VAR (high variance). The inter-block sparsity of the NIRVAR estimator reduces variance while bias is introduced whenever inter-block edges are present. Second, uncovering group structure is useful to practitioners as it reveals global properties of the underlying network. Moreover, we show that, in expectation, the block structure can be interpreted as a factor model with $K$ factors encoding the aggregated response in each community.

Each step of the NIRVAR estimation framework can be modified with ease. For instance, we also consider embedding via the SVD of the precision matrix. Although we do not pursue this direction here, one could imagine choosing the embedding dimension using, for example, the profile likelihood method of \citet{zhu2006automatic} or the ScreeNOT method of \citet{donoho2023screenot}. The motivation for using a Gaussian mixture model rather that, say, $K$-means clustering, is due to the literature on random dot product graphs \citep{RubinDelanchy22} which shows that spectral embedding yields uniformly consistent latent position estimates with asymptotically Gaussian error (up to identifiability). 

NIRVAR can model data that is represented by a multiplex network, corresponding to a network containing multiple types of edges, expressed via a graph $\mathcal{G}$ containing multiple layers of connectivity, one layer for each type of edge. 
If there are $Q$ layers, or features, then the time series data we consider consists of $NQ$ univariate series, $\{X_{i,t}^{(q)}\}_{t \in \mathbb{Z}}$, where $q \in \{1,\dots,Q\}$. 
These types of matrix time series have been considered by \citet{chang2023modelling}, who propose a one-pass estimation procedure for a tensor canonical polyadic decomposition model of the matrix time series. \citet{barigozzi2022factor} also employ a tensor-based principal component approach to a factor network autoregressive model. In contrast, to obtain latent embeddings in the case of multiple features, we unfold the tensor and compute the SVD of the concatenated sample covariance matrices across $q$ features. 
This choice is motivated by the stability properties of the unfolded adjacency spectral embedding procedure discussed in   \citet{gallagher2021spectral}. NIRVAR can also accommodate multiple lag orders in the VAR process. We detail how the methodology can be extended to higher lag orders in the supplementary material.

The rest of the paper is organised as follows. A background on SBMs is given in Section~\ref{sec:background}. The NIRVAR model is defined in Section \ref{sec:model}. Section \ref{sec:est} details our proposed estimation method and formally relates NIRVAR with the underlying weighted stochastic block model. 
Section \ref{sec:asymptotics} derives the finite and asymptotic properties of the NIRVAR estimator. The finite sample properties and the robustness of each stage of the estimation procedure are assessed through an extensive simulation study in Section \ref{sec:sim}. 
Section \ref{sec:applications} illustrates the flexibility of NIRVAR across different tasks and  in comparison with related  network time series models through three applications to financial, macroeconomic and transportation data.
The online supplementary material contains proofs, a discussion of computational complexity, further simulation studies, and additional results from the empirical analyses. All data and code are available in the Github repository \if1\blind{\href{https://github.com/bmartin9/NIRVAR}{\texttt{bmartin9/NIRVAR}}}\fi\if0\blind{\texttt{anonymised\_link}}\fi.

\subsection{Related literature} 
Many variants of the factor modelling approach of \citet{stock2002forecasting} where the large panel of time series are modelled as stemming from a relatively small number of common latent factors have been proposed in statistics and econometrics \citep[see, for example,][]{bai2003, Fan2013}. Sparse regression methods using various regularised estimation procedures have also been proposed as a way to reduce the number of model parameters; 
the Least Absolute Shrinkage and Selection Operator \citep[LASSO; ][]{lasso} is a common example.
Network time series approaches in which each univariate time series is observed on a vertex of a graph are also popular in the literature \citep[see, for example][]{dahlhaus2000graphical,eichler2007granger,ahelegbey2016bayesian}. The graph can be observed or inferred from data, with the edges of the graph encoding the co-dependence structure of the multivariate time series. For example, \citet{zhu2020multivariate} utilise an observed graph to model the multivariate time series.
Often, each edge corresponds to a parameter in the time series model. Therefore, if the graph is sparse, the number of model parameters can be smaller than the number of observations, enabling estimation and then prediction. 
In cases where the graph is inferred from data, it is itself of interest for downstream tasks such as clustering, link prediction, and estimation of the Granger causality relations between panel components \citep{lancichinetti2009community,zhang2018link}. 

There is an extensive literature on incorporating network effects into a VAR 
framework. \citet{zhu2017} and \citet{knight2016modelling} independently propose a VAR model in which the network effect on a panel component is the average of its connected neighbours in an observed network.  \citet{knight2020generalized} propose a Generalised Network Autoregressive (GNAR) model which allows for a time-varying observed network, and show the consistency of generalised least-squares estimation of the model parameters. \citet{fan2023bridging} combine the dimensionality reduction of factor modelling with the parsimony of sparse linear regression and develop a hypothesis testing framework to determine the partial covariance structure. \citet{barigozzi2023fnets} extend this approach to the setting of dynamic factors, and propose an $L_{1}$-regularised Yule-Walker method for estimating a factor adjusted, idiosyncratic VAR model. The estimated VAR coefficients are then used for network estimation. \citet{chen2023community} introduce a network VAR model 
similar to that of \citet{zhu2017} but allows for network effects between groups of panel components, 
assuming that the network adjacency matrix is observable and generated by a stochastic block model \citep[SBM;][]{holland1983stochastic}. 
\citet{gudhmundsson2021detecting} introduce a stochastic block VAR model in which the time series are partitioned into latent groups such that the spillover effects are determined by a SBM. A group detection algorithm is developed in which the VAR coefficients are estimated with the method of least-squares and used to obtain an embedding on which $K$-means clustering can be applied. 

Our proposed method differs from existing approaches in that it first detects latent groups from a latent space representation of each time series, and then estimates the VAR coefficients. Carrying out estimation in this order is a key contribution of this paper. We avoid estimating a dense VAR parameter matrix as in \citet{gudhmundsson2021detecting} or specifying tuning and thresholding parameters as in \citet{barigozzi2023fnets} and penalised regression methods such as LASSO. We also emphasise that, in contrast to \citet{zhu2017}, \citet{knight2020generalized}, \citet{chen2023community} and \citet{barigozzi2022factor}, the NIRVAR estimator does not require the underlying network to be observable, which is a realistic 
setting in many real-world applications.

\section{Background and notation}\label{sec:background}
In this section, we define the random graphs that will be used throughout the paper. We employ the random dot product graph \citep[RDPG;][]{athreya2017statistical} representation of the {SBM} since our subsequent estimation method involves constructing latent positions for each node, which are then clustered in order to recover the blocks of the {SBM}, providing a convenient mathematical framework. 

\begin{definition}[
Directed RDPG with distribution $F$ and self-loops]\label{def:RDPG} 
Let $F$ be a $d$-dimensional 
distribution with support $\supp(F) = \mathcal{H} \subset \mathbb{R}^{d}$, such that $\bm{x}^{\prime} \bm{y} \in [0,1]$ for all $\bm{x}, \bm{y} \in \mathcal{H}$. 
Let $\bm{\theta}_{1},\dots,\bm{\theta}_{N} \sim F $ be the rows of the matrix $\Theta = (\bm{\theta}_{1},\dots,\bm{\theta}_{N})^{\prime} \in \mathbb{R}^{N \times d}$ and $A$ a 
random adjacency matrix such that, conditional on $\Theta$:
\begin{align}
    \label{eq:rdpg-def}
    p(A\mid\Theta) = \prod_{i=1}^N\prod_
    {\substack{j=1 \\ j\neq i}}^N 
    (\bm{\theta}_{i}^{\prime}\bm{\theta}_{j})^{A_{ij}}(1 - \bm{\theta}_{i}^{\prime}\bm{\theta}_{j})^{1 - A_{ij}} \prod_{h=1}^N \indfun\{A_{hh}=1\},
\end{align}
where $p(\cdot)$ represents a probability mass function. 
Then we say that $A$ is the adjacency matrix of a directed RDPG with self-loops of rank at most $d$ and with latent positions given by the rows of $\Theta$, written $(A,\Theta)\sim\mathrm{RDPG}(F)$.
\end{definition}

\begin{remark}
    In contrast to the definition given by \citet{athreya2017statistical}, Definition \ref{def:RDPG} does not restrict $A$ to be symmetric and with principal diagonal elements equal to zero. 
    This is to allow for directed graphs with self loops. 
    With a slight abuse of notation, RDPG will be used in the rest of this work to denote a directed random dot product graph with self-loops.
\end{remark}

\begin{definition}[$K$-block SBM]\label{def:SBM}
    \label{eq:SBM-def}
    We say that $(A,\Theta)\sim\mathrm{RDPG}(F)$ is  
    a SBM with $K$ blocks or communities if the number of distinct rows in $\Theta$ is $K$. We define the block membership function $z : [N] \mapsto [K]$ such that $z_{i} = z_{j}$ if and only if $\bm{\theta}_{i} = \bm{\theta}_{j}$ (where $z(i) \equiv z_{i}$). Under this representation, $F(\bm{\theta})=\sum_{k=1}^K \pi_k\delta({\bm{\theta} - \bm{\nu}_k})$, where $\bm{\nu}_k\in\mathbb R^d, k\in[K]$ are community-specific latent positions chosen such that $\bm{\nu}_k^\prime\bm{\nu}_\ell \in [0,1]$ for all $k,\ell\in[K]$, and $\pi=(\pi_1,\dots,\pi_K)$ represent the prior probabilities of each node to belong to the $k$-th community, with $\pi_k\geq0$ for all $k\in[K]$ and $\sum_{k=1}^K\pi_k=1$. Here, $\delta$ is the $d$-dimensional Dirac delta function. The between-community connection probabilities can be collected in a matrix $B\in[0,1]^{K\times K}$ with $B_{k\ell}=\bm{\nu}_{k}^\prime\bm{\nu}_{\ell},\ \bm{\nu}_{k},\bm{\nu}_{\ell}\in\mathbb R^d$ for $k,\ell=1,\dots,K$, and we write $(A,\Theta)\sim\mathrm{SBM}(B,\pi)$.
    
\end{definition}

Definition~\ref{eq:SBM-def} only covers the case in which the matrix $B\in[0,1]^{K\times K}$ of between-block connection probabilities is positive semi-definite, corresponding to an assortative graph.
%
%
The motivation for this assumption is to have a multivariate time series model in which there is greater co-movement between panel components within the same community than between communities. This is common in many applications: for example, in a financial prediction context, assortative groupings can be utilised for a generalised pairs trading strategy \citep{vidyamurthy2004pairs}. The generalised RDPG \citep{RubinDelanchy22} provides an extension to the indefinite case. 
We also assume that each feature-specific graph  $\mathcal{G}^{(q)}=(\mathcal{V},\mathcal{E}^{(q)})$ with adjacency matrix $A^{(q)}$ has an associated latent position matrix $\Theta^{(q)}$ such that 
\begin{equation}
    (A^{(q)},\Theta^{(q)})\sim\mathrm{SBM}(B^{(q)},\pi^{(q)}),\quad q=1,\dots,Q,
    \label{eq:rdpg-A}
\end{equation} 
where $B^{(q)}\in[0,1]^{K\times K}$ and $\pi^{(q)}$ are feature-specific {SBM} parameters. 
The probability of an edge forming between vertices $i$ and $j$, $i\neq j$, is 
$   p_{ij}^{(q)} = B_{z_{i}z_{j}}^{(q)} = \bm{\theta}_{i}^{(q)^{\prime}} \bm{\theta}_{j}^{(q)}$, 
and the adjacency matrix corresponding to $\mathcal{G}^{(q)}$ is 
\begin{align}
   \label{eq:rdpg-A-p}
   A_{ij}^{(q)} \sim \text{Bernoulli}\left(p_{ij}^{(q)}\right).
\end{align}

\subsection{Notation}
For an integer $k$, let $[k]$ denote the set $\{1,\dots,k\}$. The $n \times n$ identity matrix is written $I_{n}$, and the indicator function is $\indfun\{\mathcal B\} = 1$ if the event $\mathcal{B}$ occurs and $0$ otherwise. For vectors, $\bm{v}_{1},\dots,\bm{v}_{n} \in \mathbb{R}^{p}$, let $M = (\bm{v}_{1},\dots,\bm{v}_{n}) \in \mathbb{R}^{p \times n}$ be the matrix whose columns are given by $\bm{v}_{1},\dots,\bm{v}_{n}$. The column space of $M$ is $\mathrm{colsp}(M)$, its rank is $\rank(M)$, and its transpose is $M^{\prime}$. We write $M_{i,:}$ and $M_{:,j}$ to denote the $i$-th row and $j$-th column of $M$ considered as vectors, respectively. The element-wise $\ell_{0}$ norm is defined as $|M|_{0} = \sum_{i=1}^{p} \sum_{j=1}^{n} \mathbbm{1}{\{ M_{ij} \neq 0 \}} $. The Frobenius norm is $\lVert M \rVert_{F} = ({\sum_{i=1}^{p} \sum_{j=1}^{n} |M_{ij}|^{2}})^{1/2}$. The spectral radius of an $ n \times n$ matrix $A$ is $\rho(A) = \max\{|\lambda_{1}|,\dots,|\lambda_{n}|\}$ where $\lambda_{1},\dots,\lambda_{n}$ are the eigenvalues of $A$. The vectorisation operator $\vecc(\cdot)$ transforms a $p \times n$ matrix $M$ into a $pn \times 1$ vector by stacking its columns. Let $\odot$ denote the Hadamard product between two matrices of the same dimensions, and $\otimes$ denote the Kronecker product between two matrices, not necessarily of the same dimension. For matrices $M_{1},\dots,M_{r}$, let $(M_{1}|\cdots|M_{r})$ and $(M_{1};\dots;M_{r})$ denote the column-wise and row-wise concatenation of the matrices, respectively. The set of $n \times d$ matrices with orthogonal columns is $\mathbb{O}(n \times d)$. For a distribution $F$ the support of $F$ is $\supp(F)$.

\section{Model} \label{sec:model}

Let $(X_{i,t}^{(q)})$ be a matrix time series consisting of $NQ$ random variables at each time $t \in \mathbb{Z}$. For example, $i \in [N]$ could label an individual and $q \in [Q]$ a particular variable or feature associated with that individual. Consider an associated multiplex network with $Q$ layers each having $N$ vertices. The random variable $X_{i,t}^{(q)}$ is observed at vertex $i$ of layer $q$ of the multiplex network. 
For each feature $q \in [Q]$ we associate a random graph $\mathcal{G}^{(q)} = (\mathcal{V},\mathcal{E}^{(q)})$ where $\mathcal{V} = [N]$. $\mathcal{G}^{(q)}$ is modelled as a {SBM}. Given $\mathcal{G}^{(q)}$, we define the NIRVAR model as follows.
\begin{definition}[NIRVAR model] \label{def:nirvar} 
    For some fixed $q\in[Q]$, let $\{\bm{X}_{t}^{(q)}\}_{t \in \mathbb{Z}}$ denote a zero mean, second order stationary stochastic process where $\bm{X}_{t}^{(q)} = (X_{1,t}^{(q)},\cdots,X_{N,t}^{(q)})^{\prime} \in \mathbb{R}^{N}$ and $q \in [Q]$. The NIRVAR model for the $q$-th feature is 
    \begin{align}
    \label{eq:nirvar-var}
    \bm{X}_{t}^{(q)} = \sum_{r=1}^{Q} (A_{q}^{(r)} \odot \Tilde{\Phi}_{q}^{(r)}) \bm{X}_{t-1}^{(r)} + \bm{\epsilon}_{t}^{(q)},
\end{align}
    in which the generic element of $A_{q}^{(r)}, r\in[Q]$ is given by Equation~\eqref{eq:rdpg-A-p} and $\Tilde{\Phi}_{q}^{(r)}, r \in [Q]$ is an $N \times N$ matrix of fixed weights. Defining $\Phi_{q}^{(r)} \coloneqq A_{q}^{(r)} \odot \Tilde{\Phi}_{q}^{(r)}$ and $\Phi_{q} \coloneqq (\Phi_{q}^{(1)}|\cdots|\Phi_{q}^{(Q)})$, we write $\bm{X}_{t}^{(q)} \sim \mathrm{NIRVAR}(\Phi_{q})$. The noise process is assumed to be white noise with $\bm{\epsilon}_{t}^{(q)}$ being continuous random vectors satisfying $\mathbb{E}(\bm{\epsilon}_{t}^{(q)}) = 0$, $\Sigma = \mathbb{E}\{\bm{\epsilon}_{t}^{(q)}(\bm{\epsilon}_{t}^{(q)})^{\prime}\}$ is non-singular, $\bm{\epsilon}_{t}^{(q)}$ and $\bm{\epsilon}_{s}^{(r)}$ are independent for $t \neq s$ and $q \neq r$, and for some finite constant $c$, $\mathbb{E}|\epsilon_{i,t}^{(q)}\epsilon_{j,t}^{(q)}\epsilon_{l,t}^{(q)}\epsilon_{m,t}^{(q)} | \leq c$ for all $t \in \mathbb{Z}$, $q \in [Q]$ and $i,j,l,m \in [N]$. 
\end{definition}
If, for each $q \in [Q]$, $\bm{X}_{t}^{(q)} \sim \mathrm{NIRVAR}(\Phi_{q})$, then $\bm{X}_{t} \mathrel{\mathop:}= (\bm{X}_{t}^{(1)^{\prime}},\dots,\bm{X}_{t}^{(Q)^{\prime}})^{\prime} \in \mathbb{R}^{NQ}$ is a VAR(1) process with coefficient matrix $\Xi \coloneqq (\Phi_{1};\dots;\Phi_{Q}) \in \mathbb{R}^{NQ \times NQ}$. Therefore, $\bm{X}_{t}^{(q)}$ will be stable, and hence stationary, if $\rho(\Xi) < 1$ \citep{lutkepohl2005new}. We also point out that cross-feature correlation is not precluded by Definition \ref{def:nirvar}; in this case, we would consider a single layer network with $NQ$ vertices.
\begin{remark}
    In Definition~\ref{def:nirvar}, we define the NIRVAR model only for the $q$-th feature, which acts as the response or endogenous variable, whereas the features $\mathcal C_{-q}=[Q]\setminus\{q\}$ are interpreted as covariates or exogenous variables. 
    Since we focus on the response of the $q$-th feature, we drop the subscript on $A_{q}^{(r)}$, $\Tilde{\Phi}_{q}^{(r)}$, $\Phi_{q}^{(r)}$, and $\Phi_{q}$ for the remainder of the paper.
\end{remark}

It will be convenient when discussing estimation of the NIRVAR model to write Equation~\eqref{eq:nirvar-var} in vectorised form. For this purpose, we recall $\bm{X}_{t} \mathrel{\mathop:}= (\bm{X}_{t}^{(1)^{\prime}},\dots,\bm{X}_{t}^{(Q)^{\prime}})^{\prime}$ and define
\begin{equation}
    \label{eq:var-notation}
    Y^{(q)} \mathrel{\mathop:}= \left(\bm{X}_{1}^{(q)},\dots,\bm{X}_{T}^{(q)}\right),  \;\; X \mathrel{\mathop:}= (\bm{X}_{0},\dots,\bm{X}_{T-1}),  \;\; U^{(q)} \mathrel{\mathop:}= \left(\bm{\epsilon}_{1}^{(q)},\dots,\bm{\epsilon}_{T}^{(q)}\right), \;\; A \mathrel{\mathop:}= \left(A^{(1)}|\cdots|A^{(Q)} \right). \notag
\end{equation}
We can then write Equation~\eqref{eq:nirvar-var} for $t=1,\dots,T$ as 
$    Y^{(q)} = \Phi X + U^{(q)}$,
or, equivalently, as 
\begin{equation}
    \label{eq:vec-nirvar}
    \bm{y}^{(q)} = \vecc(\Phi X) + \vecc\left(U^{(q)}\right) 
    = \left(X^{\prime} \otimes I_{N} \right)\bm{\beta} + \mathbf{u}^{(q)},
\end{equation}
where $\bm{y}^{(q)} \coloneqq \vecc(Y^{(q)})$, $\bm{\beta} \coloneqq \vecc(\Phi)$, and $\mathbf{u}^{(q)} \coloneqq \vecc(U^{(q)})$.
The number of non-zero elements of $\bm{\beta}$ is given by $M = | A |_{0}$.

The model can be written in terms of an unrestricted $M$-dimensional vector $\bm{\gamma}(A)$ whose elements belong to the set $\{\beta_{i} : \vecc(A)_{i} \neq 0, \mkern9mu i=1,\dots,N^{2}Q \}$. Although this vector depends on $\tilde{\Phi}$, for conciseness we write $\bm{\gamma}(A)$ instead of $\bm{\gamma}(A,\tilde{\Phi})$. Additionally, an $N^{2}Q \times M$ matrix $R(A)$ is defined via the mapping $R : \{0,1\}^{N \times NQ} \to \{0,1\}^{N^{2}Q \times M}$, where
\begin{align}
    \label{eq:R-def}
    [R(A)]_{ij} \coloneqq \vecc(A)_{i} \times \indfun \left\{\sum_{k=1}^{i-1}\vecc(A)_{k} = j-1\right\}.
\end{align}
The constraints on $\bm{\beta}$ can then be written as 
\begin{align}
    \label{eq:linear-constraints}
    \bm{\beta} = R(A)\bm{\gamma}(A).
\end{align}
The matrix $R(A)$ encodes the zero-entry locations of $\vecc(A)$ and is a standard way of representing linear constraints in VAR models \citep[see][equation (5.2.2), for example]{lutkepohl2005new}. Combining Equation~\eqref{eq:vec-nirvar} and  Equation~\eqref{eq:linear-constraints} finally yields 
\begin{align}
    \label{eq:nirvar-R}
    \bm{y}^{(q)} = \left(X^{\prime} \otimes I_{N} \right)R(A)\bm{\gamma}(A) + \mathbf{u}^{(q)}.
\end{align} 
\section{Estimation} \label{sec:est}
The NIRVAR estimation method proceeds by first imposing subset VAR restrictions through a binary matrix, $\hat{A} \in \{0,1\}^{N \times NQ}$, and then estimating $\bm{\gamma}(\hat{A})$. To obtain $\hat{A}$, a low-dimensional latent representation of each panel component is found by embedding the sample covariance matrix in some latent space. Clustering the latent positions then allows for the construction of a graph with adjacency matrix $\hat{A}$. The choice of the sample covariance matrix is motivated by the connection between the covariance matrix $\Gamma$ and the VAR coefficient matrix $\Phi$ proved in Proposition \ref{prop:cov-conn} below. Under the assumptions of the proposition, $\Gamma$, and in turn the sample covariance matrix, both inherit network properties of $\Phi$ in terms of the underlying cluster structure.

The embedding and clustering method used here are in the spirit of \citet{whiteley2022statistical} who 
provide theoretical justification for 
linear dimensionality reduction via principal component analysis, 
a subsequent embedding, and graph construction using the embedded points. The next sections describe how these steps are implemented in our setting.

\subsection{Embedding}\label{subsub:embedding}
Let ${\mathcal{X}}^{(q)} = (\bm{x}_{1}^{(q)},\dots,\bm{x}_{T}^{(q)})$ be the $N \times T$ design matrix of feature $q$, where the vector $\bm{x}_{t}^{(q)} = (x_{1,t}^{(q)},\dots,x_{N,t}^{(q)})^{\prime}$ is a realisation of the random variable $\bm{X}_{t}^{(q)}$. To construct an embedding $\hat{\bm{\psi}}_{i}^{(q)} \in \mathbb{R}^{d}$, we adopt an algebraic construction similar to the unfolded adjacency spectral embedding \citep[UASE,][]{jones2020multilayer}, which 
returns embeddings $\Psi^{(q)} \in \mathbb{R}^{N \times d}$ by considering the SVD of $A = (A^{(1)}|\cdots|A^{(Q)})$. Since we do not observe $A^{(q)}$, we instead 
consider the SVD of $S = (S^{(1)}|\cdots|S^{(Q)}) \in \mathbb{R}^{N \times NQ}$ where $S^{(q)} \coloneqq \mathcal{X}^{(q)} \mathcal{X}^{(q)^{\prime}}/T \in \mathbb{R}^{N \times N}$ is the sample covariance matrix for feature $q$. 
We can write 
$    S = UDV^{\prime} + U_{\perp}D_{\perp}V_{\perp}^{\prime}, $
where $D \in \mathbb{R}^{d \times d}$ is a diagonal matrix containing the $d$ largest singular values of $S$, and the columns of $U \in \mathbb{O}(N \times d)$ and $V \in \mathbb{O}(NQ \times d)$ are the corresponding $d$ left and right singular vectors, respectively. 
Assuming $S$ admits a low-rank approximation, 
we use only $\hat{S} = UDV^{\prime}$ to produce embeddings. In particular, the $q$-th right 
embedding is the matrix $\hat{\Psi}^{(q)} \in \mathbb{R}^{N \times d}$ obtained by dividing $\hat{\Psi} = VD^{1/2} \in \mathbb{R}^{NQ \times d}$ into $Q$ equal blocks, $\hat{\Psi} = (\hat{\Psi}^{(1)};\dots;\hat{\Psi}^{(Q)} )$. The feature-$q$ embedding of time series $i$ is thus $\hat{\bm{\psi}}_{i}^{(q)} = (\hat{\Psi}^{(q)} )_{i,:}$. 
Embedding stacked matrices
has two key stability properties proved by \citet{gallagher2021spectral} in the context of dynamic networks: it assigns the same position, up to noise, to vertices behaving similarly for a given feature (cross-sectional stability) and a constant position, up to noise, to a single vertex behaving similarly across different features (longitudinal stability). 

It is of interest to relate $\hat{\bm{\psi}}_{i}^{(q)}$ to the ground truth latent positions $\bm{\theta}_{i}^{(q)}$, since clustering of $\hat{\bm{\psi}}_{i}^{(q)}$ will be used to recover the blocks of the data generating SBM. We derive a connection between the covariance matrix $\Gamma^{(q)} = \mathbb{E}\{(\bm{X}_{t}^{(q)})(\bm{X}_{t}^{(q)})^{\prime}\}$ of a NIRVAR$(\Phi)$ process and $\bm{\theta}_{i}^{(q)}$ for the case $Q = 1$. In the special case of symmetric $\Phi$, which is a standard assumption in the RDPG literature \citep[see, for example][]{athreya2017statistical}, Proposition \ref{prop:cov-conn} shows that the rank $d$ spectral embeddings of $\Gamma^{(q)}$ and $\Phi$ are equivalent. 
Since we assume $Q=1$ in Proposition \ref{prop:cov-conn}, the superscript $q$ is dropped for readability. 

\begin{proposition}\label{prop:cov-conn}
    Let $\bm{X}_{t} \sim \mathrm{NIRVAR}(\Phi)$ and assume $\Sigma = \sigma^{2}I_{N}$. If $\Phi$ is symmetric with eigendecomposition 
$    \Phi = U_{\Phi} \Lambda_{\Phi} U_{\Phi}^{\prime} + U_{\Phi,\perp}\Lambda_{\Phi,\perp}U_{\Phi,\perp}^{\prime}, $
where $U_{\Phi} \in \mathbb{O}(N \times d)$ and $\Lambda_{\Phi}$ is a $d \times d$ diagonal matrix comprising the $d$ largest eigenvalues in absolute value of $\Phi$, then the rank $d$ truncated eigendecomposition of the covariance matrix $\Gamma = \mathbb{E}(\bm{X}_{t}\bm{X}_{t}^{\prime})$ is $\Gamma = U_{\Phi} \Lambda_{\Gamma} U_{\Phi}^{\prime}$ in which $\Lambda_{\Gamma}$ is a $d \times d$ diagonal matrix with diagonal elements $(\lambda_{\Gamma})_{i} = 1/\{1-(\lambda_{\Phi})_{i}^{2}\}$ where $(\lambda_{\Phi})_{i}$ is the corresponding diagonal entry of $\Lambda_{\Phi}$.
\end{proposition}

Under Proposition \ref{prop:cov-conn}, the eigenvectors corresponding to the $d$ largest eigenvalues of $\Gamma$ and $\Phi$ are the same. Therefore, the rank $d$ spectral embedding of $\Phi$ can be constructed, up to identifiability, from the eigenvectors and scaled eigenvalues of $\Gamma$, where the scaling is given by $(\lambda_{\Phi})_{i} = \pm \sqrt{1-1/(\lambda_{\Gamma})_{i}}$. As such, in the setting of undirected graphs and homoskedastic noise, the block structure of the SBM is contained entirely in $\Gamma$, and information from lagged autocovariance matrices is not required for community detection. 
 
\citet{gallagher2023spectral} prove a central limit theorem for the asymptotic behaviour of $\Psi_{\Phi} = U_{\Phi} \Lambda_{\Phi}^{1/2}$ where $\Phi$ is an observed symmetric matrix sampled from a weighted RDPG. They prove that, up to a sequence of orthogonal transformations, $(\Psi_{\Phi})_{i}$ is asymptotically normally distributed around the ground truth latent positions, $\bm{\theta}_{i}$. 
Proposition \ref{prop:cov-conn} relates $\Psi_{\Gamma} = U_{\Phi} \Lambda_{\Gamma}^{1/2}$ to $\Psi_{\Phi}$ and is a first step in extending existing results (which assume independent data) to the dependent data setting. We show in Section \ref{sec:sim} that, in a simulation setting, $(\Psi_{\Gamma})_{i}$ is indeed asymptotically normally distributed around $\bm{\theta}_{i}$.


\subsection{Clustering}\label{subsec:clustering}
Clustering via a Gaussian mixture model can be used to assign a label $\hat{z}_{i}^{(q)} \in \{1,\dots,K\}$ to each panel component based on $\hat{\bm{\psi}}_{i}^{(q)}$. In particular, we use Expectation-Maximisation to maximise the Gaussian mixture model incomplete log-likelihood and estimate the cluster assignments $\hat{z}_{i}^{(q)}$. We define $\hat{A} = (\hat{A}^{(1)}|\cdots|\hat{A}^{(Q)}) \in \{0,1\}^{N \times NQ}$ as 
$     \hat{A}_{ij}^{(q)} = \indfun\{\hat{z}_{j}^{(q)} = \hat{z}_{i}^{(q)}\}. $
From this definition, it can be seen that $\hat{A}$ defines a graph with $K$ cliques and therefore cannot recover inter-block edges of the data generating SBM. 

We use the following argument to motivate our choice of $K$. For a RDPG, $A_{ij}^{(q)} \sim \text{Bernoulli}(\bm{\theta}_{i}^{(q)^{\prime}} \bm{\theta}_{j}^{(q)})$ and thus $\mathbb{E}(A^{(q)}) = {\Theta^{(q)}} \Theta^{(q)^{\prime}}$, where ${\Theta^{(q)}} = (\bm{\theta}_{1}^{(q)},\dots,\bm{\theta}_{N}^{(q)})^{\prime} \in \mathbb{R}^{N \times d}$. Assuming that $\Theta^{(q)}$ is full rank, then $\text{rank}\{\mathbb{E}(A^{(q)})\} = \text{rank}({\Theta^{(q)}} \Theta^{(q)^{\prime}}) = \text{rank}({\Theta^{(q)}}) = d$. For a SBM, ${\Theta^{(q)}}$ has $K$ distinct rows (corresponding to the $K$ latent positions), and thus $\text{rank}({\Theta^{(q)}}) = K$. Since we are assuming $\mathcal{G}^{(q)}$ is a SBM, we set $K=d$.

Based on comments in \citet{chen2023community}, we note that the NIRVAR model can, in expectation, be viewed as a factor model with $K$ factors encoding the aggregated response in each community. In particular, define $Z^{(q)} \in \{0,1\}^{N \times K}$ by $Z_{ij}^{(q)} \coloneqq \indfun \{z_{i}^{(q)} = z_{j}^{(q)}\}$ and let $\mathbb{E}(\Phi_{ij}^{(q)}\mid z_{i}^{(q)} = k, z_{j}^{(q)} = l) = \breve{B}_{kl}^{(q)}$ where $\breve{B}_{kl}^{(q)} \in \mathbb{R}^{K \times K}$ is the block mean matrix. Then the expected value of the weighted adjacency matrix can be written as $\mathbb{E}(\Phi^{(q)}) = Z^{(r)} \breve{B}^{(r)}(Z^{(r)})^{\prime}$. Replacing $\Phi^{(q)}$ with $\mathbb{E}(\Phi^{(q)})$ in the NIRVAR model yields  
\begin{align}
    \label{eq:community-factors}
    \bm{X}_{t}^{(q)} = \sum_{r=1}^{Q} Z^{(r)} \breve{B}^{(r)}(Z^{(r)})^{\prime}  \bm{X}_{t-1}^{(r)} + \bm{\epsilon}_{t}^{(q)} = \sum_{r=1}^{Q} \Lambda^{(r)} \bm{F}_{t}^{(r)} + \bm{\epsilon}_{t}^{(q)},
\end{align}
where $\bm{F}_{t}^{(q)} \coloneqq (Z^{(q)})^{\prime} \bm{X}_{t-1}^{(q)}$ is a $K$-dimensional vector of community factors and $\Lambda^{(q)} \coloneqq  Z^{(r)} \breve{B}^{(r)}$ are the corresponding loadings. The estimated NIRVAR clusters can therefore be interpreted as being related to community factors. 

\subsection{Parameter estimation}
The subset VAR restrictions we impose correspond to the zero entries of $\hat{A}$. Given $\hat{A}$, there are $\widehat{M} = | \hat{A} |_{0}$ remaining unrestricted parameters that we estimate via least squares. Let $\bm{\gamma}(\hat{A})$ be the $\widehat{M}$-dimensional vector of unrestricted parameters whose elements are those in the set $\{\beta_{i} : \vecc(\hat{A})_{i} \neq 0, \mkern9mu i=1,\dots,N^{2}Q \}$. Let $R(\hat{A}) \in \{0,1\}^{N^{2}Q \times \widehat{M}}$ be the restrictions matrix corresponding to $\hat{A}$ where $R$ is defined by Equation~\eqref{eq:R-def}. 
%
The model corresponding to the estimated restrictions $\hat{A}$ is
\begin{align}
    \label{eq:nirvar-R-hat}
    \bm{y}^{(q)} = \left(X^{\prime} \otimes I_{N} \right)R(\hat{A})\bm{\gamma}(\hat{A}) + \mathbf{u}(\hat A)^{(q)}.
\end{align}
The least squares estimator of $\bm{\gamma}(\hat{A})$ minimises the objective $S\{\bm{\gamma}(\hat{A})\} = \mathbf{u}(\hat A)^{(q) \prime} (I_{T} \otimes \Sigma^{-1}) \mathbf{u}(\hat A)^{(q)}$.
Taking first derivatives and equating to zero yields the generalised least-squares estimator (GLS; see Section \ref{sec:der_gls} of the supplementary material for the derivation)
\begin{equation}
    \label{eq:GLS}
    \hat{\bm{\gamma}}(\hat{A}) = \{R(\hat{A})^{\prime}(XX^{\prime} \otimes \Sigma^{-1}) R(\hat{A})\}^{-1} R(\hat{A})^{\prime} (X \otimes \Sigma^{-1})\, \bm{y}^{(q)}.
\end{equation}
Finally, 
\begin{equation}
    \label{eq:GLS-beta}
    \hat{\bm{\beta}}(\hat{A}) = R(\hat{A}) \hat{\bm{\gamma}}(\hat{A}).
\end{equation}
In the simulation studies in Section \ref{sec:sim} and applications in Section \ref{sec:applications}, we assume that  $\Sigma = \sigma^{2}I_{N}$, implying that all the correlation among variables is explained by the network, in which case the GLS
estimator \eqref{eq:GLS} is the same as $\hat{\bm{\gamma}}_{\mathrm{OLS}}(\hat{A}) = \{R(\hat{A})^{\prime}(XX^{\prime} \otimes I_{N}) R(\hat{A})\}^{-1} R(\hat{A})^{\prime} (X \otimes I_{N})\,\bm{y}^{(q)}.$
The properties of these estimators are derived in Section~\ref{sec:asymptotics}.

\subsection{Determining the embedding dimension}
We employ tools from random matrix theory to estimate $d^{(q)}$, the rank of the RDPG $\mathcal{G}^{(q)}$. Following standard methods in the literature \citep[see, for example,][]{laloux2000random}, our estimate $\hat{d}^{(q)}$ is the number of eigenvalues of $S^{(q)}$ that are larger than the upper bound of the support of the Mar\v{c}enko-Pastur distribution \citep{marchenko1967distribution}. The Mar\v{c}enko-Pastur distribution is the limiting eigenvalue distribution of an $N \times N$ Wishart matrix whose dimension to sample size ratio is a constant, $\eta = N/T$, and is given by 
\begin{align}
f_{\text{MP}}(x;\eta,\sigma^{2}) = 
\sqrt{(x-x_{-})(x_{+}-x)} / (2\pi\sigma^{2} \eta) \indfun\left\{x_{-} \leq x \leq x_{+}\right\},
\end{align}
with $\sigma^{2}$ being the scale parameter and $x_{\pm} = \sigma^{2} (1 \pm \sqrt{\eta})^{2}$ \citep[for a derivation, see][]{bai2010spectral}. If $\lambda_{1}^{(q)},\dots,\lambda_{N}^{(q)}$ are the eigenvalues of $S^{(q)}$, then 
\begin{align}
    \hat{d}^{(q)} = \sum_{j=1}^{N} \indfun\left\{\lambda_{j}^{(q)} > x_{+}\right\}.
\end{align}
The interpretation is that all eigenvalues that are greater than $x_{+}$ cannot be attributed to random noise and are thus deemed as ``informative''.


\subsection{Alternative embeddings}
Alternatives to the sample covariance matrix could be chosen for the embedding procedure described in Section \ref{subsub:embedding}. Here, we discuss embedding the precision matrix, $\Omega^{(q)} = (S^{(q)})^{-1}$. One motivation for the use of the precision matrix is its relation to the partial correlation $\text{Cor}( Y_{i,:}^{(q)} , Y_{j,:}^{(q)} \mid Y_{-(i,j),:}^{(q)}) =\Omega_{ij}^{(q)}/(\Omega_{ii}^{(q)}\Omega_{jj}^{(q)}),
 $ 
where $Y_{-(i,j),:}^{(q)}$ denotes the set of all rows of $Y^{(q)}$ except rows $i$ and $j$. We can thus interpret $\Omega_{ij}^{(q)}$ as being proportional to the remaining correlation between time series $i$ and time series $j$ for feature $q$ after the residual effect of all other time series has been removed. The unfolded adjacency spectral embedding and Gaussian mixture model clustering steps in the estimation procedure remain unchanged using the precision matrix. However, we need to slightly modify our method for estimating $d^{(q)}$. Instead of counting the number of eigenvalues of $S^{(q)}$ that are larger than the upper bound of the support of the Mar\v{c}enko-Pastur distribution, we count the number of eigenvalues of $\Omega^{(q)}$ that are smaller than the lower bound of the support of the inverse Mar\v{c}enko-Pastur distribution, which is the distribution of the reciprocal of a Mar\v{c}enko-Pastur distributed random variable. The inverse Mar\v{c}enko-Pastur distribution, derived in Section \ref{sec:inv_marc} of the supplementary material, is given by
\begin{align}
    \label{eq:imp}
f_{\text{IMP}}(y;\eta,\sigma^{2}) = 
     (1-\eta)\sqrt{(y_{+} - y)(y - y_{-})} / (2 \pi \eta y^{2}) \indfun\left\{ y_{-} \leq y \leq y_{+}\right\}, 
\end{align} 
where 
 $   y_{\pm} = \sigma^{-2} \{(1-\eta)^{-1}(1 \pm \sqrt{\eta})\}^{2}. $
If $\zeta_{1}^{(q)},\dots,\zeta_{N}^{(q)}$ are the eigenvalues of $\Omega^{(q)}$, then our estimate of $d^{(q)}$ when using the precision matrix becomes
\begin{align}
    \hat{d}^{(q)} = \sum_{j=1}^{N} \indfun\left\{\zeta_{j}^{(q)} < y_{-}\right\}.
\end{align}

\section{Asymptotic properties}\label{sec:asymptotics}
The following proposition establishes the conditions under which $\hat{\bm{\beta}}(\hat{A})$ is an unbiased estimator of $\bm{\beta}$ and specifies the bias whenever those conditions are not met.
\begin{proposition}\label{prop:bias}
Conditional on the estimated restrictions $\hat{A}$, the NIRVAR estimator $\hat{\bm{\beta}}(\hat{A})$ given by Equation~\eqref{eq:GLS-beta} is unbiased if and only if $\mathrm{colsp}\{R(A)\} \subseteq \mathrm{colsp}\{R(\hat{A})\}$, where $A$ specifies the true restrictions and $\hat{A}$ specifies the estimated restrictions. When this condition is not satisfied, $\mathbb{E}\{\hat{\bm{\beta}}(\hat{A})|\hat{A}\} = R(\hat{A}) C \bm{\gamma}(A)$, where 
\begin{align}
    \label{eq:bias}
    C \coloneqq \{R(\hat{A})^{\prime}(XX^{\prime} \otimes \Sigma^{-1}) R(\hat{A})\}^{-1} R(\hat{A})^{\prime}(XX^{\prime} \otimes \Sigma^{-1}) R(A).
\end{align}
\end{proposition}
The structural condition $\mathrm{colsp}\{R(A)\} \subseteq \mathrm{colsp}\{R(\hat{A})\}$ is equivalent to $\mathcal{E} \subseteq \hat{\mathcal{E}}$, where $\hat{\mathcal{E}}$ is the estimated edge set. Therefore, by Proposition \ref{prop:bias}, the GLS estimator given by Equation~\eqref{eq:GLS-beta}  is biased whenever an incorrect restriction is placed on the VAR matrix, that is, whenever $\hat{A}_{ij}^{(q)} = 0$ but $A_{ij}^{(q)} = 1$, and unbiased otherwise. We now consider the asymptotic properties of $\hat{\bm{\gamma}}(\hat{A})$. 
\begin{proposition}\label{prop:asy-norm}
The NIRVAR estimator $\hat{\bm{\gamma}}(\hat{A})$ given by Equation~\eqref{eq:GLS} is a consistent estimator of $C \bm{\gamma}(A)$ where $C$ is defined by Equation~\eqref{eq:bias}, and 
\begin{align}
    \label{eq:asy-eq}
    \sqrt{T}\left\{\hat{\bm{\gamma}}(\hat{A}) - C \bm{\gamma}(A)\right\} \xrightarrow[]{d} \mathcal{N}\left(0, \left\{R(\hat{A})^{\prime}\left(\Gamma \otimes \Sigma^{-1}\right) R(\hat{A}) \right\}^{-1} \right),
\end{align}
where $\Gamma \coloneqq \mathbb{E}(\bm{X}_{t} \bm{X}_{t}^{\prime}) = \plim XX^{\prime}/T$.
\end{proposition}
It is informative to compare the asymptotic efficiency of $\hat{\bm{\gamma}}(A)$ (the GLS estimator given that the true restrictions are known) with $\hat{\bm{\gamma}}(\hat{A})$ for the case where $\mathrm{colsp}\{R(A)\} \subseteq \mathrm{colsp}\{R(\hat{A})\}$, that is, $\hat{A}$ contains no misspecified restriction, but may not include \textit{all} of the true restrictions. The following proposition proves that, in this case, $\hat{\bm{\gamma}}(A)$ is asymptotically never less efficient than $\hat{\bm{\gamma}}(\hat{A})$.
\begin{proposition}\label{prop:efficiency}
    Let $\hat{\bm{\beta}}(A) = R(A)\hat{\bm{\gamma}}(A)$ and $\hat{\bm{\beta}}(\hat{A}) = R(\hat{A})\hat{\bm{\gamma}}(\hat{A})$ be the GLS estimators given by Equation~\eqref{eq:GLS-beta} for the case where $A$ specifies the true restrictions of the data generating process and $\hat{A}$ is another set of restrictions. 
    Under the assumption that $\mathrm{colsp}\{R(A)\} \subseteq \mathrm{colsp}\{R(\hat{A})\}$, the asymptotic variance of $\hat{\bm{\beta}}(\hat{A})$ is greater than or equal to that of $\hat{\bm{\beta}}(A)$ in the sense that 
    \begin{align}
        \label{eq:eq:PSD}
        \mathrm{var} \left[ \sqrt{T}\left\{\hat{\bm{\beta}}(\hat{A}) - \bm{\beta}\right\}\right] - \mathrm{var} \left[ \sqrt{T}\left\{\hat{\bm{\beta}}(A) - \bm{\beta}\right\}\right] \succeq 0,
    \end{align}
    where $\succeq$ denotes positive semi-definite.
\end{proposition} 
The extent to which the asymptotic variance of $\hat{\bm{\gamma}}(\hat{A})$ is inflated is discussed in Section~\ref{sec:sim}.

Define $\hat{A}^{(FN)}$ to be the $N \times N$ adjacency matrix of false negative entries with elements 
\begin{align}
    (\hat{A}^{(FN)})_{ij} = \begin{cases}
        1 \quad \text{if}  \quad A_{ij} = 1, \hat{A}_{ij} = 0,\\
        0 \quad \text{otherwise}.
    \end{cases}
\end{align}
Consider the vector $\bm{\gamma}_{0}(A,\hat{A}) = R(\hat{A}^{(FN)})^{\prime}\bm{\beta}$ containing those entries of $\bm{\beta}$ which are set to zero by the estimator. Note that it may be that the estimator does not restrict any true edges to zero, in which case the estimator is unbiased by Proposition \ref{prop:bias}, and $\bm{\gamma}_{0}(A,\hat{A}) = \bm{0}$. Proposition \ref{prop:mspe} decomposes the mean squared predictive error into bias, variance, and noise. The bias and scaled variance are then bounded, in the probability limit as $T \to \infty$, explicitly in terms of $\bm{\gamma}_{0}(A,\hat{A})$ and the number of unrestricted parameters $\widehat{M}$, respectively. 
\begin{proposition}\label{prop:mspe}
Let $Q=1$, $\mathcal{D} = (X,Y)$, be a training set, $\bm{X}_{\star}$ be a test vector, independent of $\mathcal{D}$, with $\mathbb{E}(\bm{X}_{\star}) = 0$ and $\mathbb{E}(\bm{X}_{\star}\bm{X}_{\star}^{\prime}) = \Gamma$, and $\bm{\epsilon}_{\star}$ be a noise term satisfying the conditions of the noise process in Definition \ref{def:nirvar}, independent of $\mathcal{D}$ and $\bm{X}_{\star}$. Then 
    \begin{align}
        \mathbb{E}_{\mathcal{D},\bm{X}_{\star},\bm{\epsilon}_{\star}}\left\{ \lVert \hat{\Phi}\bm{X}_{\star} - (\Phi\bm{X}_{\star} + \bm{\epsilon}_{\star}) \rVert_{2}^{2} | \hat{A}\right\} &= \mathrm{Bias}_{\mathcal{D}}^{2}\{\hat{\bm{\beta}}(\hat{A}) | \hat{A} \} + \mathrm{tr}[(\Gamma \otimes I_{N}) \mathrm{var}_{\mathcal{D}}\{\hat{\bm{\beta}}(\hat{A}) | \hat{A} \}] \\ &\quad+ \mathrm{tr}(\Sigma),
    \end{align}
where $\mathrm{Bias}_{\mathcal{D}}^{2}\{\hat{\bm{\beta}}(\hat{A}) | \hat{A} \}=[\mathbb{E}_{\mathcal{D}}\{ \hat{\bm{\beta}}(\hat{A}) | \hat{A}\} - \bm{\beta}]^{\prime}(\Gamma \otimes I_{N})[\mathbb{E}_{\mathcal{D}}\{ \hat{\bm{\beta}}(\hat{A}) | \hat{A}\} - \bm{\beta}]$. Furthermore, 
\begin{align}
    \plim \mathrm{Bias}_{\mathcal{D}}^{2}\{\hat{\bm{\beta}}(\hat{A}) | \hat{A} \} \leq \kappa(\Sigma) \lambda_{1}(\Gamma) \lVert \bm{\gamma}_{0}(A,\hat{A}) \rVert_{2}^{2},
\end{align}
where $\kappa(\Sigma) = \lambda_{1}(\Sigma)/\lambda_{N}(\Sigma)$ is the condition number of $\Sigma$, and
\begin{align}
    \plim T \mathrm{tr}[(\Gamma \otimes I_{N}) \mathrm{var}_{\mathcal{D}}\{\hat{\bm{\beta}}(\hat{A}) | \hat{A} \}] \leq \lambda_{1}(\Sigma)\widehat{M}.
\end{align}
\end{proposition}
Proposition \ref{prop:mspe} shows how misclassification errors propagate to the bias of the NIRVAR estimator. In particular, let $\mathcal{E}^{(i,z_{i})} = \{(i,j) \in \mathcal{E} : z_{i} = z_{j} \}$ be the intra-block edges connected to node $i$. By Proposition \ref{prop:mspe}, if node $i$ is misclassified, the squared bias term increases by an amount proportional to $\sum_{j \in \mathcal{E}^{(i,z_{i})}} \Phi_{ij}^{2}$. Furthermore, Proposition \ref{prop:mspe} shows how misspecifying $K$ can affect the mean squared prediction error. If one were to assume balanced communities, the variance term is inversely proportional to $K$ (since $\widehat{M} \asymp N^{2}/K$), thus selecting a larger value of $K$ reduces the variance at the risk of increasing the bias.

\section{Simulation study}\label{sec:sim}
We conduct extensive simulation studies with the aim of examining the asymptotic properties, comparing NIRVAR to existing methods, and conducting robustness checks and sensitivity analyses.
We simulate the underlying network from a stochastic block model with $B_{kh}=p^{(\text{in})}\in[0,1]$ for $k=h$ and $B_{kh}=p^{(\text{out})}\in[0,1]$ for $k\neq h,\ k,h\in[K]$. 
Firstly, the ability of the NIRVAR estimator to obtain the correct restrictions should decrease as the number of edges between blocks increases. We thus count the number of errors in $\hat{A}$ as a function of the probability $p^{(\text{out})}$ of an edge forming between blocks in the simulated data generating process. Secondly, the distribution of parameter estimates $\hat{\bm{\gamma}}(\hat{A})$ obtained from a large finite sample is compared with the asymptotic one in  Proposition \ref{prop:asy-norm} and the asymptotic variance inflation given in Proposition \ref{prop:efficiency} is quantified. Lastly, we compare the NIRVAR estimated latent positions with the ground truth latent positions of the data generating SBM. 

Further studies focusing mainly on robustness checks and sensitivity analyses are described in the supplementary material. In particular, a study showing the stability properties of 
embedding stacked sample covariance 
matrices is reported in Section S3.2. Clustering under a NIRVAR model with $p>1$ is examined in Section S3.3. The sensitivity of the NIRVAR estimator to mis-specification of $d$ and $K$ is investigated in Section S3.4, whilst a cross-validation approach to choosing $K$ is discussed in Section S3.5. Section S3.6 shows that NIRVAR clustering is robust to heteroskedastic and heavy-tailed noise. Since block recovery is not affected by this heteroskedastic and heavy-tailed noise, we can use standard GLS estimation downstream to yield a robust NIRVAR estimator. In Section S3.7, we show that, when the data generating process is NIRVAR, prediction via NIRVAR outperforms an unrestricted VAR, a Bayesian VAR with a Minnesota prior \citep{doan1984forecasting}, and a LASSO regularised VAR except in very high or low-dimensional regimes. We show how to relax the inter-block hard thresholding of the NIRVAR estimator by including an $\ell_{1}$ penalty for inter-block parameters in Section S3.8, and discuss the benefits and drawbacks of doing so for practitioners. We investigate the stability of the Expectation-Maximisation algorithm for the NIRVAR clustering step in Section S3.9. Finally, in Section S3.10, we check whether NIRVAR cluster recovery and parameter estimates improve as the spectral radius of the ground truth VAR coefficient matrix $\rho(\Phi)$ increases. 

\subsection{Between block probability}\label{subsec:between}
\begin{figure}[t]
\begin{center}
\begin{subfigure}[b]{.32\textwidth}
  \includegraphics[width=.975\textwidth]{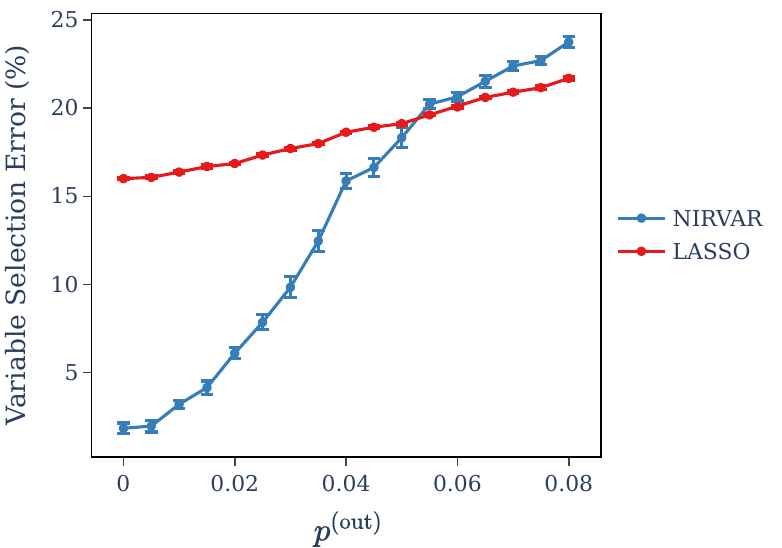}
  \caption{}
  \label{fig:sim-lasso-bias-sub1}
\end{subfigure}
\begin{subfigure}[b]{.32\textwidth}
  \includegraphics[width=.975\textwidth]{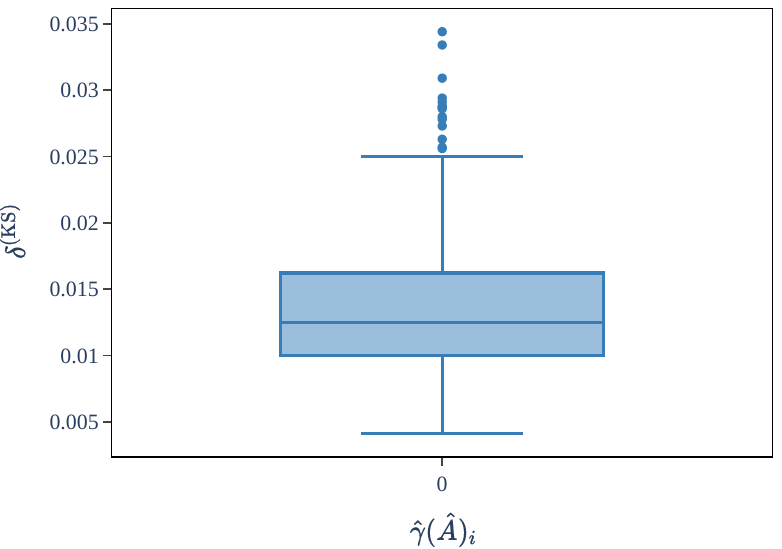}
  \caption{}
  \label{fig:sim-lasso-bias-sub2}
\end{subfigure}
\begin{subfigure}[b]{.32\textwidth}
  \includegraphics[width=.975\textwidth]{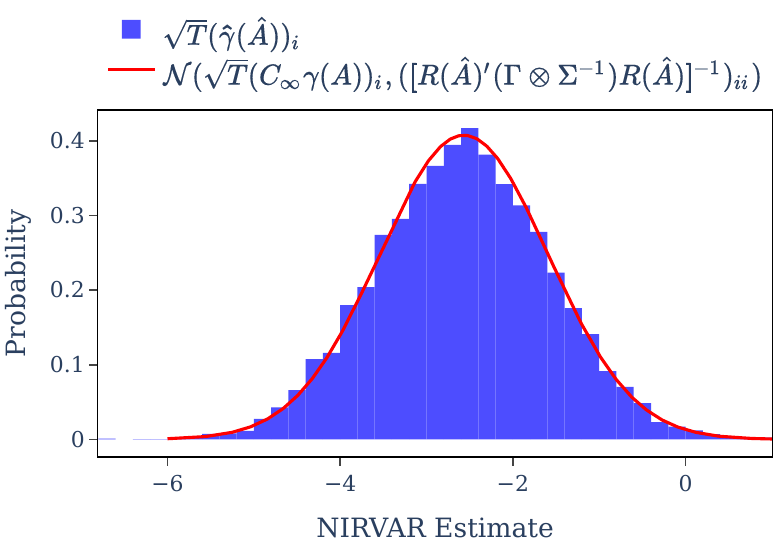}
  \caption{}
  \label{fig:sim-lasso-bias-sub3}
\end{subfigure}
\caption{(a) Percentage error in estimated restrictions as $p^{(\text{out})}$ increases. (b) KS statistic: empirical and asymptotic distribution of $\hat{\bm{\gamma}}(\hat{A})_{i}$, for $i = 1,\dots,\hat{M}$. (c) Histogram of $\hat{\bm{\gamma}}(\hat{A})_{i}$ for one fixed $i$ and  asymptotic distribution. } 
\label{fig:sim-lasso-bias}
\end{center}
\end{figure}

Fixing $N=100$, $T=1000$, $Q=1$, $K=10$, and setting $\Tilde{\Phi}^{(1)}$ by sampling each entry from a Uniform(0,1) and normalising so that $\rho(\Phi^{(1)}) = 0.9$,
we computed the percentage of incorrect entries of $\hat{A}$ as a function of $p^{(\text{out})}$. The percentage error was calculated as $100\times \sum_{i,j=1}^{N}\indfun\{ \hat{A}_{ij} \neq A_{ij}\}/N^{2}$ and compared to the percentage variable selection error of a LASSO estimator, with penalty chosen using AIC.  
Figure \ref{fig:sim-lasso-bias}(\subref{fig:sim-lasso-bias-sub1}) shows that for $p^{(\text{out})} < 0.05$ NIRVAR 
outperforms the LASSO. The percentage error increases with $p^{(\text{out})}$ and becomes larger than, though on the same scale as, that of the LASSO estimator for $p^{(\text{out})} > 0.05$. This is expected since the NIRVAR estimation procedure always constructs a graph with $K$ cliques and cannot recover edges between blocks. 

\subsection{Large sample distribution of the NIRVAR estimator}
We simulate data with $N=50$, $Q=1$, $T=5000$, $K=5$, and set $\Tilde{\Phi}^{(1)}$ by sampling each entry from a Uniform(0,1) and normalising so that $\rho(\Phi^{(1)}) = 0.9$. The probability of an edge forming within each block was set to $p^{(\text{in})} = 0.75$ and we set $p^{(\text{out})} = 0.2$. Creating 10,000 replica datasets, each having these hyperparameters, and estimating $\Phi$ for each replica provides an empirical distribution for $\sqrt{T}\hat{\bm{\gamma}}(\hat{A})_{i}$. For each $i=1,\dots,\hat{M}$, we performed a one-sample Kolmogorov-Smirnov (KS) test of whether the empirical distribution of $\sqrt{T}\hat{\bm{\gamma}}(\hat{A})_{i}$ is normal with mean $\sqrt{T}C_{\infty} \bm{\gamma}(A)$ and variance $ \{R(\hat{A})^{\prime}\left(\Gamma \otimes \Sigma^{-1}\right) R(\hat{A}) \}^{-1}$ as claimed by Proposition \ref{prop:asy-norm} to be its asymptotic distribution, where $    C_{\infty} \coloneqq \plim(C) = \{R(\hat{A})^{\prime}(\Gamma \otimes \Sigma^{-1}) R(\hat{A})\}^{-1} R(\hat{A})^{\prime}(\Gamma \otimes \Sigma^{-1})R(A).$
Figure \ref{fig:sim-lasso-bias}(\subref{fig:sim-lasso-bias-sub2}) shows a box plot of the KS statistic for each $i=1,\dots,\hat{M}$, whilst Figure \ref{fig:sim-lasso-bias}(\subref{fig:sim-lasso-bias-sub3}) shows a histogram of the observed sample for one particular $i$ plotted alongside the corresponding asymptotic normal curve. The low value of $\delta^{(\text{KS})}$ shown by the box plot in Figure \ref{fig:sim-lasso-bias}(\subref{fig:sim-lasso-bias-sub2}) indicates that the large sample distribution of $\sqrt{T}\hat{\bm{\gamma}}(\hat{A})_{i}$ mirrors the asymptotic distribution of Proposition \ref{prop:asy-norm}. Under the same simulation setting, the asymptotic variance of the NIRVAR estimator is compared with an estimator that uses all ground truth restrictions where the intra-block sparsity of the ground truth is 50\%, that is $p^{(\text{in})} = 0.5$. Defining $\alpha_{V} \coloneqq \text{tr}[\text{var}\{\bm{\hat{\beta}}(\hat{A})\}] / \text{tr}[\text{var}\{\bm{\hat{\beta}}(A)\}]$ as a measure of the extent to which the asymptotic variance of the NIRVAR estimator is inflated yields a value of $\alpha_{V} = 1.8$. The variation of $\alpha_{V}$ with different levels of sparsity and different values of $N$ is investigated (on synthetic and real data) in Section S3.1 of the supplementary material.  

\subsection{Latent position recovery}\label{subsec:latent-sim}
\begin{figure}[t]
\begin{center}
\begin{subfigure}[b]{.32\textwidth}
  \includegraphics[width=0.975\textwidth]{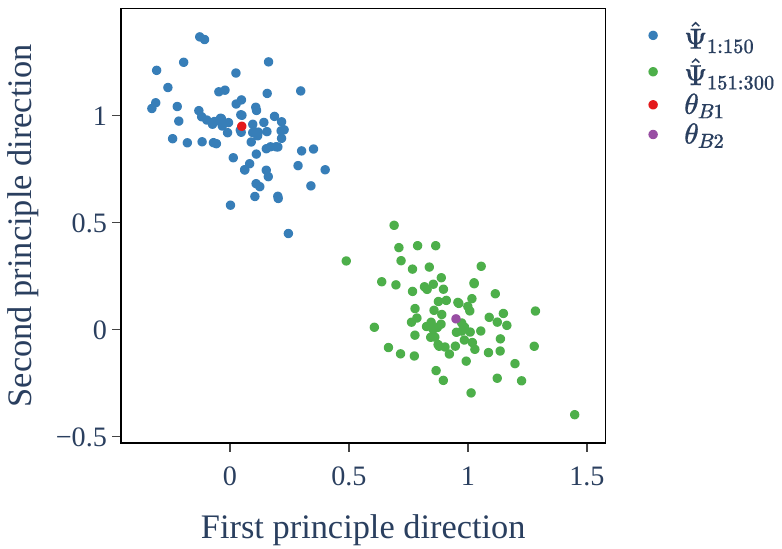}
  \caption{}
  \label{fig:sim-latent-recovery-sub1}
\end{subfigure}
\begin{subfigure}[b]{.32\textwidth}
  \includegraphics[width=0.975\textwidth]{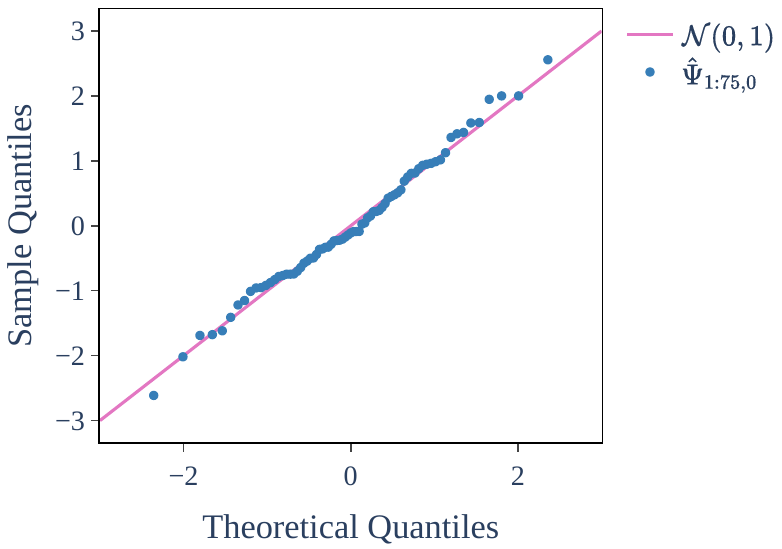}
  \caption{}
  \label{fig:sim-latent-recovery-sub2}
\end{subfigure}
\begin{subfigure}[b]{.32\textwidth}
  \includegraphics[width=0.975\textwidth]{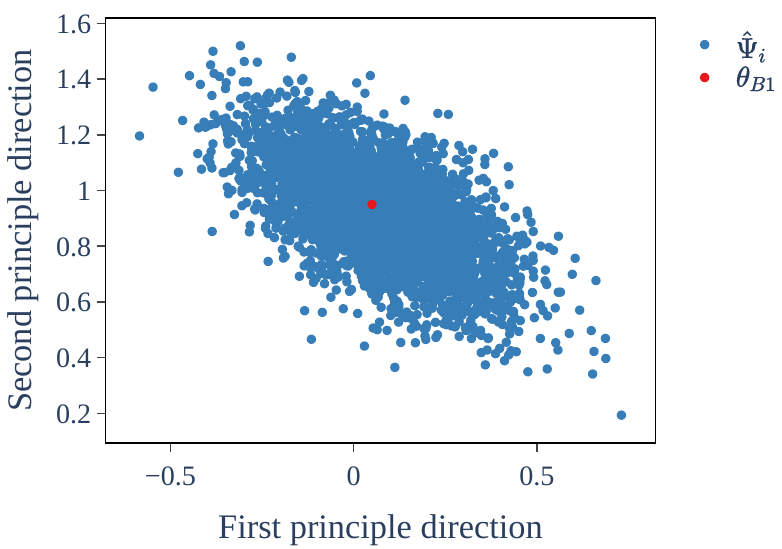}
  \caption{}
  \label{fig:sim-latent-recovery-sub3}
\end{subfigure}
\caption{(a) NIRVAR embedded latent points (scaled and rotated) and true latent positions from the simulated SBM. (b) Q-Q plot comparing the sample distribution of the NIRVAR embedded points with a normal distribution. (c) The NIRVAR embedded latent position for $i=21$ across 4000 replica samples. } 
\label{fig:sim-latent-recovery}
\end{center}
\end{figure}

We conduct a study in which we fix two ground truth latent positions $\bm{\theta}_{B1} = (0.05,0.95)^{\prime}$ and $\bm{\theta}_{B2} = (0.95,0.05)^{\prime}$ of the data generating SBM and compare the (rotated) NIRVAR embedded points $\hat{\Psi}^{(1)}$ to $\bm{\theta}_{B1}$ and $\bm{\theta}_{B2}$. The hyperparameters used to generate the data were $N=150$, $T=2000$, $Q=1$, $K=2$, $z_{1}, \dots ,z_{75} = 1$ and $z_{76}, \dots ,z_{150} = 2$. The weights $\Tilde{\Phi}$ were set to the constant $0.9/\rho(\Theta \Theta^{\prime})$ so that the spectral radius of the expected value of $\Phi$ was 0.9. With $\bm{X}_{t}^{(1)} \sim \mathrm{NIRVAR}(\Phi)$ for $t=1,\dots,T$, the left singular vectors and singular values of $S^{(1)}$ were computed, and the singular values were scaled using $(\lambda_{\Phi})_{i} = \pm \sqrt{1-1/(\lambda_{\Gamma})_{i}}$ from Proposition \ref{prop:cov-conn}. Procrustes alignment \citep{schonemann1966generalized} was then used to compare the rotated sample embedded points $\hat{\Psi}_{1:75}$ to $\bm{\theta}_{B1}$ and $\hat{\Psi}_{76:150}$ to $\bm{\theta}_{B2}$. Figure \ref{fig:sim-latent-recovery}(\subref{fig:sim-latent-recovery-sub1}) shows $\hat{\Psi}_{i}$ coloured according to the block membership $z_{i}$ for $i=1,\dots,N$. Figure \ref{fig:sim-latent-recovery}(\subref{fig:sim-latent-recovery-sub2}) shows a Q-Q plot comparing the (standardised) sample distribution $\hat{\Psi}_{1:75}$ with a standard normal distribution. The sample data is in good agreement with a normal distribution. We also conduct a further experiment in which we replicate the above procedure $4000$ times and plot $\hat{\Psi}_{i}$ for $i=21$ against $\bm{\theta}_{B1}$ across the 4000 replicas. Since $\rho(\Phi_{k})$ (with $k \in \{1,\dots,4000\}$ labelling the replica) has spectral radius less than 1 only in expectation, we only retain the first 4000 samples that have $\rho(\Phi_{k}) < 1$. 
Figure \ref{fig:sim-latent-recovery}(\subref{fig:sim-latent-recovery-sub3}) shows that the large sample distribution of $\hat{\Psi}_{21}$ is approximately normally distributed around $\bm{\theta}_{B_{1}}$. 
With the R package \texttt{MVN} \citep{korkmaz2014mvn},
we performed the Henze-Zirkler test of multivariate normality on the 4000 sample points yielding a test statistic of 0.56 and an observed $p$-value of 0.96 which supports the null hypothesis that the data is multivariate normally distributed.

\section{Applications}\label{sec:applications}

NIRVAR is applied to three data sets exhibiting different characteristics: (i) a financial dataset of excess market returns, where multiplex NIRVAR can be applied based on different embeddings, allowing us to compare several model  specifications; (ii) a vintage of macro-economic indicators typically used for predicting US industrial production; (iii) a transportation dataset of bicycle rides in London, where underlying groups are neither pre-determined nor necessarily  related to geographical coordinates, and their flows and directions are of interest. Details of each dataset are reported in the following subsections.

To position NIRVAR within the most recent literature on network models for dynamic processes, results are compared with those from three related models, namely the Factor Augmented Regression Model (FARM) of \citet{fan2023bridging}, the Factor-Adjusted Network Estimation and Forecasting for High-Dimensional Time Series (FNETS) model of \citet{barigozzi2023fnets}, and the Generalised Network Autoregressive Processes (GNAR) model of \citet{knight2020generalized}, whose specifications are reported in the supplementary materials.  
 
\subsection{Financial application to US equity market}\label{subsec:fin}
The open-to-close (OPCL) and previous close-to-close (pvCLCL) price returns of $N =648$ financial assets between 03/01/2000 and 31/12/2020 ($T = 5279$) were derived from databases provided by the Center for Research in
Security Prices, LLC. 
Using the S\&P~500 index as a proxy for the market, we construct the OPCL and pvCLCL market excess returns by subtracting the return of SPY, the exchange traded fund which tracks the  S\&P~500 index. The task is to predict the next day pvCLCL market excess returns. If the predicted next day pvCLCL market excess return of an asset has a positive (negative) sign, then a long (short) position in the asset is taken. 

We compare seven different models: NIRVAR using the covariance matrix embedding method with $Q=1$ (pvCLCL returns only) and $Q=2$ (pvCLCL and OPCL), labelled as NIRVAR C1 and NIRVAR C2, respectively; NIRVAR using the precision matrix embedding method with $Q=1$ and $Q=2$, labelled as NIRVAR P1 and NIRVAR P2, respectively; FARM with $L=1$; FNETS with $L=1$; GNAR with $L=1$ and network $\mathcal G=([N],\mathcal E)$ where $(i,j)\in\mathcal E$ if asset $i$ and asset $j$ share the same Standard Industrial Classification (SIC) division. The SIC is a four-digit 
code assigned to businesses according to their industry, 
each belonging to one the following sectors: (i) Agriculture, Forestry, and Fishing,
    (ii) Mining,
    (iii) Construction,
    (iv) Manufacturing,
    (v) Transportation and Public Utilities,
    (vi) Wholesale Trade,
    (vii) Retail Trade,
    (viii) Finance, Insurance, and Real Estate,
    (ix) Services, and
    (x) Public Administration.

We backtest each model using a rolling window between 01/01/2004 and 31/12/2020 with a look-back window of four years 
and evaluate the performance via a number of commonly used metrics in the financial literature \citep[see, for example][]{gu2020empirical}. 
Each day, we compute a measure of profit and loss, 
$    \text{PnL}_{t} = \sum_{i=1}^{N} \text{sign}(\hat{s}_{i}^{(t)}) \times s_{i}^{(t)}$,  
where $\hat{s}_{i}^{(t)}$ is the predicted return of asset $i$ on day $t$, and $s_{i}^{(t)}$ is the realised return of asset $i$ on day $t$. Note that we assign an equal portfolio weighting to each asset in our definition of $\text{PnL}_{t}$; an alternative approach would be to consider value-weighted portfolio construction methods, that account for the liquidity or market capitalization  of each asset. 

\begin{figure}[t]
\begin{center}
\begin{subfigure}{.49\textwidth}
  \includegraphics[width=1\linewidth]{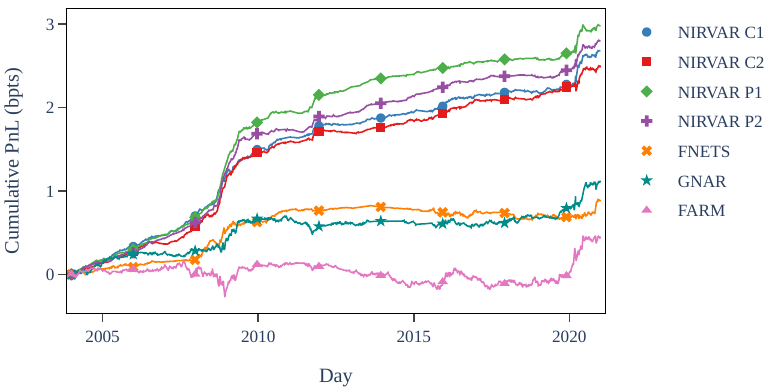}
  \caption{}
  \label{fig:CumPnL-sub1}
\end{subfigure}
\begin{subfigure}{.49\textwidth}
      \includegraphics[width=1\linewidth]{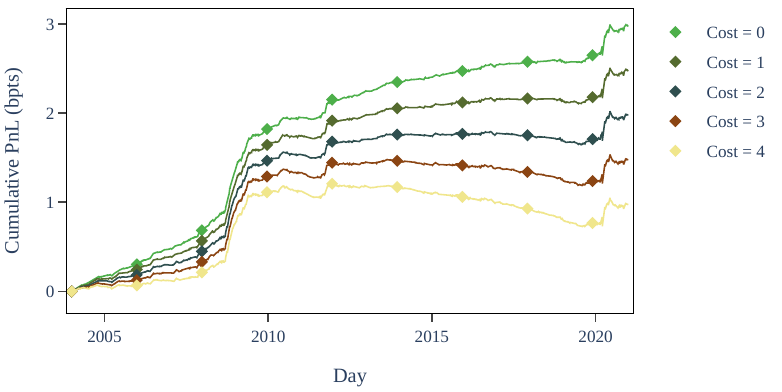}
    \caption{}
  \label{fig:CumPnL-sub2}
\end{subfigure}
\caption{(a) Cumulative PnL in bpts for the 7 models with zero transaction costs. (b) Cumulative PnL of NIRVAR P1 for different levels of transaction costs.  } 
\label{fig:CumPnL} 
\end{center}
\end{figure}

The cumulative PnL for each model is shown in Figure \ref{fig:CumPnL}(\subref{fig:CumPnL-sub1}). All NIRVAR models outperform FNETS, GNAR and FARM in terms of cumulative PnL, and NIRVAR P1 attains the highest cumulative PnL overall. We plot the cumulative PnL of NIRVAR P1 for different levels of transaction costs in Figure \ref{fig:CumPnL}(\subref{fig:CumPnL-sub2}), where a flat transaction cost (given in basis points, denoted \textit{bpts}, with 1\% = 100bpts) is applied every time we flip our position in a given asset from long (short) to short (long). From Figure \ref{fig:CumPnL}(\subref{fig:CumPnL-sub2}) we conclude that, even with a transaction cost of 4bpts, NIRVAR is profitable, especially during the global financial crisis and the COVID-19 pandemic which were periods of high market volatility.  

Table \ref{tab:returns-sharpes} reports the Mean Absolute Error (MAE) for all the estimated models, and the annualised Sharpe Ratio (SR) over 252 trading days. The SR is one of the most commonly used metrics in the financial industry, defined as 
$    \text{SR} =  \sqrt{252} \times \text{mean}(\text{PnL}) \,/\,\text{stdev}(\text{PnL})$. Although the MAE is equivalent across models, the SR is much larger for NIRVAR models, seemingly due to the best performance during crises. Buying and holding SPY across the same time period would yield a SR of 0.43, highlighting that NIRVAR models can provide greater economic benefits compared with a simple passive investment strategy. %
Within a financial context, MAE and other related metrics 
are not as representative as 
in traditional time series settings: 
since the profit or loss of a trade is assessed exclusively on 
whether it is a buy or a sell, 
different methods can attain significantly different Sharpe Ratios, 
despite having similar MAE 
\citep[see, for example][]{FinGAN}.
Further measures, reported in Tables~\ref{tab:returns-sharpes-full} and~\ref{tab:regime-sharpes} in the supplementary material, confirm that NP1 or NP2 consistently outperform the competing models. 

\begin{figure}[t]
\begin{center}
\begin{subfigure}{.32\textwidth}
  \includegraphics[width=1\linewidth]{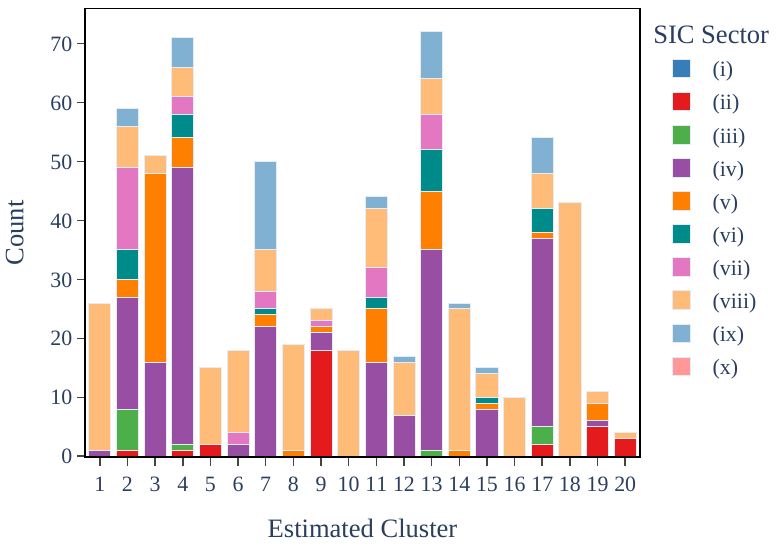}
  \caption{}
  \label{fig:returns-groups-sub1}
\end{subfigure}
\begin{subfigure}{.32\textwidth}
  \includegraphics[width=1\linewidth]{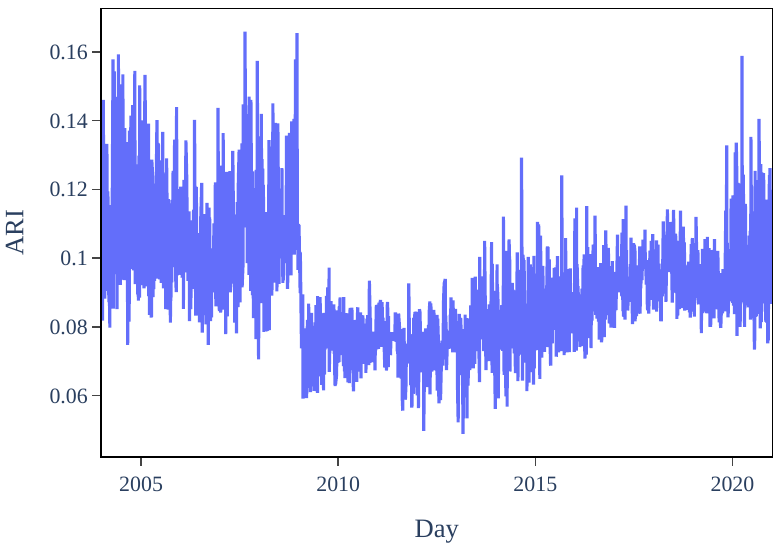}
  \caption{}
  \label{fig:returns-groups-sub2}
\end{subfigure}
\begin{subfigure}{.32\textwidth}
  \includegraphics[width=1\linewidth]{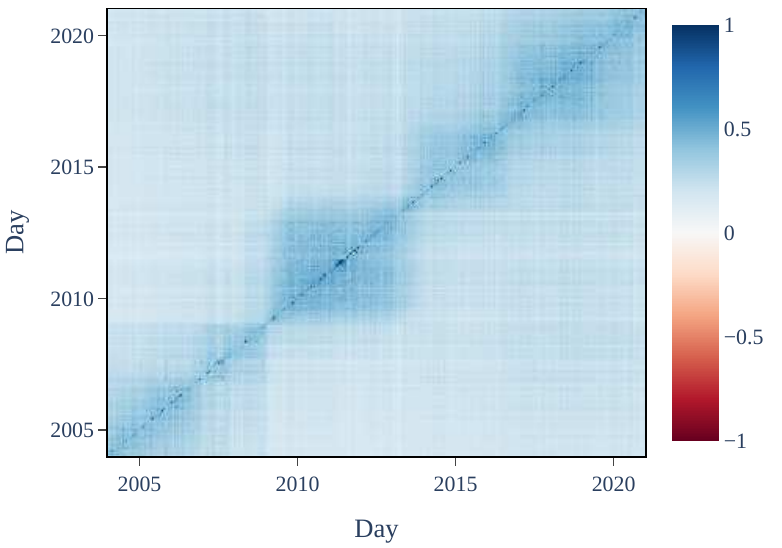}\
  \caption{}
  \label{fig:returns-groups-sub3}
\end{subfigure}
\caption{(a) Comparison of the NIRVAR estimated clusters with the SIC groups on 31/12/2020. (b) The ARI between the NIRVAR estimated clusters and the SIC groups on each backtesting day. (c) The ARI between NIRVAR estimated clusters on different days.} 
\label{fig:returns-groups}
\end{center}
\end{figure}  
%
%
%
\begin{table}[t]
\begin{tabular*}{\columnwidth}{@{\extracolsep\fill}llllllll@{\extracolsep\fill}}
\toprule
 & N C1  & N C2 & N P1 & N P2 & FARM & FNETS & GNAR\\
\midrule
Sharpe Ratio    & 2.50 & 2.34 & 2.82 & 2.69 & 0.22 & 0.78 & 0.70 \\
Mean Absolute Error & 0.012 & 0.013 & 0.012 & 0.013 & 0.012 & 0.012 & 0.012\\
\bottomrule
\end{tabular*}
\caption{Sharpe Ratio and Mean Absolute Error values for the estimated models. 
\label{tab:returns-sharpes}}%
\end{table}

Figure \ref{fig:returns-groups}(\subref{fig:returns-groups-sub1}) compares the NIRVAR estimated clusters with the 10 SIC sectors on one particular backtesting day. Figure \ref{fig:returns-groups}(\subref{fig:returns-groups-sub2}) shows the corresponding ARI for every backtesting day.  The ARI never breaches 0.17, from which we conclude that the NIRVAR estimated clusters do not agree with SIC groupings. 
The bar chart does suggest, however, that NIRVAR tends to separate Finance, Insurance, and Real Estate companies from the other SIC groups. Figure \ref{fig:returns-groups}(\subref{fig:returns-groups-sub3}) provides the ARI between the NIRVAR estimated clusters at time $t$ and the NIRVAR estimated clusters at time $s$,  where $t$ and $s$ run over every pair of backtesting days. There are two distinct blocks along the diagonal of this heatmap, the first one between 2004-2008 and the second one between 2008-2013, indicating that the NIRVAR estimated clusters underwent a shift after the global financial crisis.

\subsection{US Industrial Production}\label{subsec:FRED-MD}
FRED-MD\footnote{The database is available at \url{https://research.stlouisfed.org/econ/mccracken/fred-databases}.} \citep{mccracken2016fred} is a publicly accessible database of monthly observations of macroeconomic variables, updated in real-time. 
%
To allow direct comparison with \cite{fan2023bridging}, we choose the August 2022 vintage of the FRED-MD database which extends from January 1960 until December 2019 ($T=719$), preprocessed as discussed in \cite{mccracken2016fred}. 
Only variables with no missing values ($N = 122$) are used. 

The  task is one-step ahead prediction of the first-order difference of the logarithm of the monthly industrial production (IP) index. We backtest each model from January 2000 until December 2019 using a rolling window with a look-back window of 480 observations. The covariance matrix embedding method of NIRVAR is chosen. Following \citet{fan2023bridging}, we set $L = 24$ for FARM. For FNETS and GNAR we set $L=1$. The FRED-MD variables are divided into eight groups: (i) output and income; (ii) labour market; (iii) housing; (iv) consumption, orders, and inventories; (v) money and credit; (vi) interest and exchange rates; (vii) prices;
and (viii) stock market. The network $\mathcal G=([N],\mathcal E)$ chosen for GNAR is defined by $(i,j) \in \mathcal E$ if and only if $i$ and $j$ are in the same FRED-MD group.
\begin{figure}[t]
\centering
\begin{subfigure}{.49\textwidth}
  \includegraphics[width=\linewidth]{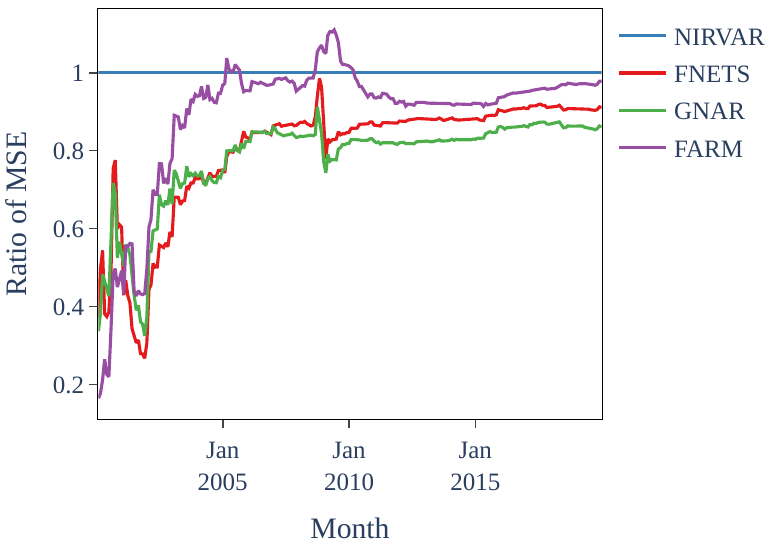}
  \caption{}
  \label{fig:fred-ratio-groups-ari-corr-sub1}
\end{subfigure}
\begin{subfigure}{.49\textwidth}
  \includegraphics[width=\linewidth]{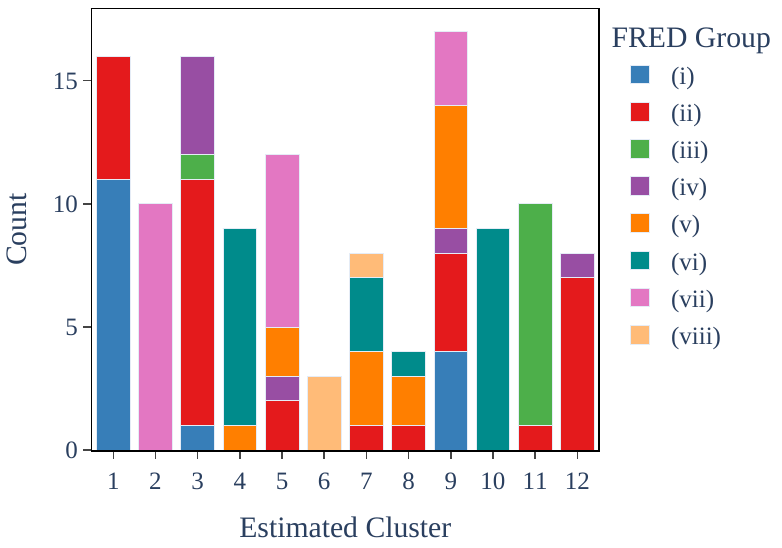}
  \caption{}
  \label{fig:fred-ratio-groups-ari-corr-sub2}
\end{subfigure}
\caption{(a) Ratio $\Delta_{t}^{(m)}$ of the cumulative MSE of NIRVAR divided by the cumulative MSE of alternative models. (b) Comparison of the NIRVAR estimated clusters with the FRED groups for January 2000.}
\label{fig:fred-ratio-groups-ari-corr}
\end{figure}

The overall MSE of the NIRVAR estimator ($0.0087$) is smaller than the MSE of  FARM ($0.0089$), which is the benchmark in this example, of  FNETS ($0.0096$) and of GNAR ($0.0101$), that is based on an imposed network that coincides with the FRED-MD groups. NIRVAR estimation and prediction is also implemented using the imposed network that coincides with the FRED-MD groups, yielding an overall MSE of 0.0097. Furthermore, a GNAR model based on a NIRVAR estimated network yields an overall MSE of 0.0096. This highlights the utility of a data driven approach to group selection and network construction. As in \citet{fan2023bridging}, we compute the ratio of cumulative MSEs between NIRVAR and each of FNETS, GNAR, and FARM. Letting $\Delta_{t}^{(m)}$ be the ratio for model $m \in \{\text{FNETS},\text{GNAR},\text{FARM}\}$ at time $t \in \{1,\dots,T\}$, we have
$    \Delta_{t}^{(m)} \coloneqq  \sum_{s =1 }^{t} \text{MSE}_{s}^{\text{(NIRVAR)}}/$ $ \sum_{s =1 }^{t} \text{MSE}_{s}^{(m)}.$ 
A value of $\Delta_{t}^{(m)}$ that is less than 1 indicates that NIRVAR is outperforming model $m$ in terms of cumulative MSE. Figure \ref{fig:fred-ratio-groups-ari-corr}(\subref{fig:fred-ratio-groups-ari-corr-sub1}) evidences that $\Delta_{t}^{(m)}$ is always below 1 for GNAR and FNETS, and, for FARM, it is below 1 the majority of the time. 

Figure \ref{fig:fred-ratio-groups-ari-corr}(\subref{fig:fred-ratio-groups-ari-corr-sub2}) compares the $K=12$ NIRVAR estimated clusters with the eight FRED-MD groups on the first month of backtesting. NIRVAR clusters do not match the FRED-MD groups closely (the ARI for each backtesting time is in the range $[0.24,0.36]$). However, NIRVAR clusters 2, 4, 10, 11, and 12 are 
in close correspondence with 
individual FRED-MD groups, while NIRVAR cluster 1 largely recovers  FRED-MD groups  (i)-(ii), that are output and income and labour market. Clusters 3, 5, 7, 8, and 9 tend to aggregate FRED-MD financial sectors (v)-(viii).  This is in line with the property of NIRVAR to model 
time series in which there is greater co-movement within communities than between communities.



The sensitivity of the estimators to extreme regimes is further explored in the supplementary materials, where we conclude that in this example, FNETS produces accurate predictions in non-crisis times while NIRVAR outperforms the competing models during crises, which is consistent with the results obtained in Section~\ref{subsec:fin}.

\subsection{Santander bicycle rides}\label{subsec:santander}
The first differences of the log daily number of bicycle rides from $N=774$ Santander stations in London from 07/03/2018 until 10/03/2020 ($T = 735$) were obtained using records from TfL Open Data (see \url{https://cycling.data.tfl.gov.uk/}). Since $N > T$, the covariance matrix is not invertible and therefore the NIRVAR precision matrix embedding method is not feasible here. FARM, FNETS and GNAR are modelled using a lag of $L = 1$. The network $\mathcal G=([N],\mathcal E)$ chosen for GNAR is defined by $(i,j)\in\mathcal E$ only if the Euclidean distance between stations $i$ and $j$ is less than 3 kilometres. 
\begin{figure}[t]
\centering
\begin{subfigure}{.32\textwidth}
  \includegraphics[width=\linewidth]{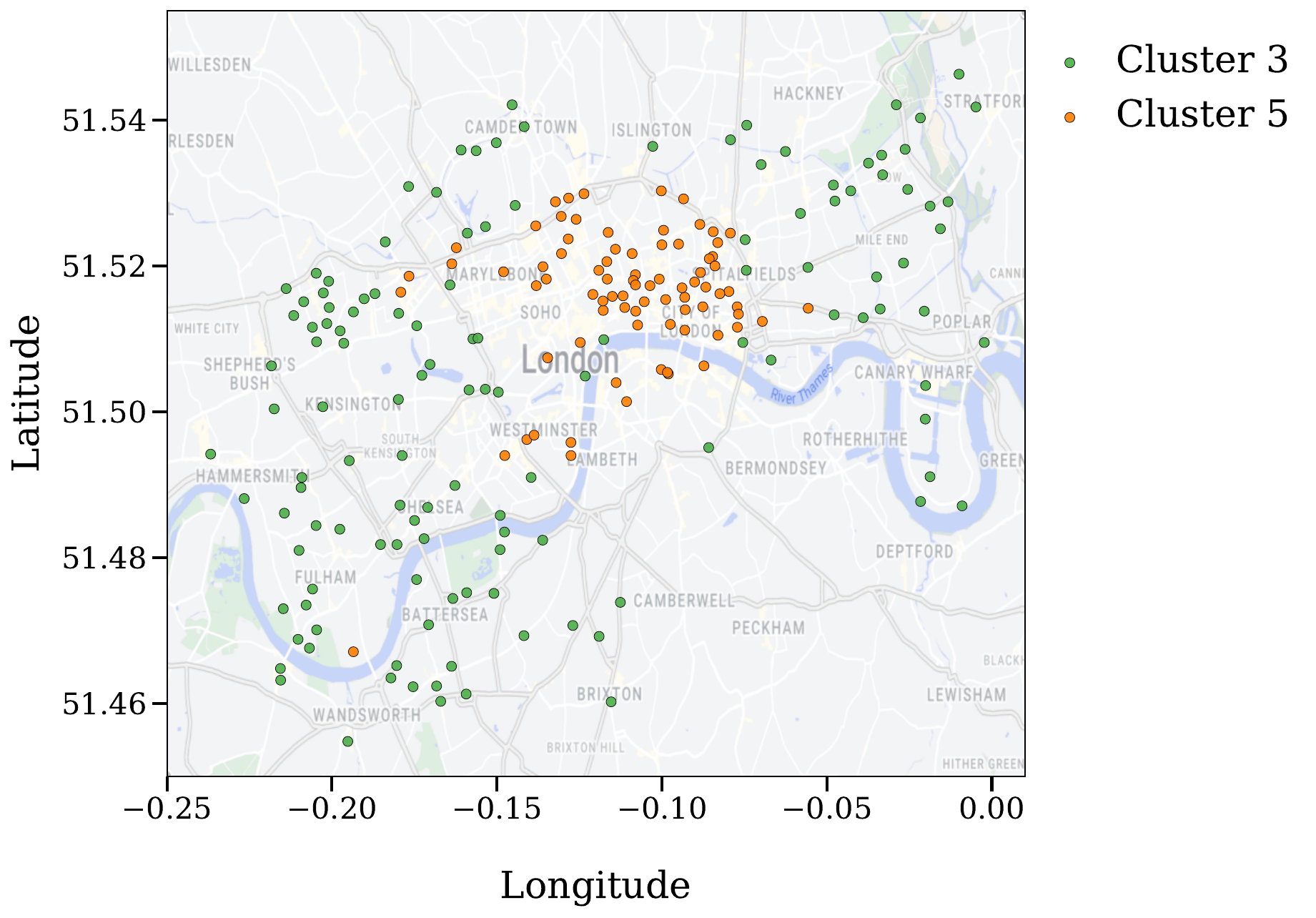}
  \caption{}
  \label{fig:santander-sub1}
\end{subfigure}
\begin{subfigure}{.32\textwidth}
  \includegraphics[width=\linewidth]{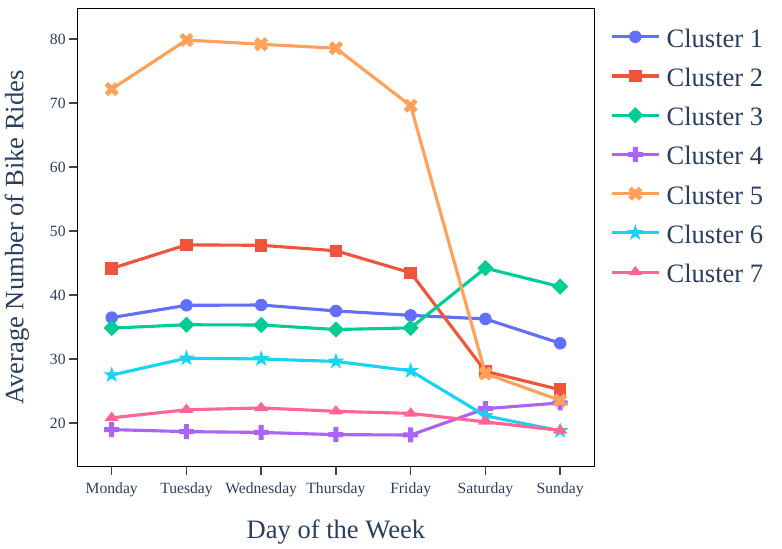}
  \caption{}
  \label{fig:santander-sub2}
\end{subfigure}
\begin{subfigure}{.32\textwidth}
  \includegraphics[width=\linewidth]{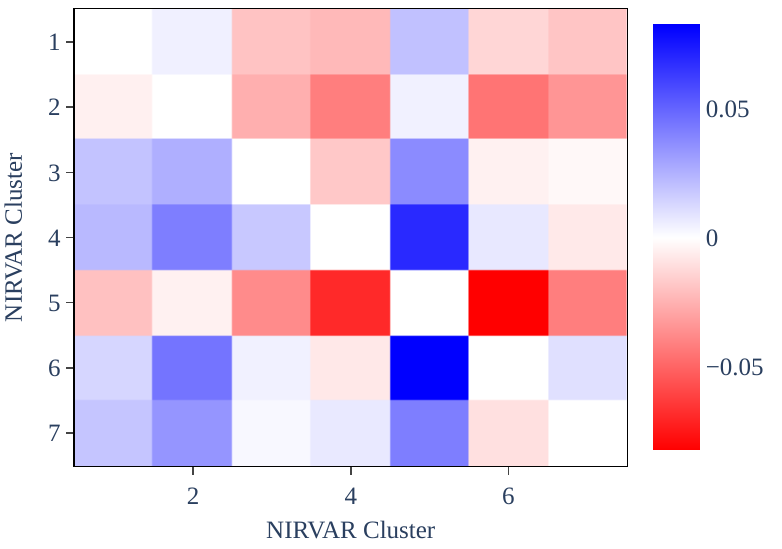}
  \caption{ }
  \label{fig:santander-sub3}
\end{subfigure}
\caption{NIRVAR on the Santander Cycles dataset. (a)  Clusters 3 and 5. (b) Average number of bicycle rides across each of the $K=7$  clusters on each weekday. (c) Flow between clusters, where each block indicates a net flow from cluster $i$ into cluster $j$.} 
\label{fig:santander}
\end{figure}

There were $K=7$ clusters estimated by NIRVAR. Figure \ref{fig:santander}(\subref{fig:santander-sub1}) shows the geographic locations of the stations in clusters 3 and 5. Cluster 5 is concentrated around central London whereas the stations in cluster 3 are located further from the center near parks and more residential areas. In contrast to the epsilon ball construction of the GNAR network, the NIRVAR clusters do not necessarily lie within some ball of a given radius. Figure \ref{fig:santander}(\subref{fig:santander-sub2}) shows the mean number of bicycle rides across each of the 7 NIRVAR clusters for every day of the week. The clusters are differentiated by their mean number of bicycle rides as well as by the change in the number of bicycle rides on weekdays compared with weekends. For example, stations in clusters 2 and 5, which are located mainly in central London, have high usage during the week and lower usage during the weekend. This likely corresponds to usage by commuters. In contrast, the usage of bicycles from stations in cluster 3 increases during weekends. This may correspond to usage for leisure by tourists or local residents, considering that many of the stations in this group are located around parks. 

To analyse the flow of bicycle rides between NIRVAR clusters, we computed a $K \times K$ matrix $F$, where $F_{ij}$ is the number of bicycle rides that start from any station in cluster $i$ and end in any station in cluster $j$. A matrix $\Tilde{F}$ with elements 
$    \Tilde{F}_{ij} = (F_{ij} - F_{ji})/(F_{ij} + F_{ji})$,  
provides a normalised measure of the imbalance in the  flow of bicycle rides between clusters. Figure \ref{fig:santander}(\subref{fig:santander-sub3}) shows $\Tilde{F}_{ij}$: rows 2 and 5 of $\Tilde{F}$ are negative (red) which corresponds to a net inflow into NIRVAR clusters 2 and 5, whose stations are located mainly in central London. 

%
To conclude, one-day-ahead predictions from NIRVAR, FARM, FNETS and GNAR are obtained using a rolling window backtesting framework from 09/02/2020 until 10/03/2020 (30 days). The overall MSE is homogeneous across models with NIRVAR achieving the lowest value (0.364), followed by FARM (0.370), GNAR (0.374) and FNETS (0.388). 

\section{Discussion}\label{sec:discussion}

We introduce NIRVAR, a framework for modelling a panel of multivariate time series as a network VAR process. Two key features characterise NIRVAR: first, the underlying network is unobserved; second, the network is reconstructed prior to estimation of the VAR model parameters. This circumvents the need to specify tuning or thresholding hyperparameters when constructing the network, and it avoids estimation of the full, dense VAR parameter matrix. Due to the regularisation induced by the network, NIRVAR is computationally fast. For example, backtesting NIRVAR on the Santander experiment in Section \ref{subsec:santander} for a single forecast took approximately 10 seconds compared with 70 seconds for a VAR\footnote{Both experiments were run using a single Intel Core i5-1145G7 CPU.}. 

A drawback of the estimation method is that it will be biased whenever there are edges between blocks of the data generating SBM. Therefore, it is only suitable for assortative graphs. One possible solution to this is to consider a mixed membership SBM \citep{airoldi2008mixed} in which each graph vertex can be associated with multiple clusters. We leave this for future research. The NIRVAR model assumes the observed random graph does not change over time. Extending the model to incorporate a time varying adjacency matrix could add flexibility and improve both the cluster assignment and the predictive performance. 

 It would be a natural extension to incorporate a factor model into the NIRVAR framework, 
 isolating 
 the idiosyncratic interactions that are due to network effects from the co-movements that are due to common factors. For example, in the financial application, we removed the market factor prior to NIRVAR estimation. 
 Incorporating a factor model would also likely improve the forecasting performance 
 on the FRED-MD dataset. \citet{mccracken2016fred} find eight factors in the sample they consider; removing these factors could give prominence to network effects, 
 improving NIRVAR estimation and prediction.


\section*{Acknowledgements}

Brendan Martin acknowledges funding from the Engineering and Physical Sciences Research
Council (EPSRC), grant number EP/S023151/1.
Francesco Sanna Passino acknowledges funding from the EPSRC, grant number EP/Y002113/1. 

\section*{Data availability}
All data and \textit{Python} code are available in the GitHub repository \href{https://github.com/bmartin9/NIRVAR}{\texttt{bmartin9/NIRVAR}}. 


\bibliographystyle{rss}
\singlespacing
\bibliography{reference}

\appendix

\vspace*{3em}

\begin{center}
{\LARGE\textbf{SUPPLEMENTARY MATERIAL}}
\end{center}
\begin{appendices}

\setcounter{section}{0}
\renewcommand{\thesection}{S\arabic{section}}

\setcounter{equation}{0}
\renewcommand{\theequation}{S.\arabic{equation}}

\setcounter{figure}{0}
\renewcommand{\thefigure}{S.\arabic{figure}}

\setcounter{table}{0}
\renewcommand{\thetable}{S.\arabic{table}}

   \section{Theory}
\subsection{Proof of Proposition 4.1}
\begin{proof}
     We prove that the covariance matrix $\Gamma = \mathbb{E}(\bm{X}_{t}\bm{X}_{t}^{{\prime}})$ can be written in terms of $U_{\Phi}$. 
     From \citetSM{lutkepohl2005new}, we have that
$
    \Gamma - \Phi \Gamma \Phi^{{\prime}} = \Sigma.
$
This is an example of a Lyapunov matrix equation and its formal solution is given by \citepSM{young1981rate}:
\begin{align}
    \Gamma = \sum_{k=0}^{\infty} \left(\Phi\right)^{k} \Sigma \left(\Phi^{{\prime}}\right)^{k},
\end{align}
which converges when 
$\rho(\Phi) < 1$ \citepSM{smith1968matrix}. Since we assume a stationary NIRVAR process, then indeed $\rho(\Phi) < 1$ and the series converges. With $\Sigma = \sigma^{2} I_{N}$ we have 
\begin{align}
\label{eq:Gamma-identity}
    \Gamma = \sigma^{2} \sum_{k=0}^{\infty}  \left(\Phi\right)^{2k}.
\end{align}
Recalling the eigendecomposition 
     \begin{align}
    \label{eq:phi-eigendecomp-supplementary} 
    \Phi = U_{\Phi} \Lambda_{\Phi} U_{\Phi}^{\prime} + U_{\Phi,\perp}\Lambda_{\Phi,\perp}U_{\Phi,\perp}^{\prime},
\end{align}
we can substitute Equation~\eqref{eq:phi-eigendecomp-supplementary} into Equation~\eqref{eq:Gamma-identity}, 
\begin{align}\label{eq:cov-conn-derivation}
    \Gamma &= \sigma^{2} \sum_{k=0}^{\infty}  (U_{\Phi} \Lambda_{\Phi} U_{\Phi}^{\prime} + U_{\Phi,\perp}\Lambda_{\Phi,\perp}U_{\Phi,\perp}^{\prime})^{2k}  \\
    &= \sigma^{2}\left( U_{\Phi} \sum_{k=0}^{\infty} \Lambda_{\Phi}^{2k} U_{\Phi}^{\prime} + U_{\Phi,\perp} \sum_{k=0}^{\infty} \Lambda_{\Phi,\perp}^{2k} U_{\Phi,\perp}^{\prime} \right)  \\
    & = \sigma^{2} \left(U_{\Phi} \Lambda_{\Gamma} U_{\Phi}^{\prime} + U_{\Phi,\perp} \Lambda_{\Gamma,\perp} U_{\Phi,\perp}^{\prime}\right),
\end{align}
where $\Lambda_{\Gamma}$ is a diagonal matrix whose diagonal entries are $(\lambda_{\Gamma})_{i} = 1/(1-(\lambda_{\Phi})_{i}^{2})$ with $(\lambda_{\Phi})_{i}$ being the $i$-th diagonal entry of $\Lambda_{\Phi}$.  Therefore, the rank $d$ eigendecomposition of  $\Gamma$ is $\Gamma = U_{\Phi} \Lambda_{\Gamma} U_{\Phi}^{\prime}$. 
 \end{proof}

 \begin{remark}\label{remark:cov-conn}
      Note that the key step in going from the first to the second line of \eqref{eq:cov-conn-derivation} relies on $U_{\Phi}^{\prime}U_{\Phi} =  I_{K}$. If we were to relax the assumption of symmetric $\Phi$ in Proposition \ref{prop:cov-conn}, then \eqref{eq:phi-eigendecomp-supplementary} would be replaced by the SVD,
   $ \Phi = U_{\Phi} \Lambda_{\Phi} V_{\Phi}^{\prime} + U_{\Phi,\perp}\Lambda_{\Phi,\perp}V_{\Phi,\perp}^{\prime}.$
Since $U_{\Phi}^{\prime}V_{\Phi} \neq I_{K}$, the infinite sum in \eqref{eq:cov-conn-derivation} would no longer have a simple form.
 \end{remark}

\subsection{Derivation of the NIRVAR generalised least-squares estimator}\label{sec:der_gls}
The generalised least-squares estimator is obtained by minimising the objective 
\begin{align}
    \label{eq:objective-ols}
    S\{\bm{\gamma}(\hat{A})\} &= \mathbf{u}(\hat A)^{(q) \prime} (I_{T} \otimes \Sigma^{-1}) \mathbf{u}(\hat A)^{(q)} \\
     &= \{\bm{y}^{(q)}- (X^{\prime} \otimes I_{N}) R(\hat{A}) \bm{\gamma}(\hat{A})\}^{\prime}(I_{T} \otimes \Sigma^{-1})\{\bm{y}^{(q)}- (X^{\prime} \otimes I_{N}) R(\hat{A}) \bm{\gamma}(\hat{A})\} \\
    &= \bm{y}^{(q)^{\prime}}(I_{T} \otimes \Sigma^{-1})\bm{y}^{(q)}+ \bm{\gamma}(\hat{A})^{\prime} R(\hat{A})^{\prime}(XX^{\prime} \otimes \Sigma^{-1}) R(\hat{A}) \bm{\gamma}(\hat{A}) -2 \bm{\gamma}(\hat{A})^{\prime} R(\hat{A})^{\prime}(X \otimes \Sigma^{-1}) \,\bm{y}^{(q)}.
\end{align}
Hence,
\begin{equation}
    \label{eq:derivative}
    \frac{\partial S\{\bm{\gamma}(\hat{A})\}}{\partial \bm{\gamma}(\hat{A})} = 2  R(\hat{A})^{\prime}(XX^{\prime} \otimes \Sigma^{-1}) R(\hat{A}) \bm{\gamma}(\hat{A}) - 2 R(\hat{A})^{\prime} (X \otimes \Sigma^{-1}) \,\bm{y}^{(q)}.
\end{equation} 
Equating to zero gives the normal equations
\begin{equation}
    \label{eq:normal-equations}
    R(\hat{A})^{\prime}(XX^{\prime} \otimes \Sigma^{-1}) R(\hat{A}) \bm{\gamma}(\hat{A}) =  R(\hat{A})^{\prime} (X \otimes \Sigma^{-1}) \,\bm{y}^{(q)}, 
\end{equation}
and, consequently, the generalised least-squared estimator is 
\begin{align}
    \label{eq:GLS-appendix}
    \hat{\bm{\gamma}}(\hat{A}) &= \{R(\hat{A})^{\prime}(XX^{\prime} \otimes \Sigma^{-1}) R(\hat{A})\}^{-1} R(\hat{A})^{\prime} (X \otimes \Sigma^{-1}) \bm{y}^{(q)}.
\end{align}
The Hessian of $S\{\bm{\gamma}(\hat{A})\}$, 
\begin{equation}
    \label{eq:Hessian} 
    \frac{\partial^{2} S\{\bm{\gamma}(\hat{A})\}}{\partial \bm{\gamma}(\hat{A}) \partial \bm{\gamma}(\hat{A})^{\prime}} = 2 R(\hat{A})^{\prime}(XX^{\prime} \otimes \Sigma^{-1}) R(\hat{A}),
\end{equation}
is positive definite which confirms that $\hat{\bm{\gamma}}(\hat{A})$ is a minimising vector.

\subsection{Proof of Proposition 5.1} 
\begin{proof}
    Substituting $\bm{y}^{(q)} = \left(X^{\prime} \otimes I_{N} \right)R(A)\bm{\gamma}(A) + \mathbf{u}^{(q)}$ into Equation~\eqref{eq:GLS-appendix} yields 
    \begin{align}
        \label{eq:estimator-re-write}
        \hat{\bm{\gamma}}(\hat{A}) &= \{R(\hat{A})^{\prime}(XX^{\prime} \otimes \Sigma^{-1}) R(\hat{A})\}^{-1} R(\hat{A})^{\prime} (X \otimes \Sigma^{-1})\{(X^{\prime} \otimes I_{N}) R(A) \bm{\gamma}(A) + \mathbf{u}^{(q)}\}  \\
        &= \{R(\hat{A})^{\prime}(XX^{\prime} \otimes \Sigma^{-1}) R(\hat{A})\}^{-1} R(\hat{A})^{\prime}(XX^{\prime} \otimes \Sigma^{-1}) R(A)\bm{\gamma}(A)  \\ &\quad + \{R(\hat{A})^{\prime}(XX^{\prime} \otimes \Sigma^{-1}) R(\hat{A})\}^{-1} R(\hat{A})^{\prime} (I_{NQ} \otimes \Sigma^{-1}) \vecc(U^{(q)}X^{\prime}). 
    \end{align}
    Noting that $\mathbb{E}\{\vecc(U^{(q)}X^{\prime})\} = \mathbb{E}\{(X \otimes I_{N})\} \mathbb{E}(\mathbf{u}^{(q)}) = 0$, we see that the bias of the estimator, given $\hat{A}$, is $\mathbb{E}\{\hat{\bm{\beta}}(\hat{A})|\hat{A}\} = R(\hat{A}) C \bm{\gamma}(A)$, where 
    \begin{align}
        \label{eq:bias-appendix}
        C \coloneqq \{R(\hat{A})^{\prime}(XX^{\prime} \otimes \Sigma^{-1}) R(\hat{A})]^{-1} R(\hat{A})^{\prime}(XX^{\prime} \otimes \Sigma^{-1}) R(A).
    \end{align}
    We have that $\mathrm{colsp}\{R(A)\} \subseteq \mathrm{colsp}\{R(\hat{A})\} \iff R(\hat{A})\bm{\gamma}(\hat{A}) = R(A) \bm{\gamma}(A) = \bm{\beta}$. This is true if and only if
        \begin{equation}
    \begin{aligned}
        \mathbb{E}\{\hat{\bm{\beta}}(\hat{A}) | \hat{A}) &= \mathbb{E}(R(\hat{A})\hat{\bm{\gamma}}(\hat{A}) | \hat{A}) \\ &= R(\hat{A}) \mathbb{E}[\{R(\hat{A})^{\prime}(XX^{\prime} \otimes \Sigma^{-1}) R(\hat{A})\}^{-1} R(\hat{A})^{\prime}(XX^{\prime} \otimes \Sigma^{-1}) R(A) \bm{\gamma}(A)| \hat{A}]  \\
        &= R(\hat{A}) \mathbb{E}[\{R(\hat{A})^{\prime}(XX^{\prime} \otimes \Sigma^{-1}) R(\hat{A})\}^{-1} R(\hat{A})^{\prime}(XX^{\prime} \otimes \Sigma^{-1}) R(\hat{A})\bm{\gamma}(\hat{A})| \hat{A}] \\ &= R(\hat{A}) \bm{\gamma}(\hat{A}) = \bm{\beta}. 
    \end{aligned}
    \end{equation} 
    \qedhere 
\end{proof}

\subsection{Proof of Proposition 5.2}
\begin{proof}
The NIRVAR estimator,
\begin{align}
    \label{eq:estimator-re-write-supplementary}
    \hat{\bm{\gamma}}(\hat{A}) &= \{R(\hat{A})^{\prime}(XX^{\prime} \otimes \Sigma^{-1}) R(\hat{A})\}^{-1} R(\hat{A})^{\prime}(XX^{\prime} \otimes \Sigma^{-1}) R(A)\bm{\gamma}(A)  \\ &\quad + \{R(\hat{A})^{\prime}(XX^{\prime} \otimes \Sigma^{-1}) R(\hat{A})\}^{-1} R(\hat{A})^{\prime} (I_{NQ} \otimes \Sigma^{-1}) \vecc(U^{(q)}X^{\prime}),
\end{align}
can be written as 
\begin{align}
    \label{eq:asy-form}
    \sqrt{T}\left\{\hat{\bm{\gamma}}(\hat{A}) - C \bm{\gamma}(A)\right\} = \left\{R(\hat{A})^{\prime}\left(\frac{XX^{\prime}}{T} \otimes \Sigma^{-1}\right) R(\hat{A}) \right\}^{-1} R(\hat{A})^{\prime} (I_{NQ} \otimes \Sigma^{-1}) \frac{1}{\sqrt{T}} \vecc(U^{(q)}X^{\prime}).
\end{align}
By Lemma 3.1 of \citetSM{lutkepohl2005new}, $\Gamma$ exists and is nonsingular. Also by Lemma 3.1 of \citetSM{lutkepohl2005new}, 
\begin{align}
        \label{eq:lemma31}
        \frac{1}{\sqrt{T}} \vecc(U^{(q)}X^{\prime}) \xrightarrow[]{d} \mathcal{N}(0, \Gamma \otimes \Sigma).
\end{align}
By Equation~\eqref{eq:asy-form}, 
\begin{align}
    \plim\left\{\hat{\bm{\gamma}}(\hat{A}) - C \bm{\gamma}(A)\right\} = \plim \left[  \left\{R(\hat{A})^{\prime}\left(\frac{XX^{\prime}}{T} \otimes \Sigma^{-1}\right) R(\hat{A}) \right\}^{-1} R(\hat{A})^{\prime} (I_{NQ} \otimes \Sigma^{-1})  \vecc\left(\frac{U^{(q)}X^{\prime}}{T}\right) \right]  \\
    = \plim\left[ \left\{R(\hat{A})^{\prime}\left(\frac{XX^{\prime}}{T} \otimes \Sigma^{-1}\right) R(\hat{A}) \right\}^{-1} R(\hat{A})^{\prime} (I_{NQ} \otimes \Sigma^{-1})\right] \plim\left\{\vecc\left(\frac{U^{(q)}X^{\prime}}{T}\right) \right\},
\end{align} 
where the second line follows from Slutsky's Theorem, and the continuous mapping theorem. But Equation~\eqref{eq:lemma31} implies that 
\begin{align}
    \plim\left\{\vecc\left(\frac{UX^{\prime}}{T}\right) \right\} = 0. 
\end{align}
Thus $\plim\{\hat{\bm{\gamma}}(\hat{A}) - C \bm{\gamma}(A)\} = 0$ (weak consistency). 
For convenience, we define 
\begin{align}
    \label{eq:hatG}
    \hat{G} \coloneqq  \left\{R(\hat{A})^{\prime}\left(\frac{XX^{\prime}}{T} \otimes \Sigma^{-1}\right) R(\hat{A}) \right\}^{-1} R(\hat{A})^{\prime} (I_{NQ} \otimes \Sigma^{-1}).
\end{align}
Then 
\begin{align}
    G \coloneqq \plim(\hat{G}) =  \left\{R(\hat{A})^{\prime}\left(\Gamma \otimes \Sigma^{-1}\right) R(\hat{A}) \right\}^{-1} R(\hat{A})^{\prime} (I_{NQ} \otimes \Sigma^{-1}).
\end{align}
By Proposition C.15(1) of \citetSM{lutkepohl2005new}, Equation~\eqref{eq:lemma31} implies that 
\begin{align}
        \label{eq:C15}
        \hat{G}\frac{1}{\sqrt{T}} \vecc(U^{(q)}X^{\prime}) \xrightarrow[]{d} \mathcal{N}(0, G (\Gamma \otimes \Sigma)G^{\prime}).
\end{align}
Thus, $\sqrt{T}\{\hat{\bm{\gamma}}(\hat{A}) - C \bm{\gamma}(A)\}$ is asymptotically normal with asymptotic variance
\begin{equation}
    \resizebox{\textwidth}{!}{$
\begin{aligned}
    &G (\Gamma \otimes \Sigma)G^{\prime} =  \\ &\left[ \left\{R(\hat{A})^{\prime}\left(\Gamma \otimes \Sigma^{-1}\right) R(\hat{A}) \right\}^{-1} R(\hat{A})^{\prime} (I_{NQ} \otimes \Sigma^{-1})\right] (\Gamma \otimes \Sigma) \left[ \left\{R(\hat{A})^{\prime}\left(\Gamma \otimes \Sigma^{-1}\right) R(\hat{A}) \right\}^{-1} R(\hat{A})^{\prime} (I_{NQ} \otimes \Sigma^{-1})\right]^{\prime}  \\ 
    &= \left\{R(\hat{A})^{\prime}\left(\Gamma \otimes \Sigma^{-1}\right) R(\hat{A}) \right\}^{-1} R(\hat{A})^{\prime} (I_{NQ} \otimes \Sigma^{-1}) (\Gamma \otimes \Sigma) (I_{NQ} \otimes \Sigma^{-1}) R(\hat{A}) \left\{R(\hat{A})^{\prime}\left(\Gamma \otimes \Sigma^{-1}\right) R(\hat{A}) \right\}^{-1}   \\ 
    &= \left\{R(\hat{A})^{\prime}\left(\Gamma \otimes \Sigma^{-1}\right) R(\hat{A}) \right\}^{-1} \left\{R(\hat{A})^{\prime}\left(\Gamma \otimes \Sigma^{-1}\right) R(\hat{A}) \right\} \left\{R(\hat{A})^{\prime}\left(\Gamma \otimes \Sigma^{-1}\right) R(\hat{A}) \right\}^{-1}  \\
    &= \left\{R(\hat{A})^{\prime}\left(\Gamma \otimes \Sigma^{-1}\right) R(\hat{A}) \right\}^{-1}.
\end{aligned}
    $}
\end{equation} \qedhere 
\end{proof}

\subsection{Proof of Proposition 5.3} 
\begin{proof}
    By Proposition C.15(1) and Proposition 5.1 of \citetSM{lutkepohl2005new}, 
    \begin{align}
        \label{eq:true-var-appendix}
        \mathrm{var} \left\{ \sqrt{T}\left(\hat{\bm{\beta}}(A) - \bm{\beta}\right)\right\} = R(A)\left[R(A)^{\prime}\left\{\Gamma \otimes \Sigma^{-1}\right\} R(A) \right]^{-1}R(A)^{\prime}.
    \end{align}
    By Proposition C.15(1) of \citetSM{lutkepohl2005new} and Proposition 5.2 of this paper,
    \begin{align}
        \label{eq:estimated-var-appendix}
        \mathrm{var} \left( \sqrt{T}\left(\hat{\bm{\beta}}(\hat{A}) - \bm{\beta}\right)\right) = R(\hat{A})\left\{R(\hat{A})^{\prime}\left(\Gamma \otimes \Sigma^{-1}\right) R(\hat{A}) \right\}^{-1}R(\hat{A})^{\prime}.
    \end{align}
    By Lemma 3.1 of \citetSM{lutkepohl2005new}, $\Gamma$ is positive definite. By assumption, $\Sigma$ is positive definite. Given this and the fact that both $\Gamma$ and $\Sigma$ are symmetric, $\Gamma \otimes \Sigma^{-1}$ is positive definite and symmetric. By Observation 7.1.8 of \citeSM{horn2012matrix}, since $\rank{(R(A))} = M$, then  $R(A)^{\prime}\left\{\Gamma \otimes \Sigma^{-1}\right\} R(A)$ is positive definite with rank $M$. By inspection, $R(A)^{\prime}\left\{\Gamma \otimes \Sigma^{-1}\right\} R(A)$ is symmetric which implies $\left[R(A)^{\prime}\left\{\Gamma \otimes \Sigma^{-1}\right\} R(A) \right]^{-1}$ is symmetric and of rank $M$. Using Observation 7.1.8 of \citeSM{horn2012matrix} again, the matrix $R(A)\left[R(A)^{\prime}\left\{\Gamma \otimes \Sigma^{-1}\right\} R(A) \right]^{-1}R(A)^{\prime}$ is positive definite with rank $M$. By the same argument, the matrix $R(\hat{A})\left\{R(\hat{A})^{\prime}\left(\Gamma \otimes \Sigma^{-1}\right) R(\hat{A}) \right\}^{-1}R(\hat{A})^{\prime}$ is positive definite with rank $\widehat{M}$. The assumption that $\mathrm{colsp}\{R(A)\} \subseteq \mathrm{colsp}\{R(\hat{A})\}$ implies that
    \begin{equation}
        \mathrm{colsp}\left[R(A)\left\{R(A)^{\prime}\left(\Gamma \otimes \Sigma^{-1}\right) R(A) \right\}^{-1}R(A)^{\prime}\right] \subseteq \mathrm{colsp}\left[R(\hat{A})\left\{R(\hat{A})^{\prime}\left(\Gamma \otimes \Sigma^{-1}\right) R(\hat{A}) \right\}^{-1}R(\hat{A})\right].
    \end{equation}
    Thus 
    \begin{equation}
        P \coloneqq R(\hat{A})\left\{R(\hat{A})^{\prime}\left(\Gamma \otimes \Sigma^{-1}\right) R(\hat{A}) \right\}^{-1}R(\hat{A})^{\prime} - R(A)\left\{R(A)^{\prime}\left(\Gamma \otimes \Sigma^{-1}\right) R(A) \right\}^{-1}R(A)^{\prime}
    \end{equation}
    has $\widehat{M} - M$ nonzero eigenvalues with all other eigenvalues being 0. Hence, $P$ is positive semi-definite and
         \begin{align}
        \label{eq:PSD-supplementary}
        \mathrm{var} \left[ \sqrt{T}\left\{\hat{\bm{\beta}}(\hat{A}) - \bm{\beta}\right\}\right] - \mathrm{var} \left[ \sqrt{T}\left\{\hat{\bm{\beta}}(A) - \bm{\beta}\right\}\right] &\succeq 0.\qedhere
    \end{align} 

\end{proof}

\subsection{Proof of Proposition \ref{prop:mspe}}
Conditional on $\hat{A}$, the mean squared error can be decomposed into a bias, variance, and noise term:
\begin{align}
\mathbb{E}_{\mathcal{D},\bm{X}_{\star},\bm{\epsilon}_{\star}}&\left\{ \lVert \hat{\Phi}\bm{X}_{\star} - (\Phi\bm{X}_{\star} + \bm{\epsilon}_{\star}) \rVert_{2}^{2} | \hat{A}\right\} = \mathbb{E}_{\mathcal{D},\bm{X}_{\star}}\{\lVert \hat{\Phi}\bm{X}_{\star} - \Phi\bm{X}_{\star}\rVert_{2}^{2} |\hat{A}\} + \mathbb{E}_{\bm{\epsilon}_{\star}}(\bm{\epsilon}_{\star}^{\prime}\bm{\epsilon}_{\star}) \\
&= \mathbb{E}_{\mathcal{D},\bm{X}_{\star}}\{(\hat{\bm{\beta}}(\hat{A}) - \bm{\beta})^{\prime}(\bm{X}_{\star}\bm{X}_{\star}^{\prime} \otimes I_{N})(\hat{\bm{\beta}}(\hat{A}) - \bm{\beta}) | \hat{A} \} + \text{tr}(\Sigma) \\ 
&= [\mathbb{E}_{\mathcal{D}}\{ \hat{\bm{\beta}}(\hat{A}) | \hat{A}\} - \bm{\beta}]^{\prime}(\Gamma \otimes I_{N})[\mathbb{E}_{\mathcal{D}}\{ \hat{\bm{\beta}}(\hat{A}) | \hat{A}\} - \bm{\beta}] \\ 
&\qquad + \mathbb{E}_{\mathcal{D}}([\hat{\bm{\beta}}(\hat{A}) - \mathbb{E}_{\mathcal{D}}\{ \hat{\bm{\beta}}(\hat{A}) | \hat{A}\}]^{\prime}(\Gamma \otimes I_{N})[\hat{\bm{\beta}}(\hat{A}) - \mathbb{E}_{\mathcal{D}}\{ \hat{\bm{\beta}}(\hat{A}) | \hat{A}\}] | \hat{A}) + \text{tr}(\Sigma) \\
&= \text{Bias}_{\mathcal{D}}^{2}\{\hat{\bm{\beta}}(\hat{A}) | \hat{A} \} + \text{tr}[(\Gamma \otimes I_{N}) \text{var}_{\mathcal{D}}\{\hat{\bm{\beta}}(\hat{A}) | \hat{A} \}] + \text{tr}(\Sigma).
\end{align}
We now provide upper bounds on $\plim \text{Bias}_{\mathcal{D}}^{2}\{\hat{\bm{\beta}}(\hat{A}) | \hat{A} \}$ and $\plim T\text{tr}[(\Gamma \otimes I_{N}) \text{var}_{\mathcal{D}}\{\hat{\bm{\beta}}(\hat{A}) | \hat{A} \}]$. Since $\Sigma$ is a non-singular covariance matrix, $\Sigma \preceq \lambda_{1}(\Sigma)I_{N}$, implying $I_{N} \preceq \lambda_{1}(\Sigma) \Sigma^{-1}$. Therefore, 
\begin{align}\label{eq:bias-bound}
    \text{Bias}_{\mathcal{D}}^{2}\{\hat{\bm{\beta}}(\hat{A}) | \hat{A} \} \leq \lambda_{1}(\Sigma)[\mathbb{E}_{\mathcal{D}}\{ \hat{\bm{\beta}}(\hat{A}) | \hat{A}\} - \bm{\beta}]^{\prime}(\Gamma \otimes \Sigma^{-1})[\mathbb{E}_{\mathcal{D}}\{ \hat{\bm{\beta}}(\hat{A}) | \hat{A}\} - \bm{\beta}].
\end{align}
We can write $\bm{\beta} = R(\hat{A})\bm{\gamma}_{1}(A,\hat{A}) + R(\hat{A}^{(FN)})\bm{\gamma}_{0}(A,\hat{A})$ where $\bm{\gamma}_{1}(A,\hat{A}) = R(\hat{A})^{\prime}\bm{\beta} \in \mathbb{R}^{\widehat{M}}$. Using this expression for $\bm{\beta}$ and Proposition \ref{prop:bias}, 
\begin{align}\label{eq:gamma0}
    \mathbb{E}_{\mathcal{D}}&\{ \hat{\bm{\beta}}(\hat{A}) | \hat{A}\} - \bm{\beta} = R(\hat{A})\{R(\hat{A})^{\prime}(XX^{\prime} \otimes \Sigma^{-1}) R(\hat{A})\}^{-1} R(\hat{A})^{\prime}(XX^{\prime} \otimes \Sigma^{-1})\bm{\beta} - \bm{\beta} \\
    &= R(\hat{A})\{R(\hat{A})^{\prime}(XX^{\prime} \otimes \Sigma^{-1}) R(\hat{A})\}^{-1} R(\hat{A})^{\prime}(XX^{\prime} \otimes \Sigma^{-1})\{R(\hat{A})\bm{\gamma}_{1}(A,\hat{A}) + R(\hat{A}^{(FN)})\bm{\gamma}_{0}(A,\hat{A})\} \\ &\quad - R(\hat{A})\bm{\gamma}_{1}(A,\hat{A}) - R(\hat{A}^{(FN)})\bm{\gamma}_{0}(A,\hat{A}) \\ 
    &= R(\hat{A})\{R(\hat{A})^{\prime}(XX^{\prime} \otimes \Sigma^{-1}) R(\hat{A})\}^{-1} R(\hat{A})^{\prime}(XX^{\prime} \otimes \Sigma^{-1})R(\hat{A}^{(FN)})\bm{\gamma}_{0}(A,\hat{A}) - R(\hat{A}^{(FN)})\bm{\gamma}_{0}(A,\hat{A}).
\end{align}
Substituting \eqref{eq:gamma0} into \eqref{eq:bias-bound} and using $\plim XX^{\prime}/T = \Gamma$, 
\begin{align}
    \plim \text{Bias}_{\mathcal{D}}^{2} &\{\hat{\bm{\beta}}(\hat{A}) | \hat{A} \} \leq \lambda_{1}(\Sigma) \bm{\gamma}_{0}(A,\hat{A})^{\prime} R(\hat{A}^{(FN)})^{\prime} [ (\Gamma \otimes \Sigma^{-1}) \\ &\qquad- (\Gamma \otimes \Sigma^{-1})  R(\hat{A})\{R(\hat{A})^{\prime}(\Gamma \otimes \Sigma^{-1}) R(\hat{A})\}^{-1} R(\hat{A})^{\prime} (\Gamma \otimes \Sigma^{-1})]R(\hat{A}^{(FN)}) \bm{\gamma}_{0}(A,\hat{A}) \\ 
    &\leq \lambda_{1}(\Sigma) \bm{\gamma}_{0}(A,\hat{A})^{\prime} R(\hat{A}^{(FN)})^{\prime} (\Gamma \otimes \Sigma^{-1}) R(\hat{A}^{(FN)}) \bm{\gamma}_{0}(A,\hat{A}) \\
    &\leq \lambda_{1}(\Sigma) \lambda_{1}(\Gamma \otimes \Sigma^{-1}) \lVert \bm{\gamma}_{0}(A,\hat{A}) \rVert_{2}^{2} \\
    &= \kappa(\Sigma) \lambda_{1}(\Gamma) \lVert \bm{\gamma}_{0}(A,\hat{A}) \rVert_{2}^{2}.
\end{align}
Next, we consider the variance term. We start by writing 
\begin{align}
\text{var}_{\mathcal{D}}\{\hat{\bm{\beta}}(\hat{A}) | \hat{A} \} = \frac{1}{T}R(\hat{A}) \{R(\hat{A})^{\prime}(XX^{\prime}/T \otimes \Sigma^{-1}) R(\hat{A})\}^{-1}R(\hat{A})^{\prime}.
\end{align}
Using $\plim XX^{\prime}/T = \Gamma$, the identity $I_{N} \preceq \lambda_{1}(\Sigma) \Sigma^{-1}$, and the cyclic property of $\text{tr}$ gives 
\begin{align}
    \plim T \text{tr}[(\Gamma \otimes I_{N}) \text{var}_{\mathcal{D}}\{\hat{\bm{\beta}}(\hat{A}) | \hat{A} \}] &=  \text{tr}[(\Gamma \otimes I_{N}) R(\hat{A}) \{R(\hat{A})^{\prime}(\Gamma \otimes \Sigma^{-1}) R(\hat{A})\}^{-1}R(\hat{A})^{\prime}] \\ 
    &\leq \lambda_{1}(\Sigma) \text{tr}[(\Gamma \otimes \Sigma^{-1}) R(\hat{A}) \{R(\hat{A})^{\prime}(\Gamma \otimes \Sigma^{-1}) R(\hat{A})\}^{-1}R(\hat{A})^{\prime}] \\ 
    &= \lambda_{1}(\Sigma) \text{tr}[R(\hat{A})^{\prime}(\Gamma \otimes \Sigma^{-1}) R(\hat{A}) \{R(\hat{A})^{\prime}(\Gamma \otimes \Sigma^{-1}) R(\hat{A})\}^{-1}] \\
    &= \lambda_{1}(\Sigma) \text{tr}(I_{\widehat{M}}) \\
    &= \lambda_{1}(\Sigma) \widehat{M}. 
\end{align}
\qed

\subsection{Derivation of the Inverse Mar\v{c}enko-Pastur Distribution}\label{sec:inv_marc}
Suppose $X \sim f_{\text{MP}}(\eta,\sigma^{2})$ where 
\begin{align}
f_{\text{MP}}(x;\eta,\sigma^{2}) = 
\begin{cases} 
\frac{1}{2\pi\sigma^{2} \eta} \sqrt{(x-x_{-})(x_{+}-x)} & \text{if } x_{-} \leq x \leq x_{+}, \\ 
0 & \text{otherwise},
\end{cases}
\end{align}
is the Mar\v{c}enko-Pastur distribution with $N/T \xrightarrow[]{} \eta \in (0,1)$ as $N,T \xrightarrow[]{} \infty$, $\sigma^{2}$ being the variance, and
\begin{align}
    x_{\pm} = \sigma^{2} (1 \pm \sqrt{\eta})^{2}
\end{align}
being the upper and lower bounds of the support of the distribution.

\begingroup
\allowdisplaybreaks
Let $Y = 1/X$. Then the distribution, $f_{\text{IMP}}$, of $Y$, called the inverse Mar\v{c}enko-Pastur distribution, has finite support over which its density is
\begin{align}
    f_{\text{IMP}}(y;\eta,\sigma^{2}) &= f_{\text{MP}}(y^{-1};\eta,\sigma^{2}) \frac{1}{y^{2}} = \frac{1}{2\pi\sigma^{2} \eta y^{-1}} \sqrt{(y^{-1}-x_{-})(x_{+} - y^{-1})} \frac{1}{y^{2}} \\
    &= \frac{1}{2\pi\sigma^{2} \eta y^{2}} \sqrt{[1 - y\sigma^{2} (1 - \sqrt{\eta})^{2}][y\sigma^{2} (1 + \sqrt{\eta})^{2} - 1)]} \\
    &= \frac{1}{2\pi\sigma^{2} \eta y^{2}} \sqrt{\sigma^{4}(1-\eta)^{2} \left(\frac{1}{\sigma^{2}} \left[ \frac{1+\sqrt{\eta}}{1-\eta}\right]^{2} - y \right) \left( y -\frac{1}{\sigma^{2}} \left[ \frac{1-\sqrt{\eta}}{1-\eta}\right]^{2} \right)}  \\
    &= \frac{1-\eta}{2 \pi \eta y^{2}} \sqrt{(y_{+} - y)(y - y_{-})},
\end{align} 
where 
\begin{align}
    y_{\pm} = \frac{1}{\sigma^{2}} \left( \frac{1 \pm \sqrt{\eta}}{1-\eta}\right)^{2}.
\end{align}
Therefore, the distribution of $Y$ is 
\begin{align}
    \label{eq:imp-supplementary}
    f_{\text{IMP}}(y;\eta,\sigma^{2}) = 
    \begin{cases}
        (1-\eta) \sqrt{(y_{+} - y)(y - y_{-})} / (2 \pi \eta y^{2}) & \text{if } y_{-} \leq y \leq y_{+}, \\ 
0 & \text{otherwise}.
    \end{cases}
\end{align}
\endgroup

\section{Extension to multiple lags}\label{sec:NIRVAR(p)}
The NIRVAR model can be extended to incorporate multiple lag orders by assuming that the VAR coefficient matrices at each lag share the same block structure. In particular, a lag $p$ NIRVAR model is given by
\begin{align}
    \label{eq:nirvar-var-lagp}
    \bm{X}_{t}^{(q)} = \sum_{j=1}^{p}\sum_{r=1}^{Q} (A_{q}^{(r,j)} \odot \Tilde{\Phi}_{q}^{(r,j)}) \bm{X}_{t-j}^{(r)} + \bm{\epsilon}_{t}^{(q)}, 
\end{align}
with $(A_{q}^{(r,j)},\Theta_{q}^{(r,j)})\sim\mathrm{SBM}(B^{(r,j)},\pi^{(r)})$ and where we assume that $\mathbb{E}(A_{q}^{(r,j)} \odot \Tilde{\Phi}_{q}^{(r,j)}) = U\Lambda^{(r,j)}U^{\prime}$. As such, the eigenvectors of $\mathbb{E}(\Phi_{q}^{(r,j)})$ are the same for each $j \in [p]$, while we allow the eigenvalues to be different. This is useful in a VAR$(p)$ modeling context where the spectral norms of the VAR coefficient matrices differ as $j \in [p]$ increases.

We extend the NIRVAR estimation procedure to the setting  $p>1$ by employing the companion matrix representation  of a VAR($p$). For ease of notation, we set $Q=1$. The case of $Q>1$ is exactly as described in Section \ref{sec:est} upon writing the VAR$(p)$ in companion form.

Let $\breve{\bm{X}}_{t} \coloneqq (\bm{X}_{t}^{\prime},\dots,\bm{X}_{t-p+1}^{\prime})^{\prime}$, $\breve{\bm{\epsilon}}_{t} \coloneqq (\bm{\epsilon}_{t}^{\prime},0,\dots,0)^{\prime}$, and 
\begin{equation}
    \breve{\Phi} \coloneqq \begin{bmatrix}
\Phi^{(1)} & \Phi^{(2)} & \cdots & \Phi^{(p-1)} & \Phi^{(p)} \\
I_{N}    & 0        & \cdots & 0          & 0       \\
0        & I_{N}    & \cdots & 0          & 0       \\
\vdots   & \vdots   & \ddots & \vdots     & \vdots  \\
0        & 0        & \cdots & I_{N}      & 0
\end{bmatrix}.
\end{equation}
Then the NIRVAR model with $Q=1$ for $p>1$ is $\breve{\bm{X}}_{t} = \breve{\Phi}\breve{\bm{X}}_{t-1} + \breve{\bm{\epsilon}}_{t}$. Let $\breve{\Gamma} \coloneqq \mathbb{E}(\breve{\bm{X}}_{t} \breve{\bm{X}}_{t}^{\prime})$ be the  $Np \times Np$ covariance matrix and $\breve{S}_{T}$ be the corresponding sample covariance matrix with eigendecomposition $\breve{S}_{T} = U_{\breve{S}}\Lambda_{\breve{S}}U_{\breve{S}}^{\prime}$. NIRVAR estimation for $p>1$ consists of setting the $N$ spectral embedded points to be $\hat{\Psi} = (U_{\breve{S}})_{N(p-1)+1:Np, 1:dp}$. In words, we extract the top $dp$ eigenvectors of $U_{\breve{S}}$ and use as embeddings the bottom $N$ rows of these eigenvectors. The embedded points are clustered using a Gaussian mixture model and the resulting block-restricted VAR parameters are estimated via OLS \citep[for details of restricted VAR($p$) estimation via OLS, see][Chapter 5]{lutkepohl2005new}.

Justification for looking at the bottom $N$ rows of $U_{\breve{S}}$ comes from Proposition \ref{prop:companion} below, which proves that the bottom $N$ rows of $U_{\breve{\Gamma}_{E}}$ consists of columns that span the same subspace as the columns of $U$. Here, $U_{\breve{\Gamma}_{E}}$ are the eigenvectors of $\breve{\Gamma}_{E}$ where $\breve{\Gamma}_{E} = \mathbb{E}\{(\breve{\bm{X}}_{t}^{E}) (\breve{\bm{X}}_{t}^{E})^{\prime}\}$ with $\breve{\bm{X}}_{t}^{E} = \mathbb{E}(\breve{\Phi}) \breve{\bm{X}}_{-1}^{E} + \breve{\bm{\epsilon}_{t}}$. In words, $\breve{\Gamma}_{E}$ is the covariance matrix that results from replacing $\Phi^{(j)}$ with $\mathbb{E}(\Phi^{(j)})$ in the NIRVAR model with $p>1$. Now the eigenvectors of $\Phi^{(j)}$ will concentrate around the eigenvectors $U$ of $\mathbb{E}(\Phi^{(j)})$ as long as the SBM sparsity is greater than order $\log{N}/N$ \citep{lei2015consistency}. We expect, although do not prove, that $\breve{S}_{T}$ will concentrate around $\breve{\Gamma}$ (as $T \to \infty$) and $\breve{\Gamma}$ will concentrate around $\breve{\Gamma}_{E}$ (as $N \to \infty$). Therefore, it is reasonable to expect that the subspace spanned by $(U_{\breve{S}})_{N(p-1)+1:Np, 1:dp}$ is close to the subspace spanned by $U_{:,1:d}$, which encodes the community structure of the SBM. ``Close'' here can be quantified by defining a metric measuring the distance between subspaces \citep[see][for definitions of metrics measuring the distance between two subspaces]{chen2021spectral}.

\begin{proposition}\label{prop:companion}
    Let $P_{j} \coloneqq \mathbb{E}(\Phi^{(j)})$ have eigendecomposition $U\Lambda^{(j)}U^{\prime}$. Let $P \coloneqq \mathbb{E}({\breve{\Phi}})$. Then $P$ is diagonalisable with eigendecomposition denoted by $V_{P}D_{P}V_{P}^{-1}$ and whereby the span of the columns of $(V_{P})_{N(p-1)+1:Np,1:Np}$ equals the span the columns of $U$ (considered as vectors). Furthermore, assuming stationarity and homoskedastic errors ($\Sigma = \sigma^{2}I_{N}$), the eigenvectors of $\breve{\Gamma}_{E}$ are also given by $V_{P}$. 
\end{proposition}

\begin{proof}
    First we show that $P$ is diagonalisable. Let $\breve{U} \coloneqq \text{diag}(U,\dots,U)$ be a $Np \times Np$ block diagonal matrix. Then 
\begin{equation}
    \breve{U}^{\prime}P\breve{U} = \begin{bmatrix}
\Lambda^{(1)} & \Lambda^{(2)} & \cdots & \Lambda^{(p-1)} & \Lambda^{(p)} \\
I_{N}    & 0        & \cdots & 0          & 0       \\
0        & I_{N}    & \cdots & 0          & 0       \\
\vdots   & \vdots   & \ddots & \vdots     & \vdots  \\
0        & 0        & \cdots & I_{N}      & 0
\end{bmatrix}.
\end{equation}
It will be useful for us to perform a change of basis so that $\breve{U}^{\prime}P\breve{U}$ is in block diagonal form. As such, we introduce the $Np \times Np$ permutation matrix $\Pi$ whose elements are defined by 
\begin{equation}
    \Pi_{ml} = \indfun\{m = (i-1)p+j \quad \text{and} \quad l = (j-1)N + i, \quad i \in [N], j \in [p]\}.
\end{equation}
Note that $\Pi^{-1} = \Pi^{\prime}$. Then $\Pi^{\prime}\breve{U}^{\prime}P\breve{U} \Pi = \bigoplus_{i=1}^{N} \breve{\Lambda}_{i}$ where 
\begin{equation}
    \breve{\Lambda}_{i} = \begin{bmatrix}
\lambda_{i}^{(1)} & \lambda_{i}^{(2)} & \cdots & \lambda_{i}^{(p-1)} & \lambda_{i}^{(p)} \\
1    & 0        & \cdots & 0          & 0       \\
0        & 1    & \cdots & 0          & 0       \\
\vdots   & \vdots   & \ddots & \vdots     & \vdots  \\
0        & 0        & \cdots & 1      & 0
\end{bmatrix}.
\end{equation}
Denote $\breve{\Lambda} \coloneqq \bigoplus_{i=1}^{N} \breve{\Lambda}_{i}$ for convenience. The eigenvalues of $\breve{\Lambda}_{i}$ are found by solving the characteristic equation $\mu^{p} - \lambda_{i}^{(1)} \mu^{p-1} - \lambda_{i}^{(2)} \mu^{p-2} - \dots - \lambda_{i}^{(p)} = 0 $. Denote the $j$-th eigenvalue of $\breve{\Lambda}_{i}$ by $\mu_{i,j}$. The eigenvector corresponding to $\mu_{i,j}$ is $\bm{w}(\mu_{i,j}) \coloneqq (\mu_{i,j}^{p-1},\mu_{i,j}^{p-2},\dots,1)^{\prime}$ which can easily be verified by noting that $\breve{\Lambda}_{i}\bm{w}(\mu_{i,j}) = \mu_{i,j} \bm{w}(\mu_{i,j})$. Let $W_{i} \coloneqq (\bm{w}(\mu_{i,1}),\dots,\bm{w}(\mu_{i,p}))$ be the $p \times p$ matrix of eigenvectors of $\breve{\Lambda}_{i}$, so that $\breve{\Lambda}_{i}W_{i} = W_{i}D_{i}$ with $D_{i} \coloneqq \text{diag}(\mu_{i,1},\dots,\mu_{i,p})$. Let $W \coloneqq \bigoplus_{i=1}^{N} W_{i}$ and $D \coloneqq \bigoplus_{i=1}^{N} D_{i}$. Then $\breve{\Lambda}W = WD$. Recalling that $\breve{\Lambda} = \Pi^{\prime}\breve{U}^{\prime}P\breve{U} \Pi$ we see that $P = VDV^{-1}$ with $V \coloneqq \breve{U}\Pi W$. This proves that $P$ is diagonalisable. 

For the second claim, we examine $\breve{U}\Pi W$: the $(ij)$-th column of $V$ is $\bm{v}_{i,j} =  (\mu_{i,j}^{p-1}\bm{u}_{i}^{\prime},\mu_{i,j}^{p-2}\bm{u}_{i}^{\prime},\dots,\bm{u}_{i}^{\prime})^{\prime}$ where $\bm{u}_{i}$ is the $i$-th column of $U$. Thus, the columns of the bottom $N$ rows of $V$ consist of $\bm{u}_{1},\dots,\bm{u}_{N}$, repeated $p$ times each. Therefore the span of the columns of $(V_{P})_{N(p-1)+1:Np,1:Np}$ equals the span the columns of $U$, considered as vectors. 

For the final claim, since $P$ is diagonalisable, $\rho(P) < 1$ (under the stationarity assumption), and $\breve{\Gamma}_{E} - P\breve{\Gamma}_{E}P^{\prime} = \sigma^{2}I_{Np}$, then it follows directly from the proof of Proposition \ref{prop:cov-conn} that $\breve{\Gamma}_{E}$ and $P$ share the same eigenspace. 
\end{proof}

\section{Simulation study} 
This section contains further simulation experiments exploring robustness, sensitivity to mis-specification of model parameters, the predictive performance of NIRVAR compared with existing methods, as well as a modified version of the NIRVAR estimator that uses penalised regression for inter-block edges in order to reduce estimation bias.

\subsection{Large sample distribution of the NIRVAR estimator}
\begin{figure}
\centering
\begin{subfigure}{.49\textwidth}
  \includegraphics[width=\linewidth]{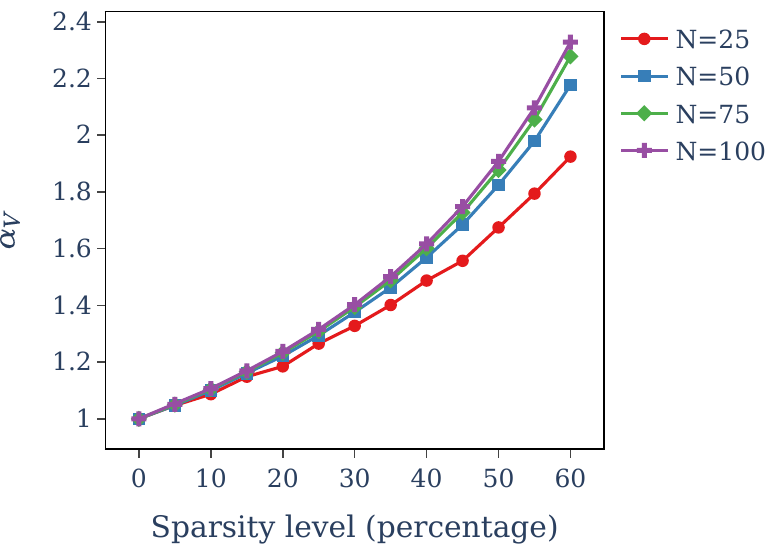}
  \caption{}
  \label{fig:variance-sub1}
\end{subfigure}
\begin{subfigure}{.49\textwidth}
  \includegraphics[width=\linewidth]{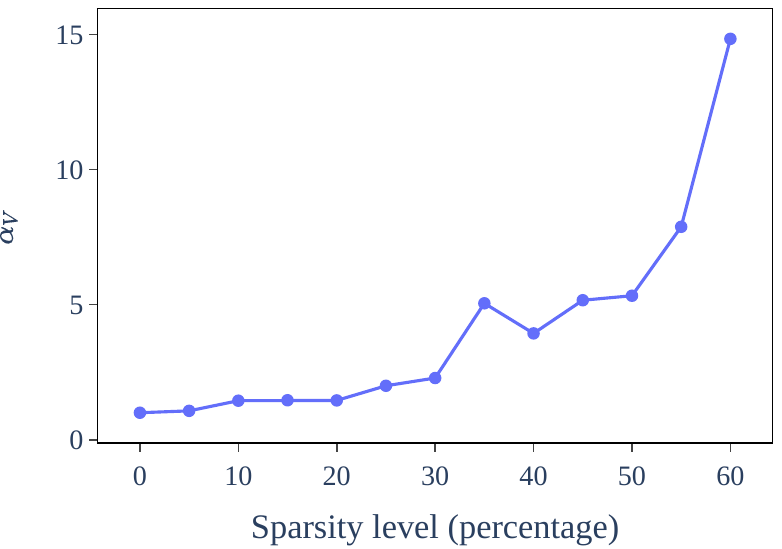}
  \caption{}
  \label{fig:variance-sub2}
\end{subfigure}
\caption{Study of asymptotic variance. (a) Comparing the asymptotic variances of a NIRVAR estimator and an estimator which imposes restrictions within the blocks of the VAR coefficient matrix for different levels of intra-block sparsity and different values of $N$. (b) The same comparison as (a) but using FRED-MD data.}
\label{fig:variance}
\end{figure}

This section reports further results relating to Proposition 5.3 and briefly discussed at the end of Section 6.2. Using the same simulation setting as in Section 6.2, we computed $\alpha_{V}$ for different levels of intra-block sparsity of the ground truth network. Figure \ref{fig:variance}(\subref{fig:variance-sub1}) shows that the ratio $\alpha_{V}$ of the asymptotic variance of a NIRVAR estimator to that of an estimator which uses all restrictions grows at a super-linear rate. This rate increases with the number of panel components, $N$. We also conducted the same experiment using the FRED-MD dataset from Section 7.2. In this case, it was necessary to estimate the asymptotic variances in Equation~\eqref{eq:true-var-appendix} and Equation~\eqref{eq:estimated-var-appendix} using the sample covariance matrix $\mathcal{X}\mathcal{X}^{\prime}/T$. Figure \ref{fig:variance}(\subref{fig:variance-sub2}) shows that in this setting of real data, $\alpha_{V}$ grows much faster with increased intra-block sparsity compared with the simulation setting. 

\subsection{Stability properties of embeddings of stacked sample covariance matrices}
\begin{figure}
\centering
\begin{subfigure}{.49\textwidth}
  \includegraphics[width=\linewidth]{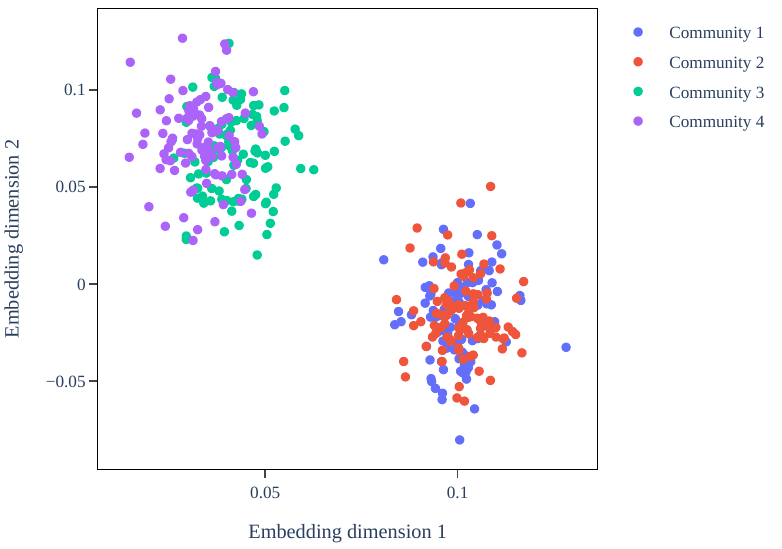}
  \caption{}
  \label{fig:multipe-features-sub1}
\end{subfigure}
\begin{subfigure}{.49\textwidth}
  \includegraphics[width=\linewidth]{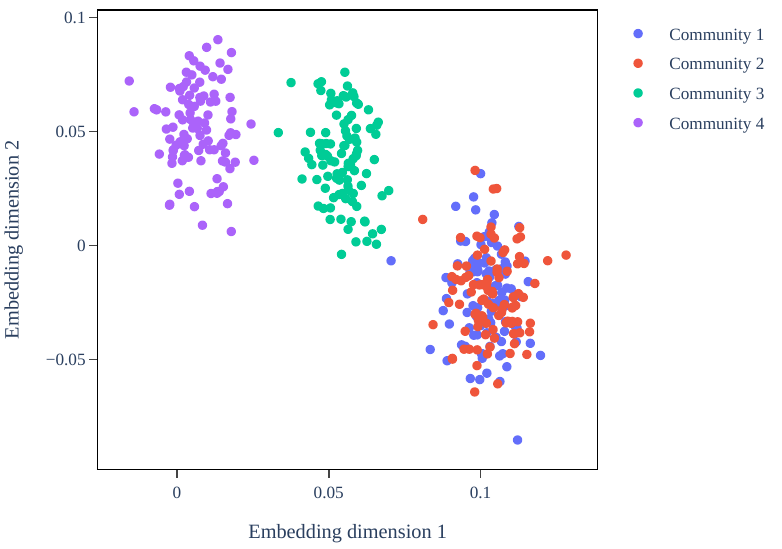}
  \caption{}
  \label{fig:multipe-features-sub2}
\end{subfigure}
\caption{First two dimensions of the SVD embedding of stacked sample covariance matrices for multiple features. (a) Embeddings of feature 1.  (b) Embeddings of feature 2. }
\label{fig:multipe-features}
\end{figure}

We simulate data from a NIRVAR model with $Q=2$, $N=400$, $T=20,000$, $K = 4$, $\rho(\Xi) = 0.9$, and $\Tilde{\Phi} \propto I_{KN}$. We set $A_{1}^{(1)}, A_{1}^{(2)} \sim \text{Bernoulli}(B_{1})$ and $A_{2}^{(1)}, A_{2}^{(2)} \sim \text{Bernoulli}(B_{2})$ where 
\begin{equation}
B_{1} = \begin{pmatrix}
0.8  & 0.8  & 0.05 & 0.05 \\
0.8  & 0.8  & 0.05 & 0.05 \\
0.05 & 0.05 & 0.8  & 0.8  \\
0.05 & 0.05 & 0.8  & 0.8 
\end{pmatrix}
\quad
B_{2} = \begin{pmatrix}
0.8  & 0.8  & 0.05 & 0.05 \\
0.8  & 0.8  & 0.05 & 0.05 \\
0.8  & 0.05 & 0.05 & 0.01 \\
0    & 0    & 0.01 & 0.05 
\end{pmatrix}.
\end{equation}
We will refer to row $k$ of $B_{1}, B_{2}$ as community $k$ for $k = 1,\dots,4$. Figure \ref{fig:multipe-features} shows the first two dimensions of the embedding of $(S^{(1)} | S^{(2)})$, with the embeddings for feature 1 and 2 given by Figure \ref{fig:multipe-features}(\subref{fig:multipe-features-sub1}) and \ref{fig:multipe-features}(\subref{fig:multipe-features-sub2}), respectively. Communities 1 and 2 are non-identifiable since the first two rows of $B_{1}$ and $B_{2}$ are identical. 
Indeed Figure \ref{fig:multipe-features} shows that the latent positions of vertices in communities 1 and 2 are similar for a given feature and across features, demonstrating both cross-sectional and longitudinal stability. In contrast, row 4 of $B_{1}$ is different from row 4 of $B_{2}$, leading to different latent positions for community 4 across features. Moreover, rows 3 and 4 of $B_{2}$ are different, leading to a separation of communities 3 and 4 in Figure \ref{fig:multipe-features}(\subref{fig:multipe-features-sub2}). 

\subsection{Clustering under a NIRVAR model with multiple lags} \label{subsec:multilag-clustering}
We simulate data from a NIRVAR model with lag $p=3$, $N=75$, $T=2000$, $Q=1$, $K=2$, $p^{(\text{in})} = 0.9$, and $p^{(\text{out})} = 0.1$. To set $\Tilde{\Phi}^{(j)}, j \in [p]$, we sampled each entry $\Tilde{\Phi}_{lm}^{(j)}$ independently from a Uniform(0,1) distribution and then normalised so that $\rho(\breve{\Phi}) \in \{0.5,0.55,\dots,0.95\}$. The objective is to compare spectral clustering using $S_{T}$ versus $\breve{S}_{T}$ (defined in Section \ref{sec:NIRVAR(p)}). The ARI between the two clustering methods and the ground truth clusters were computed for different values of $\rho(\breve{\Phi})$. For each $\rho(\breve{\Phi})$, 20 replications were produced and the corresponding standard error of the ARI calculated. Figure \ref{fig:sim-p-K-d} shows that clustering using $\breve{S}_{T}$ instead of $S_{T}$ leads to a higher ARI when the data generating process is a NIRVAR model with $p>1$. Figure \ref{fig:sim-p-K-d} also shows the ARI calculated from embedding $\sum_{t=2}^{T} (\bm{x}_{t}^{\prime}, \bm{x}_{t-1}^{\prime})^{\prime} (\bm{x}_{t}^{\prime}, \bm{x}_{t-1}^{\prime})$, labeled as ``lag 1'' since it contains information from the lag 0 and lag 1 sample autocovariances. Proposition \ref{prop:companion} suggests that, under a NIRVAR($p$) model, ``lag $j$'' and ``lag $k$'' estimated clusters may have a high ARI for $j,k \in [p]$  and a low ARI for $j \in [p]$ and $k>p$. This stability of clusterings for $j \leq p$ may be used to select the lag order under a NIRVAR($p$) model. 

\subsection{Sensitivity of NIRVAR estimation to $K$ and $d$}
\begin{figure}[t]
\centering
\begin{subfigure}{.49\textwidth}
  \includegraphics[width=\linewidth]{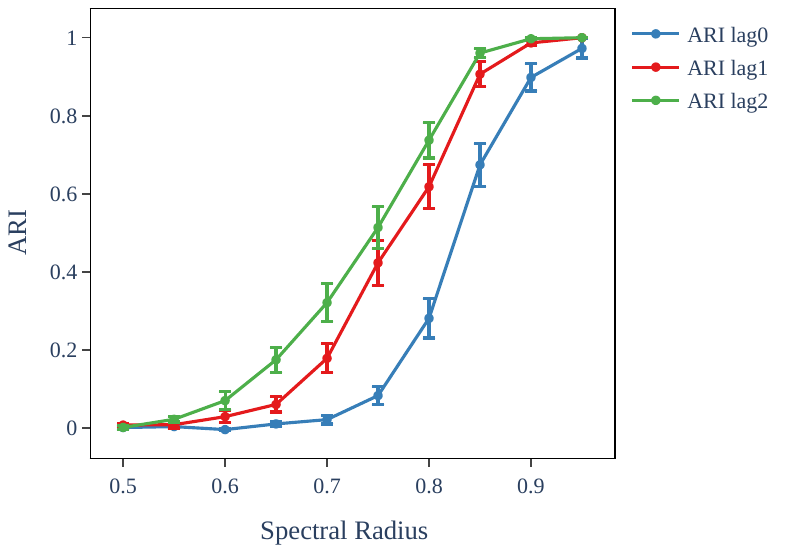}
  \caption{}
  \label{fig:sim-p-K-d-sub1}
\end{subfigure}
\begin{subfigure}{.49\textwidth}
  \includegraphics[width=\linewidth]{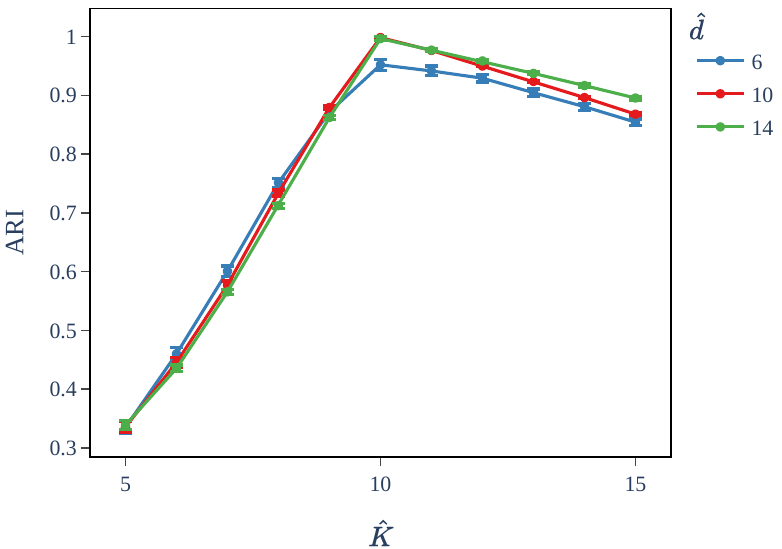}
  \caption{}
  \label{fig:sim-p-K-d-sub2}
\end{subfigure}
\begin{subfigure}{.49\textwidth}
  \includegraphics[width=\linewidth]{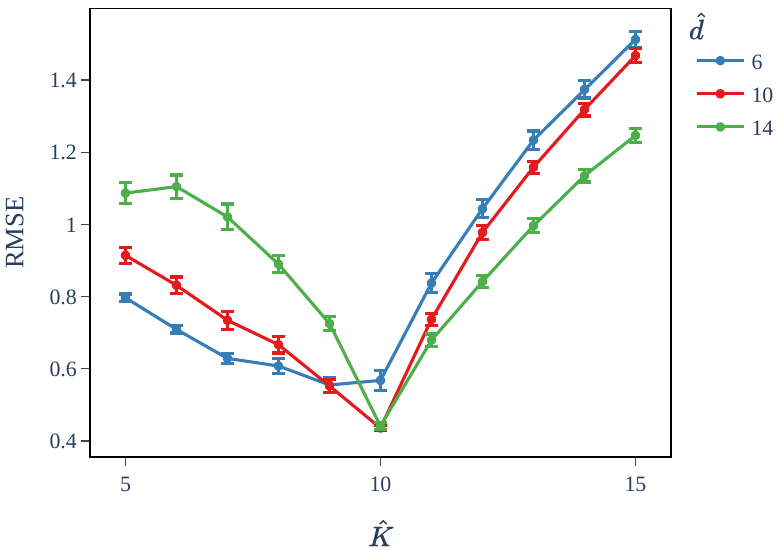}
  \caption{ }
  \label{fig:sim-p-K-d-sub3}
\end{subfigure}
\begin{subfigure}{.49\textwidth}
  \includegraphics[width=\linewidth]{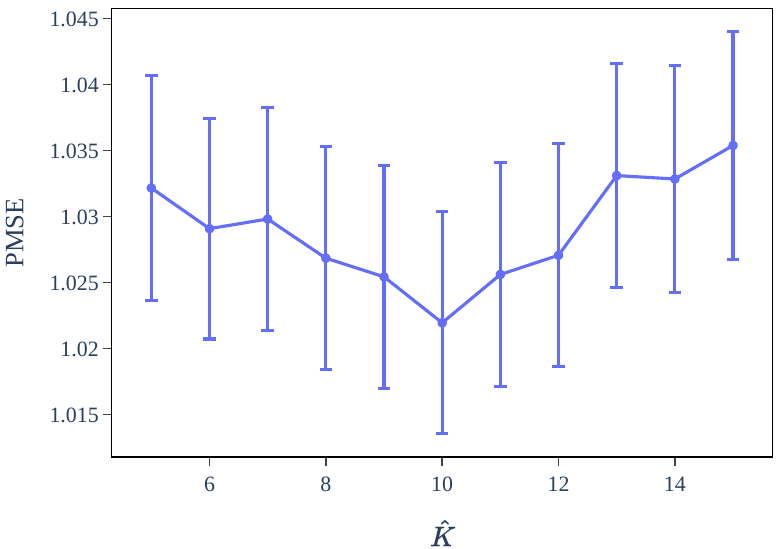}
  \caption{ }
  \label{fig:sim-p-K-d-sub4}
\end{subfigure}
\caption{(a) Spectral clustering using different autocovariance matrices under a NIRVAR data generating process with $p=3$. The sensitivity of (b) ARI and (c) RMSE to different values of $\hat{d}$ and $\hat{K}$. (d) Selecting $\hat{K}$ as the value which minimises the one-step-ahead PMSE over a validation set.  } 
\label{fig:sim-p-K-d}
\end{figure}

We simulate data from a NIRVAR model with $N=100$, $T=5000$, $Q=1$, $K=10$, $d=10$, $\rho(\Phi) = 0.9$, $p^{(\text{in})} = 1$, and $p^{(\text{out})} = 0$. For different values of $\hat{K}$ and $\hat{d}$, we compute the ARI between the NIRVAR estimated clusters and the ground truth clusters, as well as $\text{RMSE} = \lVert \hat{\Phi} - \Phi \rVert_{F}$. For each value of $\hat{K}$ and $\hat{d}$, we generate 45 replica datasets and calculate the corresponding standard errors of the ARI and RMSE values. Figure \ref{fig:sim-p-K-d} (\subref{fig:sim-p-K-d-sub2}) shows that the ARI is more sensitive to under-specifying $\hat{K}$ than over-specifying $\hat{K}$, while the ARI is not very sensitive to $\hat{d}$. The ARI is less sensitive to over-specifying $\hat{K}$ because the Gaussian mixture model can create clusters with very few nodes, with the majority of nodes still correctly clustered. Figure \ref{fig:sim-p-K-d} (\subref{fig:sim-p-K-d-sub3}) shows that the RMSE is sensitive to both under and over-specifying $K$. Under-specifying $\hat{K}$ reduces bias with a corresponding increase in variance of $\hat{\Phi}$. In the extreme case with $\hat{K} = 1$, $\hat{\Phi}$ is unbiased and a low RMSE can be achieved given large enough $T/N$. In this simulation, $T/N = 50$ is large, hence the RMSE for $\hat{K} < K$ increases at a similar rate (or slightly slower) than the RMSE for $\hat{K}>K$. We note, however, that as $T/N$ decreases, the increase in variance of $\hat{\Phi}$ will be larger for the $\hat{K} < K$ regime than for the $\hat{K}>K$ regime.

\subsection{Selecting the number of clusters via cross validation}
Instead of choosing the embedding dimension using the Mar\v{c}enko-Pastur distribution and setting $K=d$, one could instead choose $K$ and $d$ using a cross-validation approach. This would involve splitting your data into training and validation sets, backtesting over the validation set using a sliding window approach, computing some statistic of interest for each value of $(K,d)$ in your grid search on each backtesting day, and averaging over the validation set. As an example, we generate data from a NIRVAR model with $K=10$, and use cross validation to choose the value of $\hat{K} \in \{5,6,\dots,15\}$ that minimises the average one-step-ahead predictive mean squared error (PMSE) over the validation set. The hyperparameters of the NIRVAR data generating process were $N=75$, $T=2300$, $Q=1$, $K=10$, $p^{(\text{in})} = 1$,  $p^{(\text{out})} = 0$, $\rho(\Phi) = 0.7$, and $\Tilde{\Phi} \propto \bm{1}\bm{1}^{\prime}$. The validation set was chosen to have 400 observations, with the look-back window of the backtesting experiment having 1900 observations. Figure \ref{fig:sim-p-K-d} (\subref{fig:sim-p-K-d-sub4}) shows that the value of $\hat{K}$ that minimises the PMSE is $\hat{K}=10$, and thus the number of clusters selected using cross validation matches the ground truth. 

\subsection{Robustness of NIRVAR clustering to heteroskedastic and non-diagonal error matrices}\label{subsec:robust-clustering}
\begin{figure}[ht]
\centering
\begin{subfigure}{.32\textwidth}
  \centering
  \includegraphics[width=.95\linewidth]{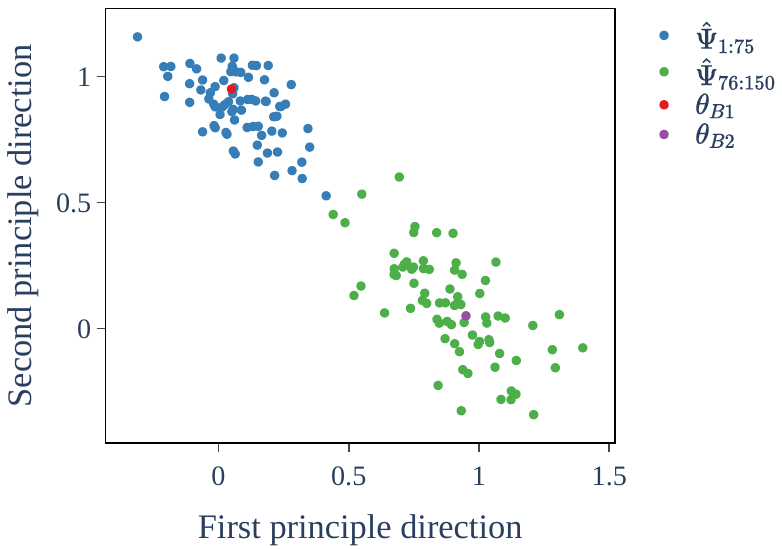}
  \caption{}
  \label{fig:sim-hetero-cluster-sub1}
\end{subfigure}
\begin{subfigure}{.32\textwidth}
  \centering
  \includegraphics[width=.95\linewidth]{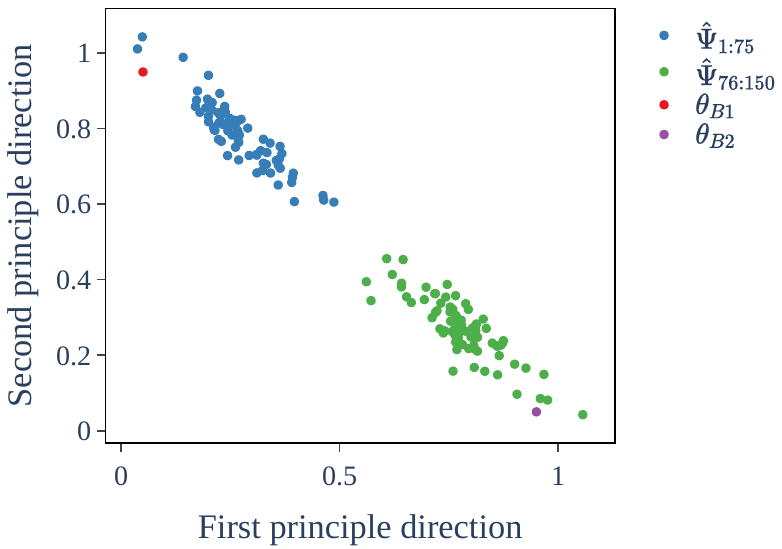}
  \caption{}
  \label{fig:sim-hetero-cluster-sub2}
\end{subfigure}
\begin{subfigure}{.32\textwidth}
  \centering
  \includegraphics[width=.95\linewidth]{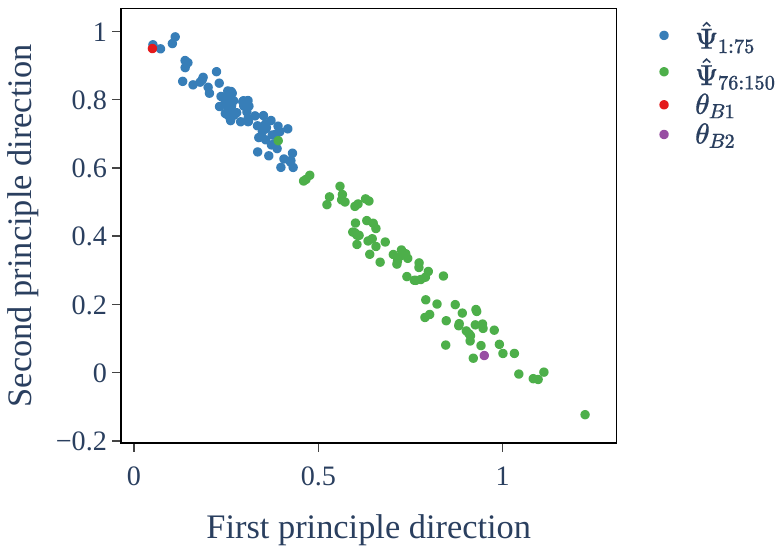}
  \caption{}
  \label{fig:sim-hetero-cluster-sub3}
\end{subfigure}
\begin{subfigure}{.32\textwidth}
  \centering
  \includegraphics[width=.95\linewidth]{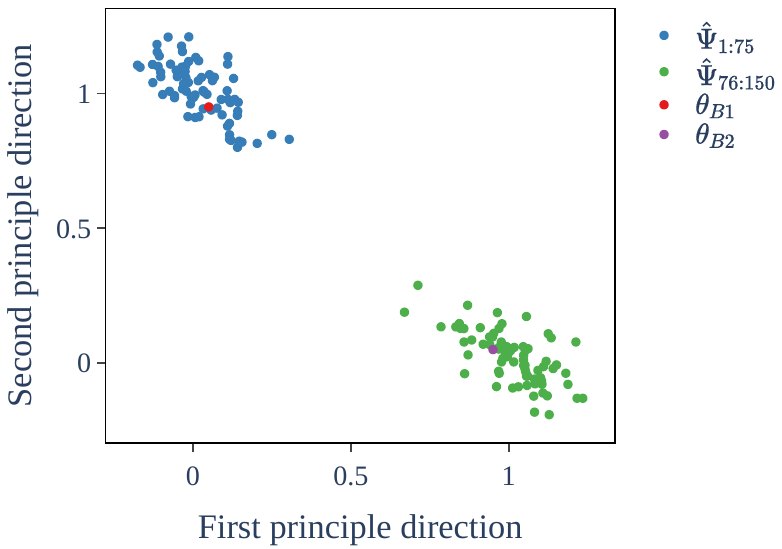}
  \caption{}
  \label{fig:sim-hetero-cluster-sub4}
\end{subfigure}
\begin{subfigure}{.32\textwidth}
  \centering
  \includegraphics[width=.95\linewidth]{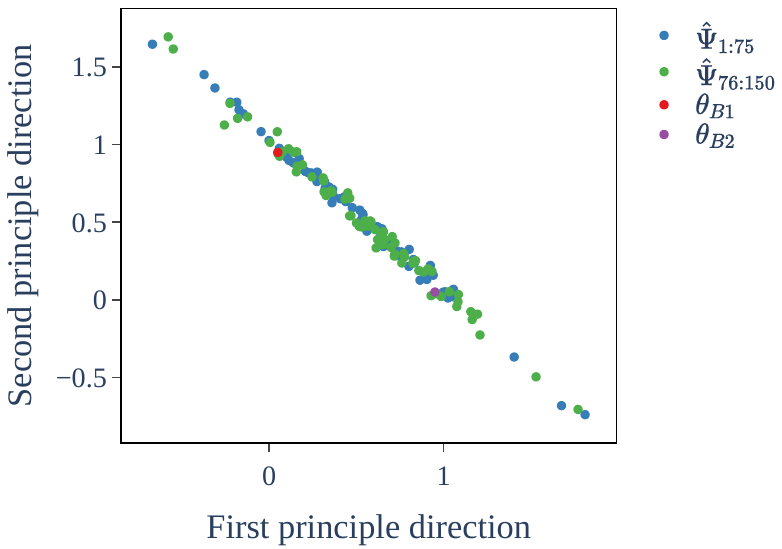}
  \caption{}
  \label{fig:sim-hetero-cluster-sub5}
\end{subfigure}
\begin{subfigure}{.32\textwidth}
  \centering
  \includegraphics[width=.95\linewidth]{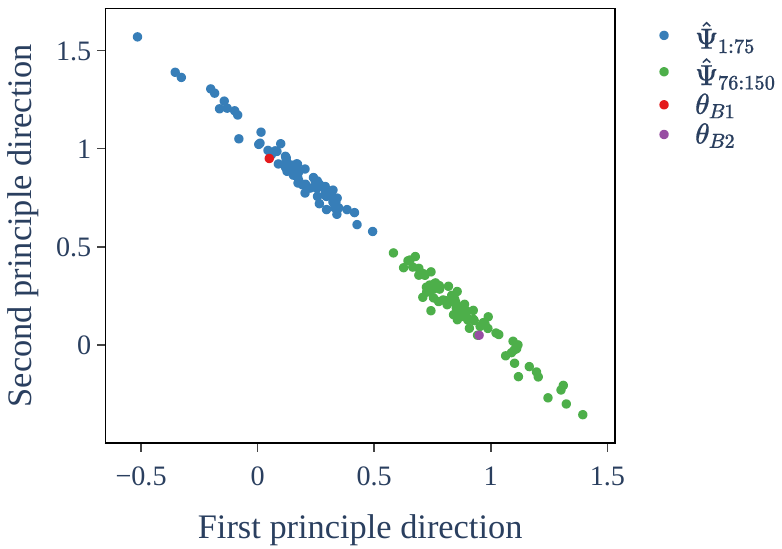}
  \caption{}
  \label{fig:sim-hetero-cluster-sub6}
\end{subfigure}
\caption{NIRVAR embedded latent points (scaled and rotated) and true latent positions from the simulated SBM under different types of error covariances.}
\label{fig:sim-hetero-cluster}
\end{figure}

In order to assess whether NIRVAR clustering can recover community structure when the error covariance matrix is not proportional to $I_{N}$, we repeat the simulation study of Section \ref{subsec:latent-sim} with the following choices for $\Sigma$: (i) $\Sigma_{ij} = \sigma_{i}^{2}\indfun\{i=j\}$ where $\sigma_{i} \sim \text{Uniform}(0.5,1)$, (ii) $\Sigma_{ii} = \sigma^{2}$ for $i=j$ and $\Sigma_{ij} = \tau$ where $\tau = 0.25$ for $i \neq j$, (iii) $\Sigma_{ii} = \sigma_{i}^{2}$ for $i=j$ and $\Sigma_{ij} = \tau$ for $i \neq j$, (iv) $\Sigma_{ii} = \sigma^{2}$ for $i=j$ and $\Sigma_{ij} = \tau^{|i-j|}$ for $i \neq j$. Figure \ref{fig:sim-hetero-cluster}(\subref{fig:sim-hetero-cluster-sub1}) shows that the NIRVAR embedded points are centered around the ground truth latent positions under the diagonal heteroskedastic error matrix (i). 

Figure \ref{fig:sim-hetero-cluster}(\subref{fig:sim-hetero-cluster-sub2}) and Figure \ref{fig:sim-hetero-cluster}(\subref{fig:sim-hetero-cluster-sub3}), corresponding to (ii) and (iii), show that the presence of off-diagonal entries in $\Sigma$ pulls the NIRVAR embedded cluster centers away from the ground truth latent positions. Crucially, however, the clusters remain distinct, and therefore downstream NIRVAR estimation tasks will be unaffected by the presence of off-diagonal entries in the error covariance matrix. Figure \ref{fig:sim-hetero-cluster}(\subref{fig:sim-hetero-cluster-sub4}), corresponding to the Toeplitz matrix (iv), shows that if the off-diagonal entries of $\Sigma$ decay exponentially as one moves away from the main diagonal, then the NIRVAR embedded points recover the ground truth latent positions. In all examples of heteroskedastic and non-diagonal $\Sigma$, the clusters remain well separated. 

We also assess the robustness of NIRVAR clustering to heavy-tailed noise. We sample the noise process from a multivariate Student's-t distribution with 7 degrees of freedom and choose $\Sigma$ to be equal to (iii) above. Figure \ref{fig:sim-hetero-cluster}(\subref{fig:sim-hetero-cluster-sub5}) shows that with $T=2000$, the clusters are not well separated. However, Figure \ref{fig:sim-hetero-cluster}(\subref{fig:sim-hetero-cluster-sub6}) shows that if we increase to $T=4000$, the clusters become well separated. Thus, in the presence of heavy tailed errors, we require a larger value of $T/N$ to recover the underlying community structure.

\subsection{Predictive performance of NIRVAR compared with existing methods}\label{subsec:predict-sim}
We generate data from a NIRVAR model and compare the predictive performance of NIRVAR with an unrestricted VAR, a Bayesian VAR model with a Minnesota prior, and a LASSO-regularised VAR model where the $\ell_{1}$ penalty is selected via cross-validation. The hyperparameters of the NIRVAR data generating process were $N=100$, $Q=1$, $K=10$, $p^{(\text{in})} = 1$, $p^{(\text{out})} = 0.1$, $\rho(\Phi) = 0.9$, and $\Tilde{\Phi}$ is drawn from a Wishart distribution. The number of training points is $T \in \{50,75,100,200,300,10000\}$. We performed a backtesting experiment for 50 time-steps with a sliding look-back window equal to $T$. The mean squared prediction error (MSPE) over the backtesting period along with the corresponding standard error is reported for each prediction model in Table \ref{tab:NIRVAR-horserace} for different values of $T/N$. Note that a VAR(1) needs $T>N$ for estimation to be possible.

NIRVAR achieves the lowest MSPE compared with the other models in every case except when $T/N = 0.5$ and when $T/N = 10,000$. In very high dimensional settings such as $T/N = 0.5$, LASSO benefits from tuning the regularisation penalty to induce sparsity. This increases bias but reduces variance, which is the dominant contributing factor to the overall MSPE in such high dimensions. In contrast, the NIRVAR estimator does not have a penalisation term that can be tuned so as to optimise the bias-variance trade-off for prediction. 

In the low-dimensional regime $T/N = 100$, it is the bias which is the driver of MSPE. Now an unrestricted VAR(1) is unbiased, the LASSO penalisation can be chosen so that the bias is removed, and the tightness of the Minnesota prior can also be chosen so that parameter shrinkage and therefore bias, is minimised. In contrast, the NIRVAR estimator remains biased since there are inter-block edges present in the data generating process. This explains why NIRVAR has the highest MSPE in the low dimensional regime. For all other values of $T/N$, NIRVAR achieves the lowest MSPE. 

For practitioners interested in forecasting in a low-dimensional regime, corresponding to $T = O(N^{2})$, we recommend adding a penalisation term for inter-block parameters to the NIRVAR estimator so as to reduce bias. This modification to the NIRVAR estimator is described in detail in Section \ref{subsec:relaxing}. In very high dimensional settings ($T \ll N$), practitioners should be aware that NIRVAR does not have tuning parameters that can be adapted so as to reduce variance. Therefore, alternatives such as LASSO and Bayesian VAR, whose shrinkage hyperparameters can be tuned, should be considered. 

\begin{table}[t]
\centering
\scalebox{0.9}{
\begin{tabular}{c|cccccc|cccccc}
    \toprule
    Metric
    & \multicolumn{6}{c|}{MSPE} 
    & \multicolumn{6}{c}{Standard error} \\
    \midrule
    \diagbox{Model}{$T/N$}
    & 100 & 3 & 2 & 1 & 0.75 & 0.5 & 100 & 3 & 2 & 1 & 0.75 & 0.5  \\
    \midrule
    NIRVAR & 0.71 & \textbf{0.78} & \textbf{0.79} & \textbf{0.84} & \textbf{0.93} & 1.17 & 0.12 & 0.12 & 0.13 & 0.13 & 0.14 & 0.15 \\
    VAR(1) & \textbf{0.68} & 1.14 & 1.55 & NA & NA & NA & 0.12 & 0.15 & 0.18 & NA & NA & NA \\
    LASSO & \textbf{0.68}  & 0.82 & 0.85 & 0.91 & 0.95 & \textbf{1.00} & 0.12 & 0.12 & 0.13 & 0.13 & 0.14 & 0.14 \\
    Bayesian VAR & \textbf{0.68} & 0.86 & 0.97 & 2.05 & 2.15 & 1.74 & 0.12 & 0.13 & 0.14 & 0.20 & 0.21 & 0.19  \\
    \bottomrule
\end{tabular}
}
\caption{The MSPE and corresponding standard error of four different prediction models on data generated from a NIRVAR model. The backtesting experiment was run for different values of $T/N$. }
\label{tab:NIRVAR-horserace}
\end{table}

\subsection{Relaxing hard-thresholding in NIRVAR estimation via penalisation}\label{subsec:relaxing} 

\begin{figure}[t]
\centering
\begin{subfigure}{.32\textwidth}
  \includegraphics[width=\linewidth]{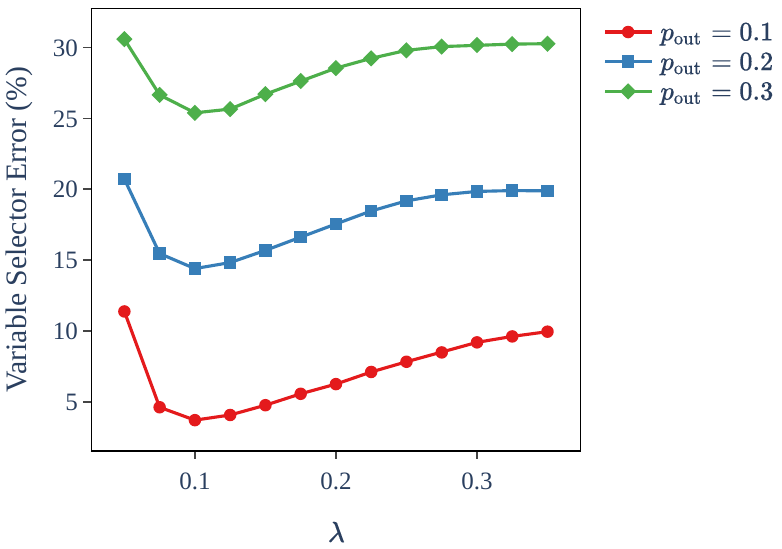}
  \caption{}
  \label{fig:sim-penalise-sub1}
\end{subfigure}
\begin{subfigure}{.32\textwidth}
  \includegraphics[width=\linewidth]{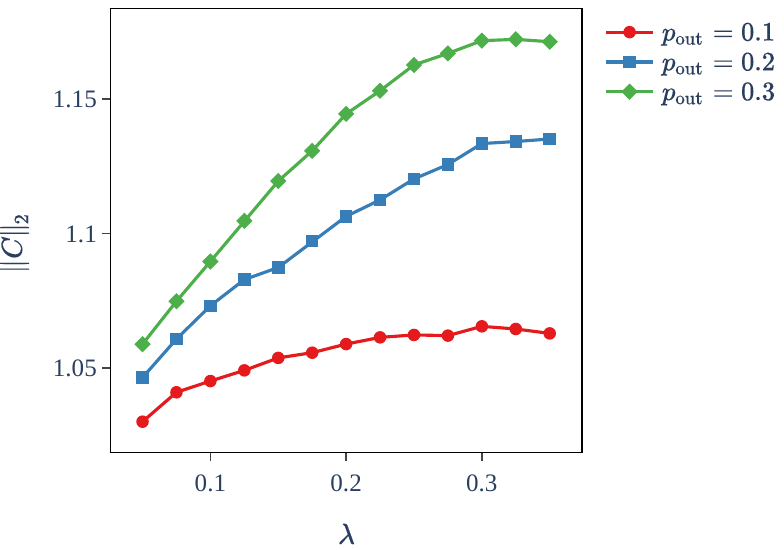}
  \caption{}
  \label{fig:sim-penalise-sub2}
\end{subfigure}
\begin{subfigure}{.32\textwidth}
  \includegraphics[width=\linewidth]{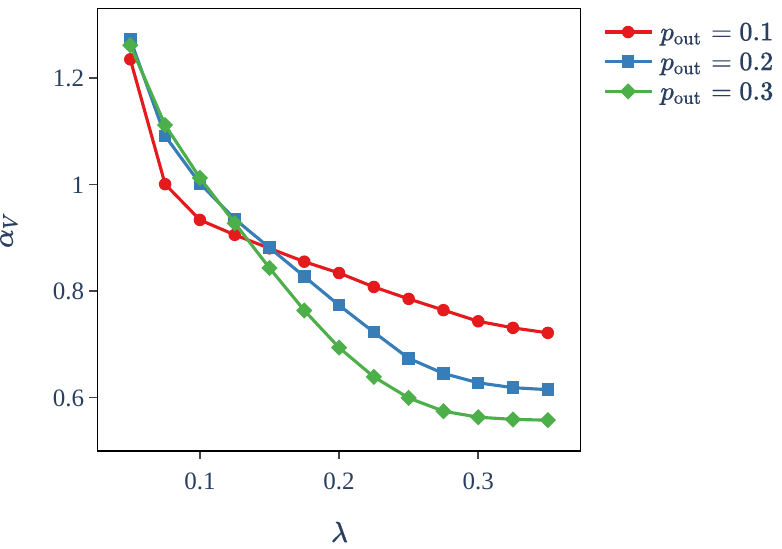}
  \caption{ }
  \label{fig:sim-penalise-sub3}
\end{subfigure}
\caption{The effect of adding an $\ell_{1}$ penalty $\lambda$ to inter-block edges on (a) the percentage of incorrect inter-block edges selected, (b) the bias of the estimator, and (c) the variance of the estimator relative to an estimator that uses the correct ground truth restrictions.} 
\label{fig:sim-penalise}
\end{figure}

The NIRVAR estimation procedure restricts inter-block entries of the VAR coefficient matrix to be zero. We can relax this assumption by adding an $\ell_{1}$ penalty to inter-block entries of $\Phi$. The estimator then becomes 
\begin{equation}
    \hat{\Phi} = \argmin_{\Phi} \text{tr} \{(Y^{(q)} - \Phi X)^{\prime} (Y^{(q)} - \Phi X)\} + \sum_{i,j=1}^{N} \lambda(1 - \hat{A}_{ij}) |\Phi_{ij}| 
\end{equation}
Note that if $\lambda \to \infty$, then we recover the NIRVAR estimator, while $\lambda = 0$ gives unrestricted OLS estimation. In order to investigate the benefits and drawbacks of including inter-block penalisation, we simulate from a NIRVAR model with $N = 50$, $T=10000$, $K=5$, $Q=1$, $p^{(\text{in})} = 1$, and $p^{(\text{out})} \in \{0.1,0.2,0.3\}$. We then calculate the variable selection error, bias $C$, and variance $\alpha_{V}$ for different values of $\lambda$. The variable selection error is defined as $100 \times \sum_{i,j : \hat{z}_{i} \neq \hat{z}_{j}} \indfun\{\hat{A}_{ij} \neq A_{ij}\}/N_{\text{inter}}$ where $N_{\text{inter}}$ is the number of inter-block entries of $\Phi$. The bias $C$ is defined in Equation \eqref{eq:bias} while $\alpha_{V} \coloneqq \text{tr}[\text{var}\{\bm{\hat{\beta}}(\hat{A})\}] / \text{tr}[\text{var}\{\bm{\hat{\beta}}(A)\}]$ measures the extent to which the asymptotic variance of the estimator is inflated. Figure \ref{fig:sim-penalise}(\subref{fig:sim-penalise-sub1}) shows that including a penalisation term can reduce the variable selection error by about 5\% for each value of $p^{(\text{out})} \in \{0.1,0.2,0.3\}$. Figure \ref{fig:sim-penalise}(\subref{fig:sim-penalise-sub2}) shows that the bias increases with $\lambda$ while Figure \ref{fig:sim-penalise}(\subref{fig:sim-penalise-sub3}) shows a corresponding decrease in variance with $\lambda$. Practitioners must note this bias-variance trade-off when deciding whether or not to include this soft-thresholding via penalisation in the NIRVAR estimation framework. In practice, $\lambda$ can be chosen using cross validation or some information criterion, for example. 

\subsection{Stability of Expectation-Maximisation for 
GMM 
clustering}
\begin{figure}[t]
\centering
\begin{subfigure}{.32\textwidth}
  \includegraphics[width=\linewidth]{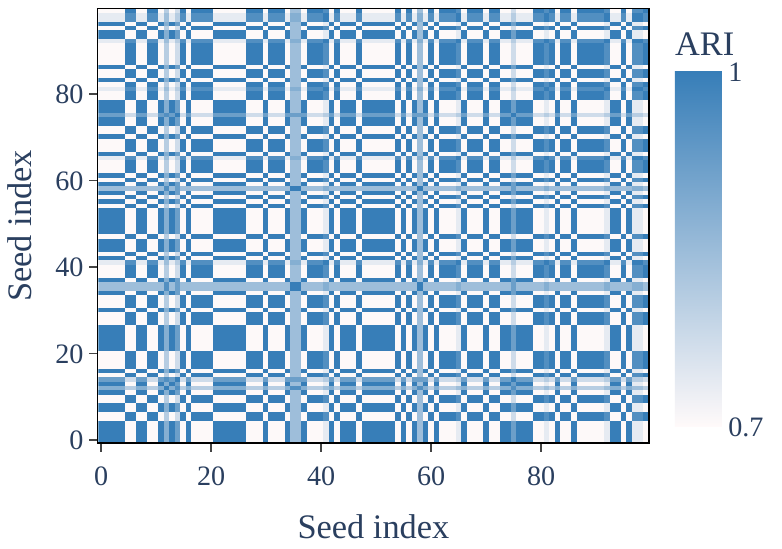}
  \label{fig:sim-EM-sub1}
\end{subfigure}
\begin{subfigure}{.32\textwidth}
  \includegraphics[width=\linewidth]{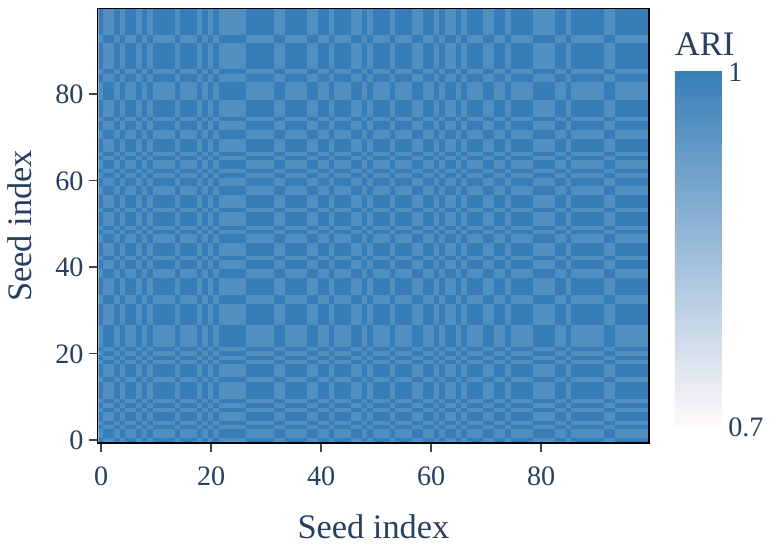}
  \label{fig:sim-EM-sub2}
\end{subfigure}
\begin{subfigure}{.32\textwidth}
  \includegraphics[width=\linewidth]{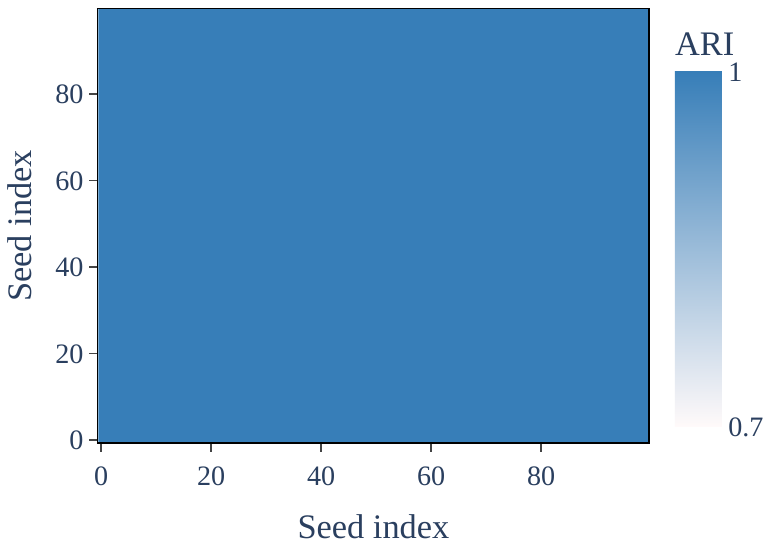}
  \label{fig:sim-EM-sub3}
\end{subfigure}
\begin{subfigure}{.32\textwidth}
  \includegraphics[width=\linewidth]{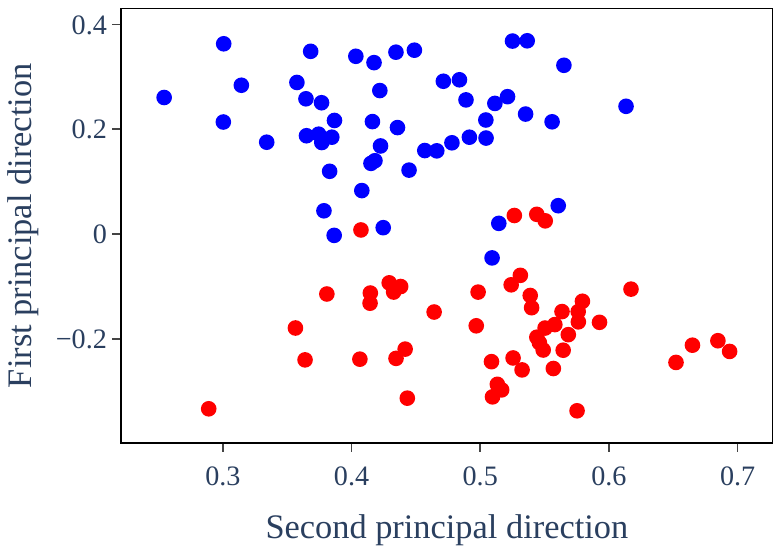}
  \caption{$T=1000$}
  \label{fig:sim-EM-sub1}
\end{subfigure}
\begin{subfigure}{.32\textwidth}
  \includegraphics[width=\linewidth]{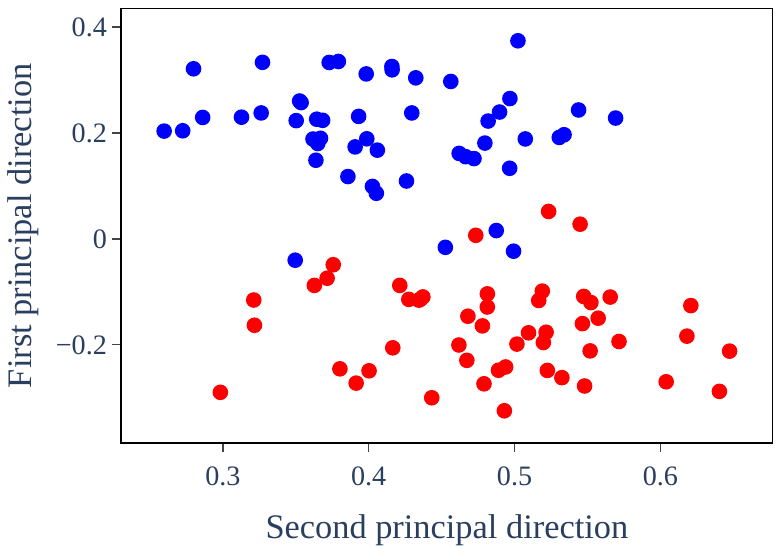}
  \caption{$T=1200$}
  \label{fig:sim-EM-sub2}
\end{subfigure}
\begin{subfigure}{.32\textwidth}
  \includegraphics[width=\linewidth]{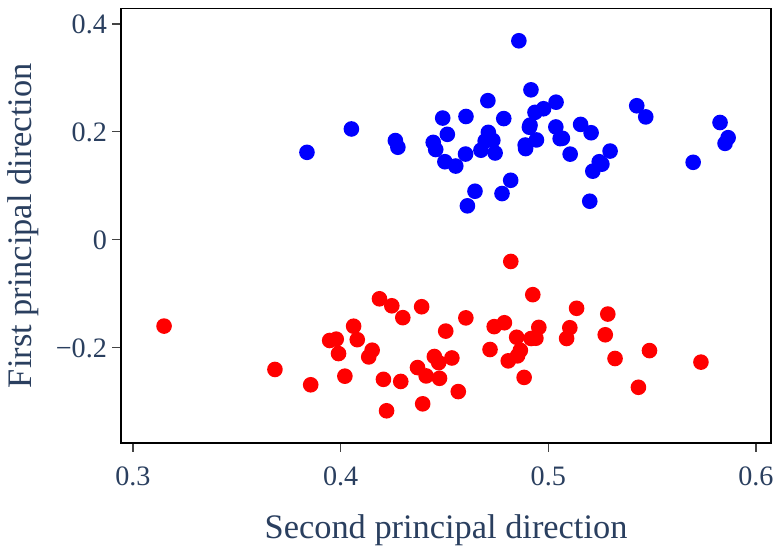}
  \caption{$T=3000$ }
  \label{fig:sim-EM-sub3}
\end{subfigure}
\caption{The ARI between 100 different clusterings of NIRVAR generated data corresponding to different seed initialisations of the Expectation-Maximisation algorithm for Gaussian mixture modelling. Also shown are the corresponding latent embeddings, coloured by the ground truth cluster labels. } 
\label{fig:sim-EM}
\end{figure}

In order to assess the stability of the Expectation-Maximisation algorithm, we simulate data from a NIRVAR model and compare the ARI of clusterings obtained using different seeds to initialise the algorithm. The data generating process used $N=100$, $T \in \{1000,1200,3000\}$, $Q=1$, $K=2$, $p^{(\text{in})} = 0.9$, $p^{(\text{out})} = 0.1$, and $\rho(\Phi) = 0.9$.  Figure \ref{fig:sim-EM} shows heatmaps of the ARIs for 100 different seed initialisations. As $T$ increases, the 100 different seed initialisations converge to a common clustering. The mean (and corresponding sample standard deviation) ARI between the 100 clusterings and the ground truth clusters was $0.81 (0.01)$ for $T=1000$, $0.82 (0.02)$ for $T=1200$, and $1$ for $T=3000$. Figure \ref{fig:sim-EM} also shows the embedded points coloured by their ground truth cluster labels. Looking at the embedded points, we see that for $T=1000$ there is overlap between the two groups whereas for $T=3000$ the clusters are well separated. Practitioners can make the clustering step of NIRVAR more robust by calculating the mean number of times vertices $i$ and $j$ appear in the same cluster over $N_{\text{seed}}$ initialisations of the Expectation-Maximisation algorithm and restricting $\Phi_{ij}$ to zero only when the mean drops below a pre-defined threshold. 

\subsection{Spectral radius}
\begin{figure}[t]
\centering
\begin{subfigure}[b]{.24\textwidth}\includegraphics[width=\linewidth]{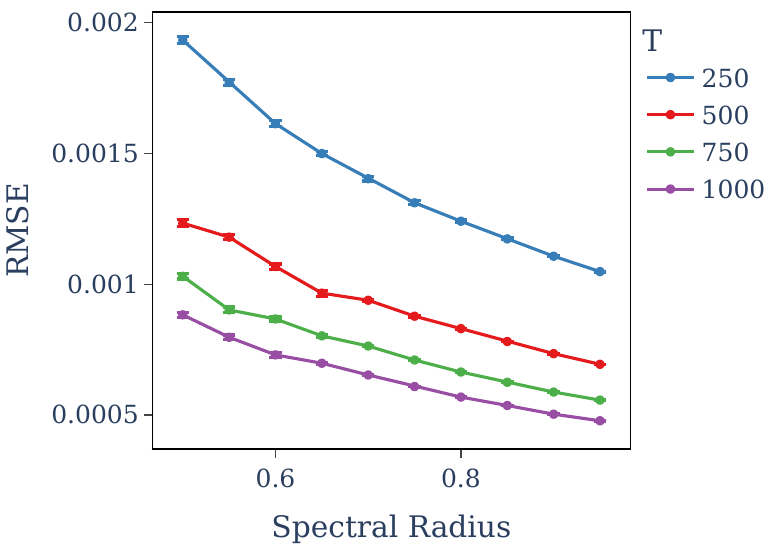}
  \caption{RMSE, $K=2$}
  \label{fig:sim-spectral-sub1}
\end{subfigure}
\begin{subfigure}[b]{.24\textwidth}\includegraphics[width=\linewidth]{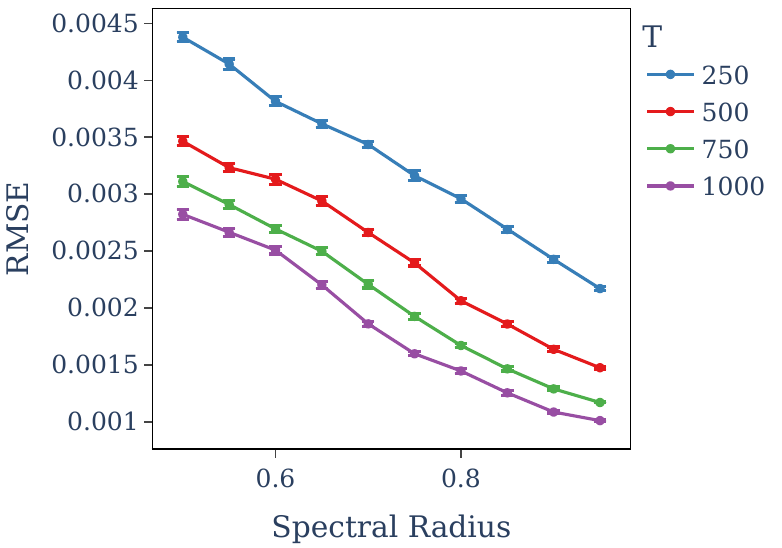}
  \caption{RMSE, $K=10$}
  \label{fig:sim-spectral-sub2}
\end{subfigure}
\begin{subfigure}[b]{.24\textwidth}\includegraphics[width=\linewidth]{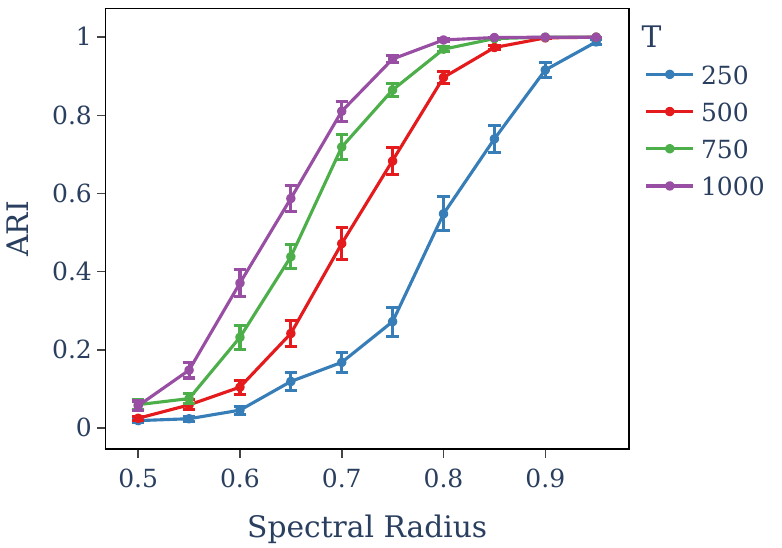}
  \caption{ARI, $K=2$}
  \label{fig:sim-spectral-sub3}
\end{subfigure}
\begin{subfigure}[b]{.24\textwidth}\includegraphics[width=\linewidth]{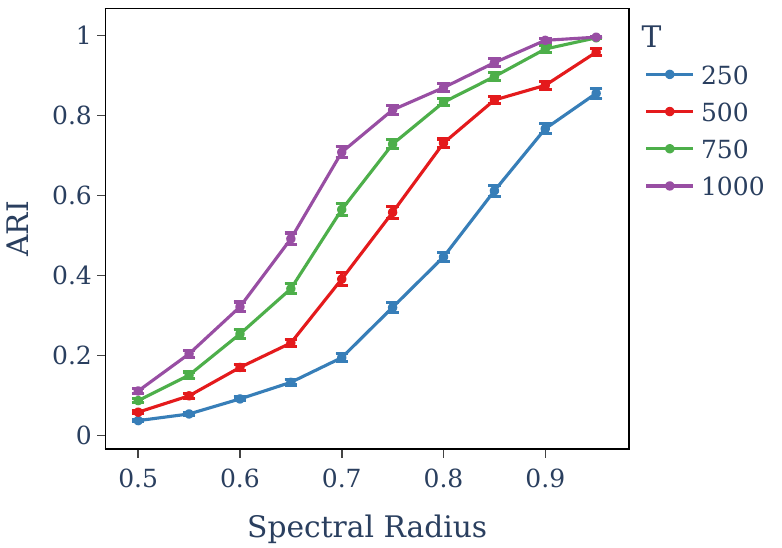}
  \caption{ARI, $K=10$}
  \label{fig:sim-spectral-sub4}
\end{subfigure}
\caption{(a)-(b) Normalised RMSE of $\hat{\Phi}$ and (c)-(d) ARI as $\rho(\Phi)$ increases.}
\label{fig:sim-spectral}
\end{figure}

In this study, we check whether NIRVAR cluster recovery and parameter estimates improve as the spectral radius of the ground truth VAR coefficient matrix $\rho(\Phi)$ increases. 
We expect this to be the case since $\rho(\Phi)$ controls the signal-to-noise ratio of the data generating process.
The hyperparameters used to generate data from the NIRVAR model were $N=100$, $Q=1$, $K \in \{2,10\}$, and $T \in \{250,500,750,1000\}$. To set $\Tilde{\Phi}$, we sampled each entry $\Tilde{\Phi}_{ij}$ independently from a Uniform(0,1) distribution and then normalised such that $\rho(\Phi) \in \{0.5,0.55,0.6,0.65,0.7,0.75,0.8,0.85,0.9,0.95\}$.
For each combination of $\rho(\Phi)$ and $T$, we sampled $\bm{X}_{t}^{(q)} \sim \mathrm{NIRVAR}(\Phi)$, obtained an estimate $\hat\Phi$ of $\Phi$ from the simulated data, and computed the normalised root mean squared error 
$\text{NRMSE} = \lVert \hat{\Phi} - \Phi \rVert_{F}/(\widehat{M} \rho(\Phi))$ and the Adjusted Rand Index \citep[ARI,][]{hubert1985comparing}, a measure of similarity between two clusterings of data. ARI ranges from -1 to 1, with higher values indicating stronger similarity. 
Averages over 15 replications are obtained, with the corresponding standard errors.  
Figure \ref{fig:sim-spectral} (\subref{fig:sim-spectral-sub1}) and (\subref{fig:sim-spectral-sub2}) show that the NRMSE decreases both as $\rho(\Phi)$ increases and as $T$ increases. Figure \ref{fig:sim-spectral} (\subref{fig:sim-spectral-sub3}) and (\subref{fig:sim-spectral-sub4}) show that the ARI approaches~1 as the spectral radius approaches 1, meaning that the true clusters are recovered in a high signal-to-noise regime.

\section{Applications}

This section reports some additional results that 
support the findings discussed in Section \ref{sec:applications} of the main document. 
        
\subsection{FARM, FNETS and GNAR}

For convenience, 
we report the specification of the three models used for benchmarking NIRVAR.

\begin{enumerate}
    \item \textbf{FARM} \citepSM{fan2023bridging}. In FARM, the stochastic process $X_{i,t}$ is modelled as 
    \begin{align}
        \label{eq:FARM}
        X_{i,t} = \mu_{i} + \lambda_{i} \sum_{\ell = 1}^{L} \tau_{\ell} f_{t-\ell} + \sum_{\ell=1}^{L} \sum_{j=1}^{N} (\bm{\Delta}_{\ell})_{ij} \xi_{j,t-\ell} + \epsilon_{i,t},
    \end{align}
    where $\mu_{i}$ is the sample mean of series $i$, $f_{t}$ is the leading factor at time $t$  with $\lambda_{i}$ the corresponding loading on this factor for series $i$, and $\xi_{j,t}$ is the residual value of series $i$ after subtracting the contribution of the factor. The parameters $\tau_{\ell}$ and $\bm{\Delta}_{\ell}$ are AR and VAR coefficients for the static factor model and idiosyncratic component, respectively. The factors and loadings are estimated using PCA, $\tau_{\ell}$ is estimated via OLS, and $\Delta_{j,\ell}$ is estimated using LASSO, with a penalty parameter selected via BIC. 
    \item \textbf{FNETS} \citepSM{barigozzi2023fnets}. In FNETS, the stochastic process $\bm{X}_{t}$ is modelled as a sum of two latent components: a factor-driven common component $\bm{\chi}_{t}$, modelled as a generalised dynamic factor model \citepSM{forni2000generalized}, and an idiosyncratic component $\bm{\xi}_{t}$, modelled as a VAR process. This results in the 
    model
    \begin{align}
        \label{eq:fnets}
        \bm{X}_{t} = \bm{\chi}_{t} + \bm{\xi}_{t},
        & &
        \bm{\chi}_{t} = \sum_{\ell=1}^{\infty} \bm{\Lambda}_{\ell} \bm{f}_{t-\ell}, & &
        \bm{\xi}_{t} = \sum_{l=1}^{L} \bm{\Delta}_{l} \bm{\xi}_{t-l} + \bm{\Gamma}^{1/2} \bm{\epsilon}_{t},
    \end{align}
    where $\mathbb{E}(\bm{\epsilon}_{t}) = 0$ and $\text{cov}(\bm{\epsilon}_{t}) = I_{N}$. Estimation of $\bm{\chi}_{t}$ proceeds by dynamic PCA, whilst for $\bm{\xi}_{t}$, \citeSM{barigozzi2023fnets} propose a $\ell_{1}$-regularised Yule-Walker estimator that requires second-order moments only.
    \item \textbf{GNAR} \citepSM{knight2020generalized}. In GNAR, a simplified model for $X_{i,t}$ takes the form
    \begin{align}
        \label{eq:GNAR}
        X_{i,t} = \sum_{\ell=1}^{L} \left( \alpha_{i,\ell} X_{i,t-\ell} + \beta_{\ell}\sum_{j \in \mathcal{N}(i)} X_{j,t-\ell}\right) + \epsilon_{i,t},
    \end{align}
    where $\mathcal{N}(i)$ denotes the set of vertices that are connected to vertex $i$ of a known graph and $\mathbb{E}(\epsilon_{i,t})=0$ with $\mathrm{var}(\epsilon_{i,t}) = \sigma_{i}^{2}$. The 
    full GNAR model in \citetSM{knight2020generalized} extends Equation~\eqref{eq:GNAR} allowing for a time varying known network, edge weights, multiple edge covariates, and stage-neighbours. A key difference between NIRVAR and GNAR is that GNAR assumes the network structure to be known, which is typically not the case in most real-world applications.
\end{enumerate}    

\subsection{Excess market returns}\label{subsec:excess-returns}

Tables \ref{tab:returns-sharpes-full} and  \ref{tab:regime-sharpes} report some additional results supporting the findings discussed in Section \ref{subsec:fin} of the main document. 
We first define some of the statistics used in Table \ref{tab:returns-sharpes-full}. Letting $\text{PnL}_{t}^{-} \coloneqq \{ \text{PnL}_{t} : \text{PnL}_{t} < 0 \}$ and $\text{PnL}^{-} \coloneqq \{ \text{PnL}_{t}^{-}  \}_{t=1,\dots,T}$, we define the Sortino ratio as
\begin{align}
    \label{eq:sortino}
    \text{SortR} = \frac{\text{mean}(\text{PnL})}{\text{stdev}(\text{PnL}^{-})} \times \sqrt{252}.
\end{align}
The maximum drawdown is defined as 
\begin{align}
    \label{eq:max-draw}
    \text{MaxDrawdown} = \max_{t,s}\left\{\frac{\text{CPnL}_{t} - \text{CPnL}_{s}}{\text{CPnL}_{t}} : t<s \right\},
\end{align}
where $\text{CPnL}_{t} = \sum_{r=1}^{t} \text{PnL}_{r}$. The hit ratio is the mean daily percentage of correct predictions, whereas the long ratio is the mean daily percentage of long predicted positions. The mean absolute error (MAE) and the root mean squared error (MSE) are defined as 
\begin{align}
    \label{eq:mae-rmse}
    \text{MAE} = \frac{1}{TN} \sum_{t=1}^{T} \sum_{i=1}^{N} \left|\hat{s}_{i}^{(t)} - s_{i}^{(t)}\right|, & & 
    \text{RMSE} = \sqrt{\frac{1}{TN} \sum_{t=1}^{T} \sum_{i=1}^{N} \left(\hat{s}_{i}^{(t)} - s_{i}^{(t)}\right)^{2}}.
\end{align}

\begin{table}[t]
\begin{tabular*}{\columnwidth}{@{\extracolsep\fill}llllllll@{\extracolsep\fill}}
\toprule
 & N C1  & N C2 & N P1 & N P2 & FARM & FNETS & GNAR\\
\midrule
Sharpe Ratio    & 2.50 & 2.34 & \textbf{2.82} & \underline{2.69} & 0.22 & 0.78 & 0.70 \\
Sortino Ratio    & 4.57 & 4.18 & \textbf{4.80} & \underline{4.63} & 0.36 & 1.39 & 1.13\\
Maximum Drawdown (\%)   & 111 & 170  & \underline{61} & \textbf{48} & 531 & 107 & 257 \\
Hit Ratio (\%)   & 50.6& 50.6 & \textbf{50.7} & \textbf{50.7} & 48.7 & 50.2 & 41.5 \\
Long Ratio (\%)   & \underline{50.1} & 49.8 & \underline{50.1} & \textbf{50.0} & 49.0& \underline{50.1} & 40.7  \\
Mean RMSE   & 0.018 & 0.019 & 0.018 & 0.019 & 0.018 & 0.018 & 0.018 \\
Mean Daily PnL (bpts)   & 2.68 & 2.49 & \textbf{3.00} & \underline{2.79} & 0.44 & 0.89 & 1.10 \\
Mean Absolute Error & 0.012 & 0.013 & 0.012 & 0.013 & 0.012 & 0.012 & 0.012\\
\bottomrule
\end{tabular*}
\caption{Statistics on the financial returns predictive performance of each model over the backtesting period. Note that ``N'' is an abbreviation of NIRVAR. The best performing model is shown in bold, whilst the second best performing model is underlined. 
\label{tab:returns-sharpes-full}}%
\end{table}

NIRVAR using the precision matrix embedding with or without an additional feature turn out to give the best results across models and indicators. 

\begin{table}[t]
\begin{tabular*}{\columnwidth}{@{\extracolsep\fill}llllllll@{\extracolsep\fill}}
\toprule
 & N C1  & N C2 & N P1 & N P2 & FARM & FNETS & GNAR\\
\midrule
01/01/2004 - 31/05/2007    & 4.44 & 3.09 & \textbf{5.14} & \underline{4.49} & 0.56 & 1.17 & 1.11 \\
01/06/2007 - 21/10/2010    & 3.87 & 4.28 & \textbf{5.08} & \underline{4.60}  & 0.02 & 2.05& 1.14 \\
22/10/2010 - 19/03/2014    & 1.91 & 1.41 & \textbf{3.10} & \underline{2.36} & -0.56 & 0.46& -0.22 \\
20/03/2014 - 09/08/2017    & 1.78 & \underline{2.04} & 1.41 & \textbf{2.38} & -0.37& -0.28& -0.06 \\
10/08/2017 - 31/12/2020    & \textbf{1.70} & 1.34 & \underline{1.37} & \underline{1.37} & 1.10 & 0.45& 1.07 \\
\bottomrule
\end{tabular*}
\caption{Annualised Sharpe ratios over five equally spaced periods. 
The best performing model is shown in bold, whilst the second best performing model is underlined.} \label{tab:regime-sharpes}%
\end{table}

Table \ref{tab:regime-sharpes} provides the Sharpe ratio of each model on five equally-spaced intervals of the backtesting period. NIRVAR performs particularly well around the time of the global financial crisis. We also note that NIRVAR P2 and NIRVAR C2 are the best performing models in the fourth sub-period considered, suggesting that the extra feature is beneficial in this period.

\subsection{US Industrial production}\label{subsec:appendix-fred}


Table \ref{tab:fred} reports the overall MSE, the sum of the MSE at points for which the model over (under)-estimated the realised value, and the sum of the MSE in extreme and non-extreme regimes where a realised value is defined to be extreme if its magnitude is above the $90$-th quantile.
NIRVAR, FARM and FNETS give equivalent results, with a slightly better performance of FNETS in non-crisis periods and NIRVAR both overall and in extreme regimes. 

\begin{figure}[ht]
\centering
\begin{subfigure}{.49\textwidth}
  \centering
  \includegraphics[width=.9\linewidth]{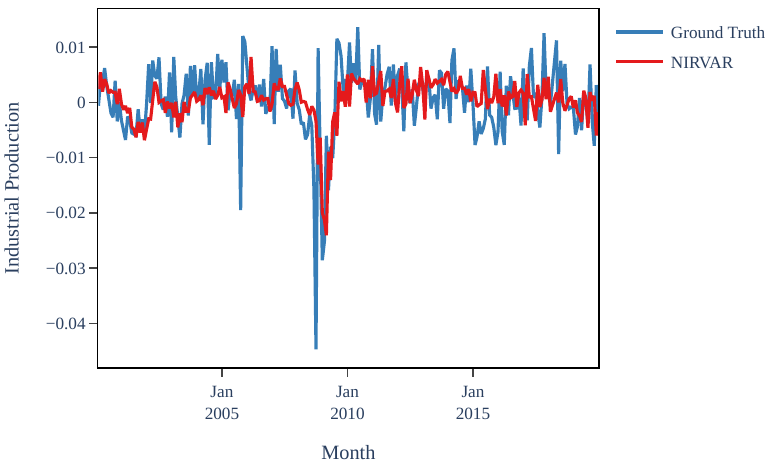}
  \caption{}
  \label{fig:fred-gt-sub1}
\end{subfigure}
\begin{subfigure}{.49\textwidth}
  \centering
  \includegraphics[width=.9\linewidth]{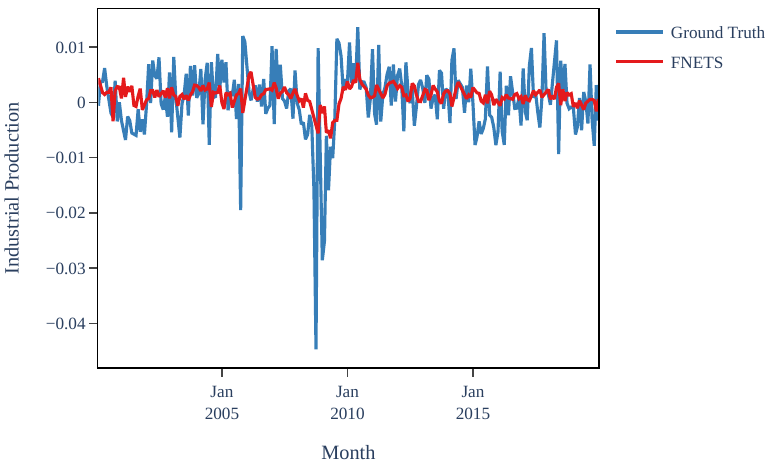}
  \caption{}
  \label{fig:fred-gt-sub2}
\end{subfigure}
\begin{subfigure}{.49\textwidth}
  \centering
  \includegraphics[width=.9\linewidth]{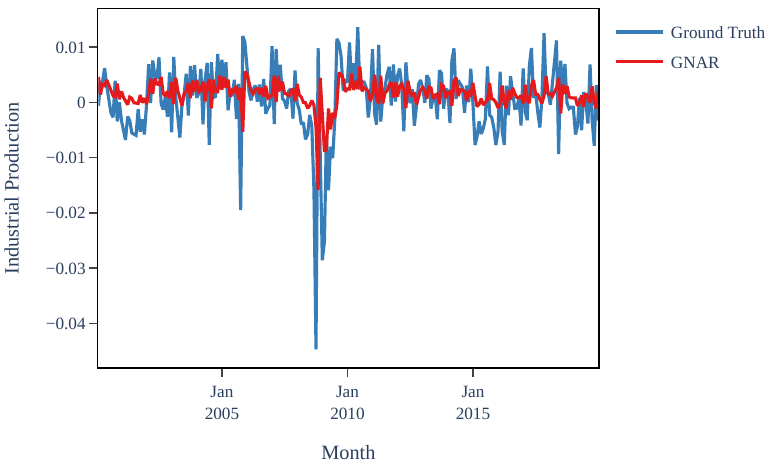}
  \caption{}
  \label{fig:fred-gt-sub3}
\end{subfigure}
\begin{subfigure}{.49\textwidth}
  \centering
  \includegraphics[width=.9\linewidth]{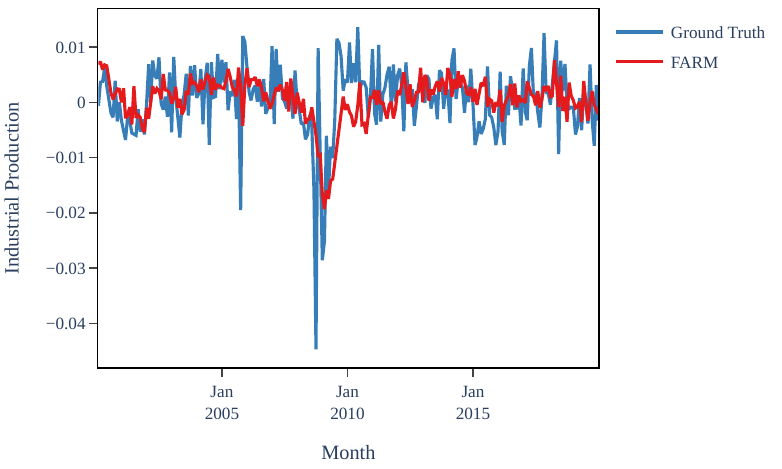}
  \caption{}
  \label{fig:fred-gt-sub4}
\end{subfigure}
\caption{Predicted (log-differenced) IP against the realised (log-differenced) IP for each model.}
\label{fig:fred-gt}
\end{figure}

Figure \ref{fig:fred-gt} shows the realised difference of the logarithm of IP against the corresponding prediction of each model. Figure \ref{fig:fred-gt}(\subref{fig:fred-gt-sub1}) and Figure \ref{fig:fred-gt}(\subref{fig:fred-gt-sub4}) show that both NIRVAR and FARM are more reactive to extreme values than FNETS and GNAR, shown in Figure \ref{fig:fred-gt}(\subref{fig:fred-gt-sub2}) and Figure \ref{fig:fred-gt}(\subref{fig:fred-gt-sub3}), respectively. 

We also compute the MSE for a VAR(12) model and find that it performs considerably worse than the network based models (the overall MSE was 0.0210). This is likely due to a lack of regularisation leading to high variance of the OLS estimator. Indeed, regularised estimation of a VAR(12) model using LASSO drastically improved the result, giving a MSE of 0.0089, which is comparable to the MSE's of FARM and NIRVAR. The LASSO hyperparameter was chosen using cross-validation.

\begin{table}[ht]
\begin{tabular*}{\columnwidth}{@{\extracolsep\fill}lllll@{\extracolsep\fill}}
\toprule
 & NIRVAR  & FARM & FNETS & GNAR\\
\midrule
Overall MSE    & \textbf{0.0087}   & \underline{0.0089}  & 0.0096 & 0.0101  \\
MSE: under-estimated    & \textbf{0.0070}   & \underline{0.0076}  & 0.0092 & 0.0087   \\
MSE: over-estimated    & 0.0017   & \underline{0.0013}  & \textbf{0.0003} & 0.0015  \\
MSE: non-extreme & 0.0042 & 0.0042 & \textbf{0.0039} & \underline{0.0041} \\
MSE: extreme & \textbf{0.0045} & \underline{0.0047} & 0.0057 & 0.0060 \\
\bottomrule
\end{tabular*}
\caption{Overall MSE and regime statistics 
for the task of forecasting US IP. The best performing model is shown in bold and the second best performing model is underlined.} \label{tab:fred}%
\end{table}

\subsection{Estimating the scale parameter of the Mar\v{c}enko-Pastur distribution} 
The scale parameter of the Mar\v{c}enko-Pastur distribution is chosen as the value of $\sigma^{2}$ that minimises the Kolmogorov-Smirnov statistic: 
\begin{align}
\label{eq:MP-best-fit}
    \sigma_{*}^{2} = \argmin_{\sigma^{2}}  \sup_{x} |F_{\text{MP}}(x;\eta,\sigma^{2}) - F_{N}(x)|,
\end{align}
where $F_{\text{MP}}(x;\eta,\sigma^{2})$ is the Mar\v{c}enko-Pastur cumulative distribution and $F_{N}(x) = k/N$, where $k$ is the number of observations less than or equal to $x$, is the observed cumulative step-function of the sample eigenvalues of the covariance matrix. The Mar\v{c}enko-Pastur cumulative distribution is obtained through numerical integration of $f_{\text{MP}}(x;\eta,\sigma^{2})$. The minimisation is done using the 
BFGS algorithm \citepSM[see, for example,][]{fletcher2000practical}. 

\begin{figure}[!h]
\begin{center}
\begin{subfigure}{.49\textwidth}
  \centering
  \includegraphics[width=.9\linewidth]{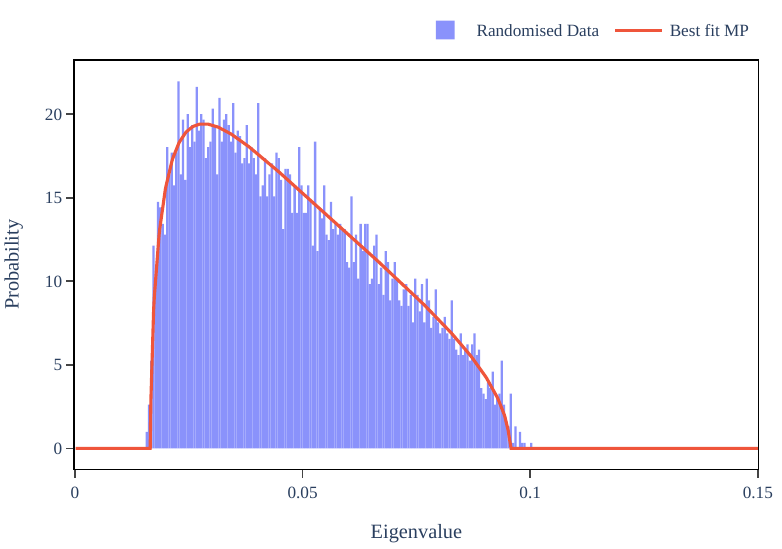}
  \caption{Covariance matrix}
  \label{fig:MP-fitting-sub1}
\end{subfigure}
\begin{subfigure}{.49\textwidth}
      \centering
      \includegraphics[width=0.9\linewidth]{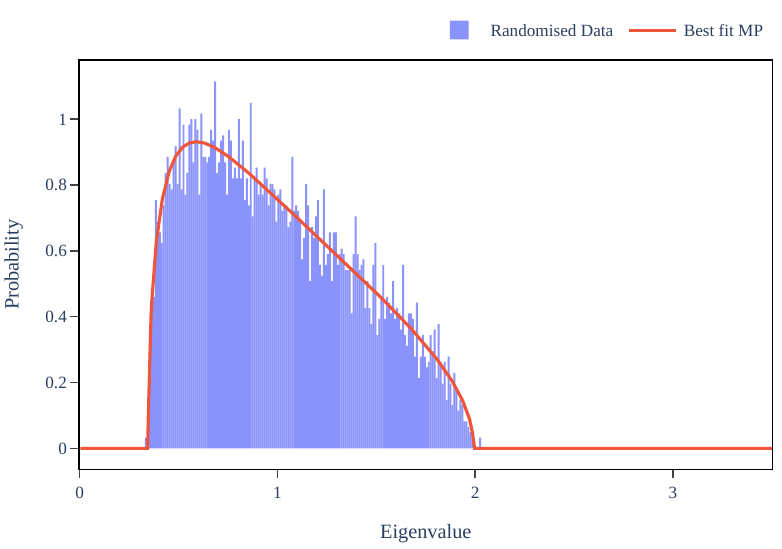}
    \caption{Correlation matrix}
  \label{fig:MP-fitting-sub2}
\end{subfigure}
\caption{(a) Best fit Mar\v{c}enko-Pastur distribution alongside the eigenvalues of the covariance matrix used in the FRED-MD application after the rows and columns of the design matrix were randomly permuted. (b) The Mar\v{c}enko-Pastur distribution with $\sigma^{2} = 1$ alongside the eigenvalues of the correlation matrix used in the FRED-MD application after the rows and columns of the design matrix were randomly permuted. } 
\label{fig:MP-fitting} 
\end{center}
\end{figure}

One can also consider embedding the correlation matrix instead of the covariance matrix. In this case, $\sigma^{2}$ is set equal to 1. As an example, Figure \ref{fig:MP-fitting}(\subref{fig:MP-fitting-sub1}) shows a histogram of the eigenvalues of the covariance matrix used in the FRED-MD application in Section \ref{subsec:FRED-MD} after the rows and columns of the design matrix were randomly permuted. The best fit Mar\v{c}enko-Pastur distribution obtained using Equation~\eqref{eq:MP-best-fit} is also plotted in Figure \ref{fig:MP-fitting}(\subref{fig:MP-fitting-sub1}) and fits the histogram well. Figure \ref{fig:MP-fitting}(\subref{fig:MP-fitting-sub2}) shows a histogram of the eigenvalues of the correlation matrix corresponding to the FRED-MD application in Section \ref{subsec:FRED-MD} after the rows and columns of the design matrix were randomly permuted. The Mar\v{c}enko-Pastur distribution with $\sigma^{2} = 1$ is also plotted in Figure \ref{fig:MP-fitting}(\subref{fig:MP-fitting-sub2}) and fits the histogram well. 


\section{Computational complexity} 
One of the advantages of NIRVAR is its computational efficiency compared with alternative methods such as penalised regression or Bayesian VAR methods which require hyperparameter tuning. Assuming balanced communities, $N_{k} \asymp N_{s}$ for $k,s \in [K]$ where $N_{k}$ is the number of vertices in block $k$, and assuming $\Sigma$ is diagonal (or block diagonal with the same block structure as $\Phi$), the overall computational complexity of the NIRVAR estimator is $O(N^{3}/K^{2} + TN^{2}/K)$ when $Q=1$.

To see this, we note that choosing the embedding dimension using the Mar\v{c}enko-Pastur method can be done using a Lanczos type algorithm in $O(TNd)$ time \citep{cullum2002lanczos}. \citet{gallagher2021spectral} show that unfolded adjacency spectral embedding is $O(TNd)$. Expectation-Maximisation with $n_{\text{iter}}$ iterations is $O(n_{\text{iter}}NKd^{2})$ assuming full covariances. Given the restrictions from the clustering step, NIRVAR estimation requires estimating $K$ decoupled VAR(1) models, each of dimension $O(N/K)$. The time complexity of each is $O(N^{3}/K^{3} + TN^{2}/K^{2})$. As such, the least-squares estimation dominates the computational complexity and the overall rate is $O(N^{3}/K^{2} + TN^{2}/K)$. In comparison, least-squares estimation of an unrestricted VAR is $O(N^{3}+ TN^{2})$.

In the case of general $\Sigma$, the complexity of feasible GLS, and therefore NIRVAR estimation is $O(N^{3} + TN^{2})$ since estimation of $\Sigma$ is $O(TN^{2})$ and inverting $\hat{\Sigma}$ is $O(N^{3})$. Therefore, NIRVAR estimation is faster than unrestricted VAR estimation only when $\Sigma$ is diagonal (as is the case in OLS, for example) or $\Sigma$ shares the same block structure as $\Phi$.

\end{appendices}

\bibliographystyleSM{rss}
\bibliographySM{reference}

\end{document}